\documentclass[bibyear]{aa}

\usepackage{aas_macros}

\usepackage{rotating}
\usepackage[section]{placeins}
\usepackage{pdfpages}
\usepackage{ulem}
\usepackage{array}
\usepackage[T1]{fontenc}
\usepackage{ae,aecompl}

\DeclareOldFontCommand{\sf}{\normalfont\sffamily}{\mathsf} 

\usepackage{color}

\DeclareGraphicsExtensions{.eps,.png,.pdf}

\usepackage{mathtools}

\usepackage{graphicx}	
\usepackage{amsmath}	
\usepackage{amssymb}	

\usepackage{adjustbox}

\definecolor{Antonio}{rgb}{1.0, 0.75, 0.0}

\definecolor{Doris}{rgb}{0.19, 0.55, 0.91}

\definecolor{Celeste}{rgb}{0.6, 0.4, 0.8}

\newcommand{\myemail}{doris.stoppacher@uam.es, dstoppacher@us.es}

\newcommand{\PUC}{Instituto de Astrof\'{i}sica, Pontificia Universidad Cat\'{o}lica de Chile, Campus San Joaqu\'{i}n, Avda. Vicu\~{n}a Mackenna 4860, Santiago, Chile}

\newcommand{\UAM}{Departamento de F\'{i}sica Te\'{o}rica, M\'{o}dulo 15, Facultad de Ciencias, Universidad Aut\'{o}noma de Madrid, Cantoblanco, 28049 Madrid, Spain}

\newcommand{\US}{Facultad de F\'{i}sicas, Universidad de Sevilla, Campus de Reina Mercedes,  Avda.\ Reina Mercedes s/n, 41012 Sevilla, Spain}

\newcommand{\IATE}{Instituto de Astronom\'{i}a Te\'{o}rica y Experimental (IATE), CONICET-UNC, Laprida 854, X5000BGR, C\'{o}rdoba, Argentina}

\newcommand{\UTFSM}{Departamento de F\'{i}sica, Universidad T\'ecnica Federico Santa Mar\'{i}a, Casilla 110-V, Avda. Espa\~{n}a 1680, Valpara\'{i}so, Chile}

\newcommand{\AB}{Universidad Andres Bello, Facultad de Ciencias Exactas, Departamento de Ciencias Físicas, Instituto de Astrofísica, Av. Fernández Concha 700, Santiago, Chile}

\newcommand{\CIAFF}{Centro de Investigaci\'{o}n Avanzada en F\'{i}sica Fundamental (CIAFF), Facultad de Ciencias, Universidad Aut\'{o}noma de Madrid, 28049 Madrid, Spain}

\newcommand{\ICRAR}{International Centre for Radio Astronomy Research, University of Western Australia, 35 Stirling Highway, Crawley, Western Australia 6009, Australia}

\newcommand{\IAGo}{Institut f\"{u}r Astrophysik, Georg-August Universit\"{a}t G\"{o}ttingen, Friedrich-Hund-Platz 1, 37077, G\"{o}ttingen, Germany}

\newcommand{\CarnObs}{Carnegie Observatories, 813 Santa Barbara Street, Pasadena, CA 91101, USA}

\usepackage{epsfig}
\usepackage{epstopdf} 
\usepackage{fancyhdr}
\usepackage{multirow} 
\usepackage{multicol}
\usepackage{rotating}
\usepackage{url,times,amsfonts,color}
\usepackage{txfonts,float}
\usepackage{titlesec}

\newcolumntype{M}[1]{>{\centering\arraybackslash}m{#1}}
\usepackage{booktabs}

\usepackage{float}
\usepackage[nodayofweek,level]{datetime}
\usepackage{xspace}%
\usepackage[nottoc]{tocbibind}

\usepackage{blindtext} 				
\usepackage{lineno}					
\usepackage{epigraph}				
\usepackage{parskip}
  
\usepackage{breakurl}

\usepackage{orcidlink}

\usepackage{natbib,twoopt}
\bibpunct{(}{)}{;}{a}{}{,}             
\makeatletter
  \newcommandtwoopt{\citeads}[3][][]{\href{http://adsabs.harvard.edu/abs/#3}%
    {\def\hyper@linkstart##1##2{}%
     \let\hyper@linkend\@empty\citealp[#1][#2]{#3}}}
  \newcommandtwoopt{\citepads}[3][][]{\href{http://adsabs.harvard.edu/abs/#3}%
    {\def\hyper@linkstart##1##2{}%
     \let\hyper@linkend\@empty\citep[#1][#2]{#3}}}
  \newcommandtwoopt{\citetads}[3][][]{\href{http://adsabs.harvard.edu/abs/#3}%
    {\def\hyper@linkstart##1##2{}%
     \let\hyper@linkend\@empty\citet[#1][#2]{#3}}}
  \newcommandtwoopt{\citeyearads}[3][][]%
    {\href{http://adsabs.harvard.edu/abs/#3}
    {\def\hyper@linkstart##1##2{}%
     \let\hyper@linkend\@empty\citeyear[#1][#2]{#3}}}
\makeatother

\newcommand{\galacticus}{\textsc{Galacticus}\xspace}

\newcommand{\MDPL}{\textsc{MDPL2}\xspace}

\newcommand{\MDG}{\textsc{The~MultiDark-Galaxies}\xspace}
\newcommand{\MDgal}{\textsc{MDPL2}-\texttt{Galacticus}\xspace}

\newcommand{\consistenttree}{\textsc{Consistent Trees}}

\newcommand{\rockstar}{\textsc{Rockstar}}

\newcommand{\sdss}{\texttt{SDSS}\xspace}
\newcommand{\boss}{\texttt{BOSS}\xspace}

\newcommand{\cmass}{\texttt{CMASS}\xspace}

\newcommand{\den}{\texttt{Gal-dens}\xspace}

\newcommand{\low}{low-SFR}

\newcommand{\lowZ}{low-\Zcold}
\newcommand{\highZ}{high-\Zcold}

\newcommand{\red}{red}
\newcommand{\passive}{passive}

\newcommand{\lowSM}{{\ifmmode{M_{*,\mathrm{low}}}\else{$M_{*,\mathrm{low}}$\xspace}\fi}}
\newcommand{\highSM}{{\ifmmode{M_{*,\mathrm{high}}}\else{$M_{*,\mathrm{high}}$\xspace}\fi}}
\newcommand{\lowHM}{{\ifmmode{M_{\mathrm{halo,low}}}\else{$M_{\mathrm{halo,low}}$\xspace}\fi}}
\newcommand{\highHM}{{\ifmmode{M_{\mathrm{halo,high}}}\else{$M_{\mathrm{halo,high}}$\xspace}\fi}}

\newcommand{\cSFRD}{\textsc{cSFRD}\xspace}
\newcommand{\SFRD}{\textsc{SFRD}\xspace}

\newcommand{\twoPCF}{\textsc{2pCF}\xspace}
\newcommand{\twoPCFs}{\textsc{2pCFs}\xspace}

\newcommand{\ri}{\textit{(r-i)}\xspace}
\newcommand{\gr}{\textit{(g-i)}\xspace}
\newcommand{\gi}{\textit{(g-i)}\xspace}

\newcommand{\Mbnd}{{\ifmmode{M_{\mathrm{bnd}}}\else{$M_{\mathrm{bnd}}$}\fi}}
\newcommand{\Mfof}{{\ifmmode{M_{\mathrm{fof}}}\else{$M_{\mathrm{fof}}$}\fi}}
\newcommand{\Mmean}{{\ifmmode{M_{\mathrm{200m}}}\else{$M_{\mathrm{200m}}$}\fi}}
\newcommand{\MBN}{{\ifmmode{M_{\mathrm{BN98}}}\else{$M_{\mathrm{BN98}}$}\fi}}
\newcommand{\Rc}{{\ifmmode{R_{\mathrm{200c}}}\else{$R_{\mathrm{200c}}$}\fi}}
\newcommand{\Vz}{{\ifmmode{V_{\mathrm{eff}}}\else{$V_{\mathrm{eff}}$\xspace}\fi}}
\newcommand{\ltsima}{$\; \buildrel < \over \sim \;$}
\newcommand{\gtsima}{$\; \buildrel > \over \sim \;$}
\newcommand{\lsim}{\lower.5ex\hbox{\ltsima}}
\newcommand{\gsim}{\lower.5ex\hbox{\gtsima}}

\newcommand{\kpc}{{\ifmmode{\mathrm{kpc}}\else{$\mathrm{kpc}$\xspace}\fi}}
\newcommand{\Mpc}{{\ifmmode{\mathrm{Mpc}}\else{$\mathrm{Mpc}$\xspace}\fi}}
\newcommand{\Gpc}{{\ifmmode{\mathrm{Gpc}}\else{$\mathrm{Gpc}$\xspace}\fi}}

\newcommand{\hkpc}{{\ifmmode{h^{-1}\mathrm{kpc}}\else{$h^{-1}\mathrm{kpc}$\xspace}\fi}}
\newcommand{\hMpc}{{\ifmmode{h^{-1}\mathrm{Mpc}}\else{$h^{-1}\mathrm{Mpc}$\xspace}\fi}}
\newcommand{\hGpc}{{\ifmmode{h^{-1}\mathrm{Gpc}}\else{$h^{-1}\mathrm{Gpc}$\xspace}\fi}}

\newcommand{\MpcCu}{{\ifmmode{\mathrm{Mpc}^3}\else{$\mathrm{Mpc}^3$\xspace}\fi}}
\newcommand{\MpcV}{{\ifmmode{\mathrm{Mpc}^{-3}}\else{$\mathrm{Mpc}^{-3}$\xspace}\fi}}

\newcommand{\hMsun}{{\ifmmode{h^{-1}\mathrm{M_{\odot}}}\else{$h^{-1}\mathrm{M_{\odot}}$}\fi}}
\newcommand{\Msun}{{\ifmmode{\mathrm{M_{\odot}}}\else{$\mathrm{M_{\odot}}$\xspace}\fi}}

\newcommand{\Msunyr}{{\ifmmode{\mathrm{M_{\odot}yr^{-1}}}\else{$\mathrm{M_{\odot}yr^{-1}}$\xspace}\fi}}
\newcommand{\Gyr}{{\ifmmode{\mathrm{Gyr}}\else{$\mathrm{Gyr}$\xspace}\fi}}
\newcommand{\yr}{{\ifmmode{\mathrm{yr}}\else{$\mathrm{yr}$\xspace}\fi}}
\newcommand{\yrmo}{{\ifmmode{\mathrm{yr}^{-1}}\else{$\mathrm{yr}^{-1}$\xspace}\fi}}

\newcommand{\kms}{{\ifmmode{\mathrm{kms}^{-1}}\else{$\mathrm{kms}^{-1}$\xspace}\fi}}
\newcommand{\Zsolar}{{\ifmmode{\mathrm{Z}_{\odot}}\else{$\mathrm{Z}_{\odot}$\xspace}\fi}}

\newcommand{\SFR}{{\ifmmode{\mathrm{SFR}}\else{$\mathrm{SFR}$\xspace}\fi}}
\newcommand{\SFRs}{{\ifmmode{SFRs}\else{$\mathrm{SFRs}$\xspace}\fi}}
\newcommand{\sSFR}{{\ifmmode{\mathrm{sSFR}}\else{$\mathrm{sSFR}$\xspace}\fi}}
\newcommand{\sSFRs}{{\ifmmode{\mathrm{sSFRs}}\else{$\mathrm{sSFRs}$\xspace}\fi}}

\newcommand{\Mstar}{{\ifmmode{M_*}\else{$M_*$\xspace}\fi}}
\newcommand{\Mhalo}{{\ifmmode{M_{\mathrm{halo}}}\else{$M_{\mathrm{halo}}$\xspace}\fi}}
\newcommand{\Mhot}{{\ifmmode{M_{\mathrm{hot}}}\else{$M_{\mathrm{hot}}$\xspace}\fi}}
\newcommand{\Mcold}{{\ifmmode{M_{\mathrm{cold}}}\else{$M_{\mathrm{cold}}$\xspace}\fi}}
\newcommand{\cgf}{{\ifmmode{M_{\mathrm{cold}}/M_{\mathrm{*}}}\else{$M_{\mathrm{cold}}/M_{\mathrm{*}}$\xspace}\fi}}
\newcommand{\Mc}{{\ifmmode{M_{\mathrm{200c}}}\else{$M_{\mathrm{200c}}$\xspace}\fi}}
\newcommand{\Mvir}{{\ifmmode{M_{\mathrm{vir}}}\else{$M_{\mathrm{vir}}$\xspace}\fi}}
\newcommand{\Mbh}{{\ifmmode{M_{\mathrm{BH}}}\else{$M_{\mathrm{BH}}$\xspace}\fi}}
\newcommand{\vmax}{{\ifmmode{V_{\mathrm{max}}}\else{$V_{\mathrm{max}}$\xspace}\fi}}
\newcommand{\vsat}{{\ifmmode{V_{\mathrm{max_{\mathrm{sat}}}}}\else{$V_{\mathrm{max_{\mathrm{sat}}}}$\xspace}\fi}}
\newcommand{\Zcold}{{\ifmmode{Z_{\mathrm{cold}}}\else{$Z_{\mathrm{cold}}$\xspace}\fi}}
\newcommand{\rhalfmass}{{\ifmmode{R_{\mathrm{1/2,DM}}}\else{$r_{\mathrm{1/2}}$\xspace}\fi}}
\newcommand{\rbulgevsdisk}{{\ifmmode{R_{\mathrm{bulge/disk}}}\else{$R_{\mathrm{bulge/disk}}$\xspace}\fi}}
\newcommand{\Mzstars}{{\ifmmode{M_{\mathrm{Z_{*}}}}\else{$M_{\mathrm{Z_{*}}}$\xspace}\fi}}
\newcommand{\Mzhot}{{\ifmmode{M_{\mathrm{Z_{\mathrm{hot}}}}}\else{$M_{\mathrm{Z_{\mathrm{hot}}}}$\xspace}\fi}}
\newcommand{\Mzgas}{{\ifmmode{M_{\mathrm{Z_{\mathrm{cold}}}}}\else{$M_{\mathrm{Z_{\mathrm{cold}}}}$\xspace}\fi}}
\newcommand{\Vbulge}{{\ifmmode{V_{\mathrm{bulge}}}\else{$V_{\mathrm{bulge}}$\xspace}\fi}}
\newcommand{\Vdisk}{{\ifmmode{V_{\mathrm{disk}}}\else{$V_{\mathrm{disk}}$\xspace}\fi}}
\newcommand{\jbar}{{\ifmmode{j_{\mathrm{bar}}}\else{$j_{\mathrm{bar}}$\xspace}\fi}}

\newcommand{\bheff}{{\ifmmode{\Mbh/\Mvir}\else{{\Mbh/\Mvir}\xspace}\fi}}
\newcommand{\SHMR}{{\ifmmode{\mathrm{SHMR}}\else{$\mathrm{SHMR}$\xspace}\fi}}
\newcommand{\SHMRs}{{\ifmmode{SHMRs}\else{$\mathrm{SHMRs}$\xspace}\fi}}
\newcommand{\CGF}{{\ifmmode{\mathrm{CGF}}\else{$\mathrm{CGF}$\xspace}\fi}}
\newcommand{\Tcons}{{\ifmmode{T_{\mathrm{cons}}}\else{$T_{\mathrm{cons}}$\xspace}\fi}}
\newcommand{\con}{{\ifmmode{C_{\mathrm{NFW}}}\else{$C_{\mathrm{NFW}}$\xspace}\fi}}

\newcommand{\n}{{\ifmmode{n_{\mathrm{gal}}}\else{$n_{\mathrm{gal}}$}\fi}}
\newcommand{\Ngal}{{\ifmmode{N_{\mathrm{gal}}}\else{$N_{\mathrm{gal}}$}\fi}}
\newcommand{\Norph}{{\ifmmode{N_{\mathrm{orphan}}}\else{$N_{\mathrm{orphan}}$}\fi}}
\newcommand{\Nxorph}{{\ifmmode{N_{\mathrm{non-orphan}}}\else{$N_{\mathrm{non-orphan}}$}\fi}}

\newcommand{\logT}{{\ifmmode{\log_{\mathrm{10}}}\else{$\log_{\mathrm{10}}$\xspace}\fi}}
\newcommand{\dex}{{\ifmmode{\mathrm{dex}^{-1}}\else{$\mathrm{dex}^{-1}$\xspace}\fi}}
\newcommand{\zstart}{{\ifmmode{z_{\mathrm{ref}}}\else{$z_{\mathrm{ref}}$\xspace}\fi}}

\newcommand{\Mr}{{\ifmmode{M_{\mathrm{r}}}\else{$M_{\mathrm{r}}$\xspace}\fi}}
\newcommand{\rp}{{\ifmmode{{{r_{\mathrm{p}}}}\else{${{r_{\mathrm{p}}}}$\xspace}\fi}}
\newcommand{\pim}{{\ifmmode{{{\pi_{\mathrm{max}}}}}\else{${{\pi_{\mathrm{max}}}}}$\xspace}\fi}}

\newcommand{\Mcut}{{\ifmmode{M_{\mathrm{cut}}}\else{$M_{\mathrm{cut}}$\xspace}\fi}}
\newcommand{\Mmin}{{\ifmmode{M_{\mathrm{min}}}\else{$M_{\mathrm{min}}$\xspace}\fi}}
\newcommand{\Muno}{{\ifmmode{M_{\mathrm{1}}}\else{$M_{\mathrm{1}}$\xspace}\fi}}
\newcommand{\sigM}{{\ifmmode{\sigma_{\log_{\mathrm{10}}\Mstar}}\else{$\sigma_{\log_{\mathrm{10}} \Mstar}$\xspace}\fi}}

\def\lesssim{\mathrel{\hbox{\rlap{\hbox{\lower4pt\hbox{$\sim$}}}\hbox{$<$}}}}
\def\gtrsim{\mathrel{\hbox{\rlap{\hbox{\lower4pt\hbox{$\sim$}}}\hbox{$>$}}}}

\newcommand{\Tab}[1]{Table~\ref{#1}}
\newcommand{\Sec}[1]{Section~\ref{#1}}
\newcommand{\App}[1]{Appendix~\ref{#1}}

\newcommand{\Eq}[1]{Eq.~(\ref{#1})}

\newcommand{\Fig}[1]{Fig.~\ref{#1}}
\newcommand{\Figsstart}[1]{Figs.~\ref{#1}}
\newcommand{\Figsend}[1]{\ref{#1}}
\newcommand{\beq}{\begin{equation}}
\newcommand{\eeq}{\end{equation}}

\def\beqa{\begin{eqnarray}}
\def\eeqa{\end{eqnarray}}

\begin{document}

   \title{A semi-analytical perspective on massive red galaxies:}

   \subtitle{I. Assembly history, environment \& redshift evolution}

   \author{D.\ Stoppacher\orcidlink{0000-0002-3281-9956}\inst{1}\fnmsep\inst{2}\fnmsep\inst{3}\fnmsep\thanks{\email{\myemail}}\and
          A.\ D.\ Montero-Dorta\inst{4}\and
            M.\ C.\ Artale\inst{5}\and
            A.\ Knebe\inst{1}\fnmsep\inst{6}\fnmsep\inst{7}\and
            N.\ Padilla\inst{8}\and     
            A.\ J.\ Benson\inst{9}\and
            C.\ Behrens\inst{10}
          }
   \institute{\UAM\
   \and\PUC\
   \and\US\
   \and\UTFSM\
   \and\AB\
   \and\CIAFF\
   \and\ICRAR\
   \and\IATE\
   \and\CarnObs\
   \and\IAGo}

   \date{Received Month Day, 2024; accepted Month Day, 2024}

  \abstract
   {The evolution of galaxies within a self-consistent cosmological context remains one of the most outstanding and challenging topics in modern galaxy formation theory. Investigating the assembly history and various formation scenarios of the most massive and passive galaxies, particularly those found in the densest clusters, will enhance our understanding of why galaxies exhibit such a remarkable diversity in structure and morphology.}
   {In this paper we simultaneously investigate the assembly history and redshift evolution of semi-analytically modelled galaxy properties of luminous and massive central galaxies between $0.56 < z < 4.15$, alongside their connection to their halos as a function of large-scale environment.}
   {We extract sub-samples of galaxies from a mock catalogue representative for the well-known \boss-\cmass\ sample, which includes the most massive and passively evolving system known today. Utilising typical galaxy properties such as star formation rate, \gi\ colour, or cold gas-phase metallicity (\Zcold), we track the redshift evolution of these properties across the main progenitor trees.}
   {We present results on galaxy and halo properties, including their growth and clustering functions, for each of our sub-samples. Our findings indicate that galaxies in the highest stellar and halo mass regimes are least metal-enriched (using \Zcold\ as a proxy) and consistently exhibit significantly larger black hole masses and higher clustering amplitudes compared to sub-samples selected by e.g.\ colour or star formation rate. This population forms later and also retains large reservoirs of cold gas. In contrast, galaxies in the intermediate and lower stellar/halo mass regimes consume their cold gas at higher redshift and were among the earliest and quickest to assemble their stellar and black hole masses. In addition, we observe a clear trend where the clustering of the galaxies selected according to their \Zcold-values (either \lowZ\ or \highZ) depends on the density of their location within the large-scale environment.}
   {We assume that in particular galaxies in the \lowZ\ and \highZ\ sub-samples form and evolve through distinct evolutionary channels, which are predetermined by their location within the large-scale environment of the cosmic web. Furthermore, their clustering dependence on the environment could be an important area for further investigation.}

   \maketitle
\nolinenumbers
\section{Introduction} \label{sec:introduction}

The mechanisms driving galaxy evolution operate across a wide range of spatial and temporal scales. These include the size of star-forming molecular clouds, a few parsecs in diameter, to tidally interacting galaxies on the cluster scale, or effects incorporating the entire network of the cosmic web such as inter-connectivity via filaments or gravitational collapse of large-scale structures. On the temporal scale, galaxy evolution encompasses both short-term star formation events lasting less than one megayear and the long-term assembly of ancient elliptical galaxies hosted by the most massive dark matter halos today. Indeed, galaxy evolution is influenced not only by various internal physical processes \citep{Kormendy79_morph_ref,Dressler80,Mannucci10,Conroy13_rev,Kalinova21_quench}, but also by the evolution of the dark matter halo where it resides in \citep{Somerville+Dave15_rev,Wechsler+Tinker18_rev} and its associated local and large-scale environment \citep{Blanton07,Shandarin09_cosmic_web_rev,Peng10,Argudo18,WangL18,Contini20,Dutta20,Rosas-Guevara22_EAGLE_voids}. 

In this context, the most massive red galaxies, typically living in the richest clusters today, are particularly interesting. They not only constitute the backbone of the cosmic web, but their formation also provides significant insights into the formation and evolution of our Universe \citep[see e.g.][]{Reid10_LRG_clustering_cosmology,Zhai23_Aemulus}. These galaxies represent the final stages of galaxy evolution and are extensively used as tracers of the large-scale structure in cosmological surveys \citep{Shandarin09_cosmic_web_rev,Conselice14_evo_rev,Inagaki15_MAH-cluster,Favole16b,Saito16}.

The evolution of massive red galaxies has been explored from multiple perspectives. Regarding their stellar mass assembly histories, the general consensus is that these galaxies form the majority of their stars early on \citep[e.g.][]{DeLucia06,Maraston09,Maraston13,Lui16_LRGs,Johnston22_SFH_obs}. However, there have been documented episodes of rejuvenation \citep{Hawarden79_reju_ref,Pandya17_reju,Remus23_reju,Zhang23_reju}.

Additionally, the relationship between the internal evolution of massive red galaxies and their local and large-scale environments has been investigated using various observational, statistical, and numerical tools for 90 years
\citep{Hubble36,Zwicky61,Dressler80,Zehavi05,Thomas10_alpha_envr,Koyama13,Luparello15,Filho15_lowZ_envr,Schaye15_EAGLE,Musso18,Pandey+Sarkar20,Santucho20,Sarkar+Pandey20,Sureshkumar21_2pCF,Alarcon23_Diffstar}. Importantly, the clustering of massive red galaxies is known to be enhanced relatively to the general galaxy population \citep[see e.g.\ pioneering work by][]{Kaiser84_clustering_ref,Efstathiou+Rees88_clustering_ref}, as they typically reside in the most massive halos \citep{Sheth+Tormen01_bias,Croton07_ref}. In general, more luminous and massive galaxies with redder colours and early-type morphology exhibit stronger clustering and tend to live in denser regions compared to their less massive, bluer, and later-type counterparts. Several studies have previously explored the connection between massive red galaxies and the so-called assembly bias effect, which refers to the secondary dependencies of halo and galaxy clustering at fixed halo mass \citep{Lin16_assembly,Montero-Dorta17_assembly,Niemiec18}.

In terms of galaxy formation, the widely accepted scenario suggests that massive galaxies in the early stage of the Universe undergo an immense starburst phase and subsequent rapid quenching \citep{Forrest20}. These galaxies are thought to belong to either a first or second wave of formation, with their bulges forming early \& fast or later \& more slowly \citep{Costantin21} where distinct events in their evolution history, such as major merger, help to drive their mass assembly \citep[e.g.][]{Lackner12_MA_envr_hydro,Hashemizadeh21,Sawicki20_MA_obs,Spavone21_MA_obs,Dolfi23_AH_lop}. This aligns well with the proposed two-phase scenario proposed by \citet{Oser10_ref}, where with in-situ star formation and a subsequent ex-situ accretion phase are responsible for the build-up the stellar mass component of a galaxy which posits that the build-up of a galaxy's stellar mass is due to an early phase of in-situ star formation followed by a later phase of ex-situ accretion.

Given the considerations mentioned above, we can identify four major drivers that control the evolution of a galaxy. These are its intrinsic properties (i.e., how many baryons were initially available to form a galaxy) and baryonic processes (such as stellar feedback, and outflows, among others), its galaxy-halo connection (the characteristics of the dark matter halo in which the galaxy resides), its assembly history (including the redshift evolution of both galaxy and halo properties), and its environment (such as the galaxy's location the galaxy in either less-dense or more-dense regions of the Universe and the of its clustering).

It is important to note that these four elements are highly correlated and interact with each other on many levels, as repeatedly reported in the literature. For instance, the connection between intrinsic properties and environment can be illustrated by the growth of black holes, which can facilitate the quenching of the star formation, generally known as AGN\footnote{AGN is an acronym for Active Galactic Nucleus}-feedback -- a process that particularly influences the fate of massive cluster galaxies \citep[see e.g.][]{Croton06,Davies21_Eagle}. Another example is that quenched galaxies tend to prefer specific environments such as the edge of filaments \citep{Song21_MA_envr}. In addition, \citet{Kim20} proposed that compact ellipticals consist of galaxies with two distinct origins depending on their local environment. Furthermore, the merging history of gas can impact galaxy evolution, as demonstrated for core-rotating early-type galaxies, which have different assembly processes compared to their core-less counterparts \citep{Krajnovic20}. By examining the stellar mass assembly histories of simulated galaxies, \citet{Gupta20_SFH_obs} found a rapid increase in the ex-situ stellar mass fraction of massive galaxies at $z < 3.5$, while this fraction remains constant for their low-mass counterparts across cosmic time. An example of how the assembly history and galaxy-halo connection jointly influence intrinsic properties is provided by \citep{Bose+Loeb21_HA_model}, who observed variations in the stellar velocity dispersion with age and halo concentration. Finally, \citet{Harada23_metal_outflow} reported a strong link between gas and metal outflow in proto-clusters, which are highly sensitive to halo mass.

Utilising observational galaxy properties presents a challenge, as it necessitates inferring formation histories and halo properties that cannot be directly extracted from a merger tree as readily available in simulations. Nonetheless, we find that studies of massive red galaxies can significantly benefit from integrating various aspects of galaxy evolution. In this context, we will simultaneously examine the assembly histories, clustering, galaxy-halo connection, and environment in this work. This approach builds on the groundwork laid by \citet[][]{Stoppacher19}, who investigated the main properties and clustering of luminous red galaxies using the \galacticus\ semi-analytical model (SAM) developed by \citet{Benson12}, which resembles the selected \boss-\cmass\ sample at $z\sim0.5$. Their study demonstrated that specific star formation rate and the cold gas fraction correlate with halo mass and large-scale environment (less-dense or more-dense regions). Furthermore, they observed a strong bimodality in the plane of cold-gas phase metallicity and specific star formation rate (see their Fig.\ 10). In this paper, we extend their analysis by examining the evolutionary history of the same galaxies in order to illuminate their diverse formation channels. We also investigate the origins of the bimodality found in the cold gas-phase metallicity and stellar mass planes. 

We utilise modelled \boss-\cmass\ galaxy properties from the aforementioned SAM in the redshift range of $0.5\gtrsim z \gtrsim 4$, following the same selection procedure of modelled CMASS-galaxies as in \citet[][]{Stoppacher19}. Their study presented a method to mimic the photometric selection of luminous and massive galaxies, producing a galaxy sample that is both quantitatively and qualitatively comparable to observations. We rely on this modelled data because SAMs methods for generating catalogues of galaxy properties and tracking the evolution of statistically significant samples. These models are usually built upon $N$-body dark matter simulations using merger trees (information on the hierarchical formation of dark matter halos). Unlike other modelling techniques, SAMs do not explicitly solve fundamental equations of for example hydrodynamics, but instead, use simplified recipes to account for baryonic physics as a post-processing step. This includes phenomenological treatments of baryonic processes and coarse-grained the properties of galaxies, allowing them to solve the system of equations more rapidly galaxy properties, enabling the system of equations to be solved more efficiently. SAMs are adjusted (or tuned) to reproduce observed observed galaxy distributions and are constrained by empirical measurements. Although the modelling of the physical processes is simplified, the advantage of SAMs lies in their ability of handling sub-grid physics efficiently and adaptively, making them an excellent tool for exploring a wide range of galaxy properties across diverse parameter spaces \citep[][]{Baugh06,Benson10,Baugh12,Somerville+Dave15_rev}. 

This work is organised as follows: In \hyperref[sec:data]{\Sec{sec:data}} we describe the parent catalogue used to extract our SAM-CMASS mock-galaxy sample, and in \hyperref[sec:method]{\Sec{sec:method}} we explain how we select sub-samples from this catalogue and track progenitors through their merger trees. Our results are presented in \hyperref[sec:results]{\Sec{sec:results}}, followed by a detailed discussion of the key findings in \hyperref[sec:discussion]{\Sec{sec:discussion}}. We summarise our work and provide an outlook to future studies in \hyperref[sec:summary]{\Sec{sec:summary}}.

The adopted cosmology in this paper is based on a flat $\Lambda$CDM model with the following cosmological parameters: $\Omega_\mathrm{m}=0.307, \Omega_\mathrm{b}=0.048, \Omega_\Lambda=0.693, \sigma_\mathrm{8}=0.823, n_\mathrm{s}=0.96$, and a dimensionless Hubble parameter $h=0.678$ \citep{Planck15}. Hereafter, $h$ is absorbed into the numerical values of properties throughout the text, as well as in all tables and figures. We use typical dependencies of the Hubble parameter, as outlined in \citet{Croton13_h}, where masses from simulations are typically scaled with $h^{-1}$.

\section{Data selection and sample evaluation} \label{sec:data}

\subsection{Galaxy catalogue and simulation details} \label{sec:sim}

Our modelled galaxy catalogue is based on the well-known \boss-\cmass\ catalogue from the Sloan Digital Sky Survey (SDSS) \citep{Alam15_SDSS}, which is well-constrained and extensively studied \citep[e.g][]{Cuesta16,Montero-Dorta16,Chuang16,Favole16b,Rodriguez-Torres16,Montero-Dorta17,Ross17,Sullivan17,Guo18,Mueller18}. This galaxy catalogue was originally designed to target the most luminous and massive galaxies in order to produce a uniformly distributed sample of galaxies at redshift $0.43<z<0.7$. The photometric selection included \gr\ and \ri\ colours \citep{Fukugita96_ugriz} to isolate only the reddest and most massive galaxies at high redshifts, while also allowing for an extension towards bluer colours, meaning that ``blue-cloud''-galaxies could still enter the \cmass-sample. For further details, we refer to the \boss\ target selection and reduction pipeline\footnote{\url{https://www.sdss.org/dr12/algorithms/boss_galaxy_ts/}}. We use data from \textsc{Data Release 12}, specifically the Large-Scale Structure (LSS) catalogue \footnote{\url{https://data.sdss.org/sas/dr12/boss/lss/}} \citep{Reid16_BOSS_DR12_LSS} from the SDSS Science Archive Server. This was cross-matched with the Portsmouth\footnote{\url{http://www.sdss.org/dr13/spectro/galaxy_portsmouth/}} passive galaxy sample to include stellar masses, based on the stellar population models of \citet{Maraston05} and \citet{Maraston09}.

The model we use in this study, the semi-analytical galaxy formation and evolution code \galacticus, developed by \citet{Benson12}, was run on the \textsc{MultiDark Planck 2} simulation \citep[hereafter \MDPL:][]{Klypin16_MD} and released as part of \MDG\ \citep{Knebe17_MD}. \MDPL\ is an $N$-body dark matter-only simulation with a side-length of 1000 \hMpc, tracking the evolution of $3840^3$ dark matter particles, each with a mass of $m_\mathrm{p} = 2.23 \times 10^9$ \Msun. Halos and sub-halos were identified using \rockstar\ \citep[][]{Behroozi13a} and merger trees were constructed with \consistenttree\ \citep[][]{Behroozi13b}. More information on the model can be found in \hyperref[app:SAM]{\App{app:SAM}}. This version of \galacticus\ was released under the name \MDgal\ and is publicly available on \url{www.cosmosim.org} and \url{www.skiesanduniverses.org}. The model adopts the same cosmology as used in this work.

\subsection{Selecting modelled galaxies from the galaxy catalogue} \label{sec:tarsel}

For the selection procedure of SAM-CMASS mock-galaxies we refer to Section 2 of our companion paper \citet[][hereafter S19]{Stoppacher19}, which outlines how to extract modelled BOSS-CMASS galaxies from the SAM galaxy catalogue. We adopt their approach, applying the same selection algorithm to the galaxy catalogue \MDgal\ \citep{Knebe17_MD}. As described in Section 3 of S19, the authors tested various selection procedures and extracted several CMASS mock-galaxy samples which are described alongside those of the observed \cmass-sample from \boss\ (see their Table 1). For reasons detailed in S19, they replicated the photometric \cmass\ target selection of \boss\ using a ``down-sampling'' approach on the modelled galaxies. This approach was thoroughly assessed and verified to produce a valid and comparable mock-galaxies sample, as demonstrated by the stellar mass functions, the galaxy-halo connection, and the clustering function, all of which show good agreement with observations ( see S19,  Fig.\ 4 and Figs.\ 6-8). For this study, we specifically choose the mock-galaxy sample called \den\footnote{We adopt the name convention of S19, where the label ``dens'' refers to the density selected sample.} as our reference (parent) sample, since the modelled sample was required to match the number density of its observational counterpart. 

\subsection{Methodology of selecting sub-samples and tracking progenitors}\label{sec:method}

\begin{figure}
	\centering
	\includegraphics[angle=0,width=0.48\textwidth]{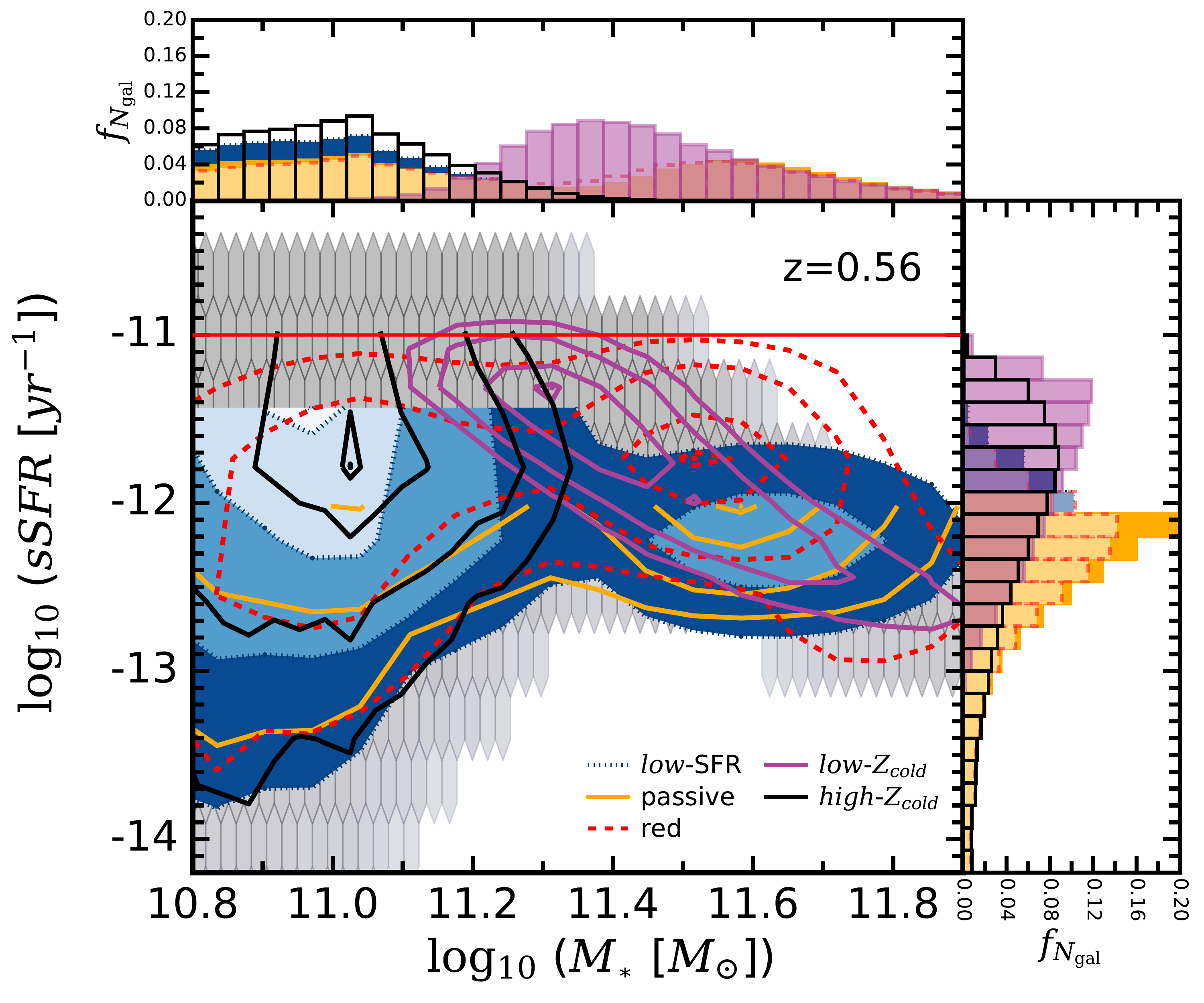}\vspace{-0.3cm}%
	\caption{Sub-samples extracted from the entire dataset of SAM-CMASS mock-galaxy catalogue, \den, as described in \hyperref[tab:subsel]{\Tab{tab:subsel}} and represented by coloured contours in the \sSFR-\Mstar\ parameter space at $\zstart=0.56$. The number density distribution of the entire dataset is depicted as, logarithmically binned hexagons in the background. The horizontal solid red line marks the classic quiescent separation, \logT(\sSFR\ [\yr])$\sim-11$ \citep{Franx08}.}\label{fig:ssfr2mstar}\vspace{-0.4cm}
\end{figure}

\begin{figure}
	\centering
	\includegraphics[angle=0,width=0.48\textwidth]{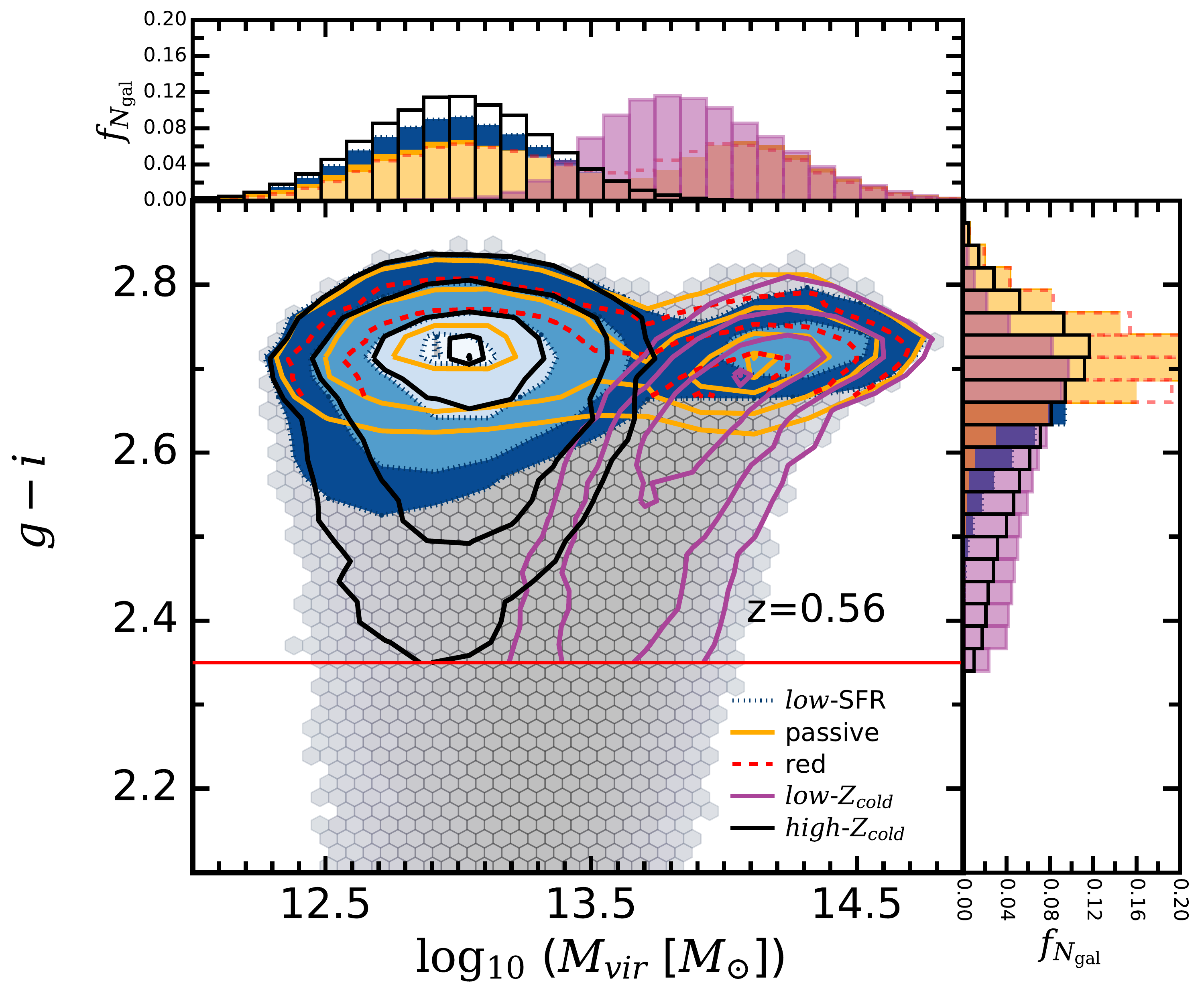}\vspace{-0.3cm}%
	\caption{Sub-samples extracted from the entire dataset of SAM-CMASS mock-galaxy catalogue, \den, as described in \hyperref[tab:subsel]{\Tab{tab:subsel}} and represented by coloured contours on the \gi-\Mvir\ parameter space at $\zstart=0.56$. The number density distribution of the entire dataset is depicted as grey, logarithmically binned hexagons in the background. The horizontal solid red line marks the classic separation of red and blue galaxies, $\gi=2.35$ \citep{Masters11}.}\label{fig:g-i2mhalo}\vspace{-0.4cm}
\end{figure}

Within this work, we aim at studying the star formation and assembly histories of distinct populations of galaxies, such as those exhibiting bimodality in the cold gas-phase metallicity as mentioned in \hyperref[sec:introduction]{\Sec{sec:introduction}}. To achieve this, we first need to establish selection criteria to guide our study. In this section, we describe this methodology and subsequently apply these criteria on our selected parent sample, \den\ -- the modelled CMASS-galaxies from \MDgal\ -- to produce what we refer to as ``sub-samples''. For clarity, \den\ represents the population of the most massive and luminous galaxies in the Universe. We also specify that our analysis includes only central galaxies\footnote{For details on the definitions of galaxy type such as ``central'' or ``satellite'', we refer to App.~2 in \citet{Knebe17_MD}.}.

\begin{table}
	\begin{center}
        \centering\caption{Overview on sub-samples used in this work.}\vspace{-0.3cm}
        \setlength{\tabcolsep}{4pt}
		\begin{tabular}{m{0.18\columnwidth}||m{0.7\columnwidth}}
			\hline
            \multicolumn{1}{c||}{\multirow{2}{*}{\parbox{0.18\columnwidth}{\centering sub-sample name}}}  &  \multicolumn{1}{c}{\multirow{2}{*}{\parbox{0.7\columnwidth}{\centering selection criterion}}}\\
            & \\
			\hline
			\hline   
			\low		& 20 \% lowest star formation rate (\SFR)		         \\\hline
			\passive	& 20 \% passive galaxies / lowest specific \SFR\ (\sSFR) \\\hline    
			\red		& 20 \% reddest galaxies / highest \gi \\\hline 
			\lowZ		& 20 \% lowest cold gas-phase metallicities	(\Zcold) and $\gi>2.35$\\\hline
			\highZ		& 20 \% highest cold gas-phase metallicities (\Zcold) and $\gi>2.35$\\
			\hline
			\hline
			(i)				& (ii)					\\
		\end{tabular}
		\tablefoot{This table compiles our choice of sub-samples extracted from the modelled SAM-CMASS mock-galaxy catalogue, \den, at $\zstart=0.56$ -- the redshift of sample selection. In the first column (i) we state the sub-sample's name which is inspired by its selection criterion shown in the second column (ii).}\label{tab:subsel}
	\end{center}\vspace{-0.4cm}
\end{table}

We define our selection criteria based on typical galaxy properties such as observed colour separation \gi, star formation rate (\SFR), or total cold gas-phase metallicity, \Zcold. \Zcold\ represents the metallicity of the cold gas available for star formation, typically with temperatures below $\sim$100 K \citep{Dave20_OH_Eagle} Thus, this property serves as an important diagnostic for various processes in galaxy evolution, including gas in- and out-flow, and star formation in cold gas clouds \citep[e.g.][]{Hughes13_OH,Lutz20_OH,Wang+Lilly21_OH}. In observational data, this property is often quantified as the ratio of the number density of oxygen atoms to that of hydrogen atoms, $12+\logT(\mathrm{O/H})$, since oxygen is the most abundant heavy element in the cosmos. As our model does not output oxygen abundances, we estimate this property using the masses of metals and normalise them by the Solar metallicity defined as $8.69+\log_{10}(\Mzgas/\Mcold)-\log_{10}(\Zsolar)$, where \Mzgas\ is the mass of metals in the cold gas-phase. We use $\Zsolar=0.0134$ \citep{Asplund09} for the Sun’s metallicity and the factor $8.69$ for its oxygen abundance \citep{Allende01}. This standard procedure is common in semi-analytical models. Additionally, we apply the same normalisation of the Oxygen abundance determined at redshift $z=0$ to normalise the prediction of our model at higher redshifts primarily for the reason to facilitate comparisons of metallicities across various sub-samples. Our goal is to ensure a consistent approach across all redshifts rather than precise measurements.

We select five sub-samples from the entire population of centrals present in the SAM-CMASS mock-galaxy sample, \den, at redshift $\zstart=0.56$ -- is our reference Redshift of sample selection -- and name them after their selection criterion. Thereby we always select 20\% of their total amount of central galaxies in \den\ (270,000) e.g.\ 20\% with lowest \SFR\ for the sub-sample addressed as ``\low'' or 20\% of those with the reddest colours \gi\ for the sub-sample addressed as ``\red'' as described in \hyperref[tab:subsel]{\Tab{tab:subsel}}. This results in approximately $\sim$50,000 galaxies per sub-sample at  $\zstart=0.56$.

The first three sub-samples listed in \hyperref[tab:subsel]{\Tab{tab:subsel}} contain only luminous red galaxies (LRGs); however, we find that the remaining two sub-samples (\lowZ\ and \highZ) extracted on the basis of their cold gas-phase metallicity include both ``red-sequence'' and ``blue-cloud'' galaxies. The latter are massive galaxies with mild star formation, which results in slightly bluer colours \citep[see e.g.][]{Eisenstein11_SDSS3}. To avoid contamination from these star-forming galaxies, we exclude the ``blue-cloud'' members from the \lowZ\ and \highZ\ sub-samples. As a result, we also require these sub-samples to meet the classic colour separation $\gi>2.35$ \citet{Masters11}. These two sub-samples are particularly interesting because they map the prominent bimodality in \Zcold\ as found by S19 (see their Fig.\ 10).

In \hyperref[fig:ssfr2mstar]{\Fig{fig:ssfr2mstar}} we show our defined sub-samples as coloured contours in the specific star formation rate (sSFR) versus stellar mass (\Mstar) parameter space together, with the entire dataset of SAM-CMASS mock-galaxies, \den, as grey, logarithmically binned hexagons in the background. Note that we use for all our contour-figures the following confidence levels expressed as percentages: [13.6, 31.74, 68.26, 95, 99.7]. Additionally, the histogram panels on the top and the right-hand marginal axes show the distribution of galaxies along the binned axes, normalised to the total number of galaxies per sub-sample, using 35 bins. The histograms show the same colour and line style keys as the contours of the corresponding sub-samples. As pointed out previously, the modelled galaxies exhibit a strong bimodality in the specific star formation rate-stellar mass plane. Therefore, we explicitly include the sub-samples \lowZ\ and \highZ\ selected on the basis of the cold gas-phase metallicity, \Zcold, in our study since they can be almost perfectly mapped onto the bimodal distribution in stellar mass. Interestingly, galaxies selected based on other properties such as colour or star formation rate can not be assigned clearly to either lower or higher stellar mass. We confirm that only passive galaxies enter our sub-samples since the contour lines are all located below the quiescent separation (red solid line) as defined by \citet{Franx08}.

In \hyperref[fig:g-i2mhalo]{\Fig{fig:g-i2mhalo}} we show the observed colour \gi\ as a function of halo mass (\Mvir) for the same sub-samples as described in \hyperref[fig:ssfr2mstar]{\Fig{fig:ssfr2mstar}}. The red solid horizontal line indicates the typical red-blue separation \gi$>2.35$ \citep{Masters11}, as mentioned before. As expected, the figure clearly distinguishes between the low and high metallicity populations, showing a similar bimodal distribution in halo masses, analogous to the bimodality observed in stellar masses in \hyperref[fig:ssfr2mstar]{\Fig{fig:ssfr2mstar}}. Specifically, the \lowZ\ sub-sample is found in halos of higher masses, while the \highZ\ sub-sample occupies halos of lower masses, a pattern that was previously noted by S19 and depicted in their Fig.\ 10. In S19, authors concluded that galaxies with either lower or higher metallicities are also associated with different environments (see their Table 2). This observation motivated the inclusion of these sub-samples in the current analysis to investigate whether the assembly and evolution of galaxies within these sub-samples occurred through separate formation channels, similar to the formation paths of luminous red galaxies (LRGs) in observations \citep[e.g.][]{Montero-Dorta17_assembly}. It is important to note that the blue-cloud population is explicitly excluded from both the \lowZ\ and \highZ\ sub-samples. Furthermore, for the purpose of narrative continuity, these sub-samples are referred to as  ``more'' and ``less'' metal-enriched. However, that does not mean that they are metal-poor. All galaxies in the sample, being luminous red galaxies, are metal-rich with cold gas-phase metallicity values of $\Zcold\gtrsim9$, as predicted by observations \citep[see e.g.][]{Maiolino+Mannucci18_rev}.

After formulating our selection criteria and identifying five sub-samples of modelled CMASS-galaxies, the evolutionary history of each galaxy in the samples needs to be determined. This is achieved by using unique identification numbers (\texttt{parentIndex}) of central dark matter halos hosting the galaxies of interest which allows tracing their main progenitor halos through their merger trees back in time. This information is provided by the halo finder and corresponding tree builder algorithm, in our case \rockstar\ \citep{Behroozi13a} and \consistenttree\ \citep{Behroozi13b}, respectively, both of which can be accessed via the \textsc{Cosmosim}-database\footnote{\url{www.cosmosim.org}}. Note that our goal is to investigate how each galaxy subsample, defined at a fixed redshift of $\zstart=0.56$, evolves over cosmic time. This means that each subset is tracked through time using the main branches of their merger trees. While galaxies may undergo changes in star formation rates and metallicity over time, they will remain in a fixed subset in our analysis. For more information on the technical aspect of tracking halos through cosmic history, we refer to \hyperref[app:M2]{\App{app:M2}}. This approach allows for the study of the true redshift evolution of galaxy and halo properties for each galaxy that was included in a particular sub-sample. It is crucial to note that, using this technique, this work focuses on galaxies that were the reddest at \zstart, but these have not necessarily evolved out the reddest at higher redshifts. This method is a common practice for examining the redshift evolution of galaxy properties in mock catalogues. A schematic representation of this selection and tracking method can be found in \hyperref[fig:cartoon_M2]{\Fig{fig:cartoon_M2}} in \hyperref[app:M2]{\App{app:M2}}.

\section{Results}\label{sec:results}

We present results on the redshift evolution and assembly history of galaxy properties from modelled massive and luminous red galaxies using the CMASS mock-galaxies from the semi-analytical model \galacticus. We apply selection criteria based on typical galaxy properties such as colour or star formation activity to select five sub-samples as discussed extensively in \hyperref[sec:data]{\Sec{sec:data}}. We remind the reader that we use only central galaxies in our analysis.

\subsection{Redshift evolution of galaxy and halo masses}\label{sec:res_mstar_mhalo}

\begin{figure}
	\centering
	\includegraphics[width=\columnwidth,angle=0]{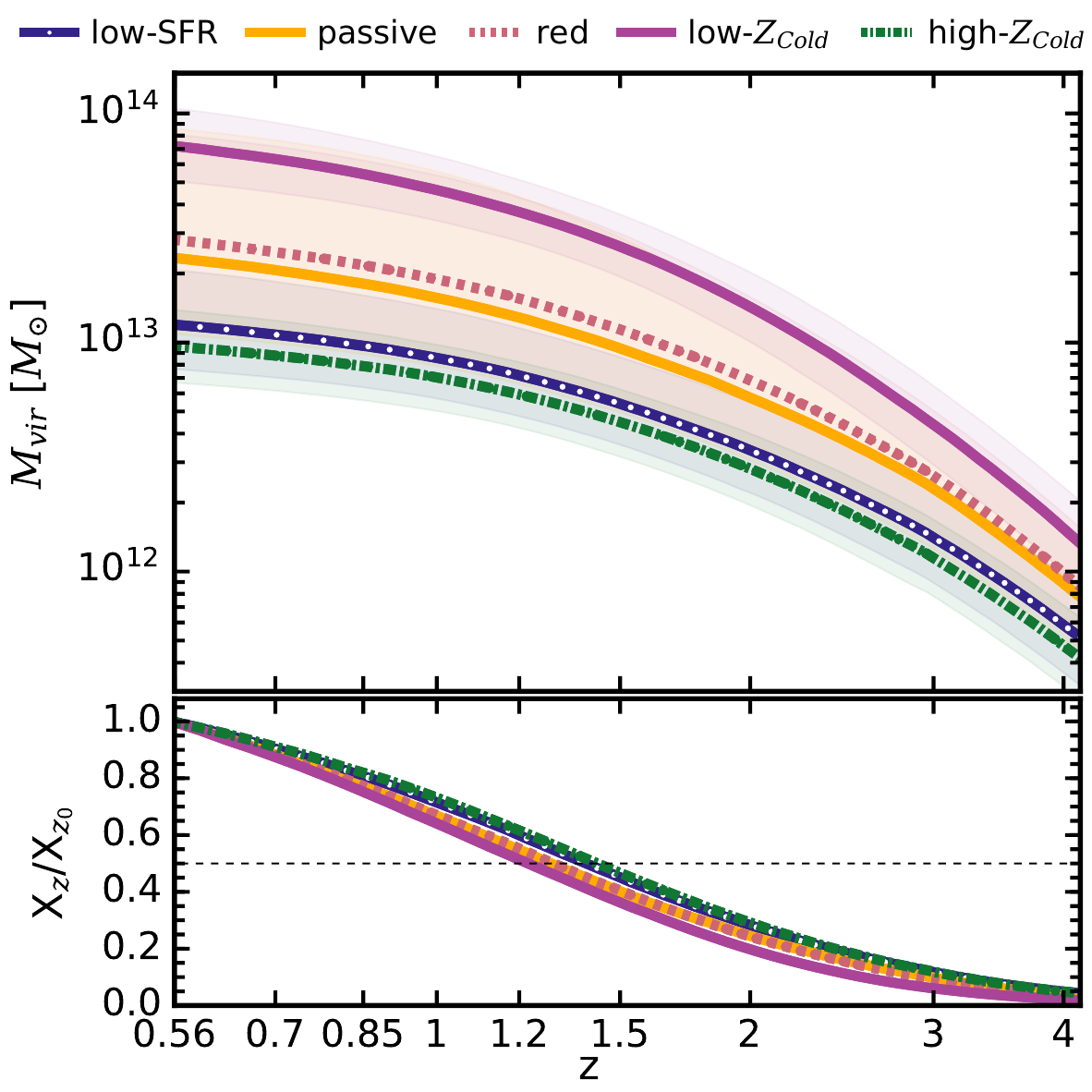}\vspace{-0.3cm}%
	\caption{In the upper panel we show the redshift evolution of median values of the halo mass, \Mvir, according to the selection procedure outlined in \hyperref[fig:cartoon_M2]{\Fig{fig:cartoon_M2}} for our five selected sub-samples: \low\ (solid blue line with white dots indicating the redshift values of each snapshot), \passive\ (solid light yellow line), \red\ (short-dashed red line), \lowZ\ (solid dark magenta line), and \highZ\ (dotted-dashed green line). The shaded regions represent the range spanning between the $32^{\mathrm{th}}$ and the $68^{\mathrm{th}}$ percentile around the median. Their corresponding mass growth history relative to the reference redshift of our study, $\zstart=0.56$, as $z_0$. In this panel, $\rm X_{z}$ denotes the values of \Mvir\ at a specific snapshot/redshift compared to the values at $z_0$ ($\rm X_{z_0}$). The horizontal dashed black line indicates the threshold of 50\% of the total halo mass at \zstart.}\label{fig:res_zevol_mvir_M2}\vspace{-0.4cm}
\end{figure}

In the upper panel of \hyperref[fig:res_zevol_mvir_M2]{\Fig{fig:res_zevol_mvir_M2}}, we show the redshift evolution of the halo mass, \Mvir\footnote{The \galacticus\ model uses the mass definitions provided by Eq.(6) in \citet{Bryan+Norman98} to define the dark matter halo mass. For further details, see Sec.\ 2.5 and Eqs.\ (7) and (8) in \citet{Knebe17_MD}}, for our five selected sub-samples: \low\ (solid blue line with white dots indicating the redshift values of each snapshot), \passive\ (solid light yellow line), \red\ (short-dashed red line), \lowZ\ (solid dark magenta line), and \highZ\ (dotted-dashed green line). In this figure and the following, we show median values of all galaxies present in each sub-sample along with the range spanning between the $32^{\mathrm{th}}$ and the $68^{\mathrm{th}}$ percentile, shown as shaded coloured regions using the above-defined colour and line style keys. The lower panel of the same figure corresponds to their halo mass growth history (X$_{z}$/X$_{z_0}$) with X$_z$ being the halo mass at a specific snapshot/redshift and X$_{z_0}$ being the halo mass they hold at the reference redshift of our study $\zstart=0.56$. The lower panel utilises the same colour scheme, line style keys, and statistical methods as in the upper panel.

Our results indicate that all defined sub-samples exhibit comparable evolutionary tracks but reach slightly different halo masses at \zstart. The \low\ and \highZ\ sub-samples, as well as \passive\ and \red\ sub-samples, show very well-aligned evolution, reaching the lowest ($\Mvir\sim10^{13}$ \Msun) and intermediate ($\Mvir\sim2\times10^{13}$ \Msun) mass regimes, respectively. In contrast, \lowZ\ galaxies are exceptions, acquiring significantly higher halo masses of around $\Mc\sim10^{14}$ \Msun\ compared to galaxies from the other four sub-samples. However, all galaxies still assemble half of their masses at similar redshifts between $1.2 < z < 1.4$.

The redshift evolution of the stellar mass, \Mstar, mirrors the trends observed for halo mass evolution, therefore a separate plot is not provided. Instead, the following results are reported: aligned with results on the halo mass, the \lowZ\ galaxies consist also of the most massive ones in stellar mass which assemble half of their \Mstar\ at $z\sim1.2$, while \low\ and \highZ\ galaxies completed half of their mass assembly already at $z\sim1.5$. Furthermore, the \lowZ\ (\highZ) galaxies show the highest (lowest) stellar-to-halo mass ratio, $\mathrm{SHMR}~=~\Mstar/\Mvir$. Other sub-samples show intermediate values, with \low-galaxies holding values comparable to \highZ\ galaxies, and \red- and \passive-galaxies values similar to \lowZ. Interestingly, the evolution of the SHMR as a function of redshift peaks at $z\sim3.5$ with \SHMR$~\sim~0.01$ for the \low\ and \highZ\ samples. A similar peak can be found for the rest of the sub-samples but slightly later. Furthermore, from $z\sim1.5$ to lower redshifts the \SHMR\ evolution is almost constant across all sub-samples.

\subsection{Redshift evolution of the cold gas and black hole masses}\label{sec:res_mcold_mbh}
but assemble half of their final halo mass either later at  $z\sim0.9$ (\low, \passive, \red) or slightly earlier at $z\sim1.2$ (\low\ and  \highZ). Notably
\begin{figure}
	\centering
	\includegraphics[width=\columnwidth,angle=0]{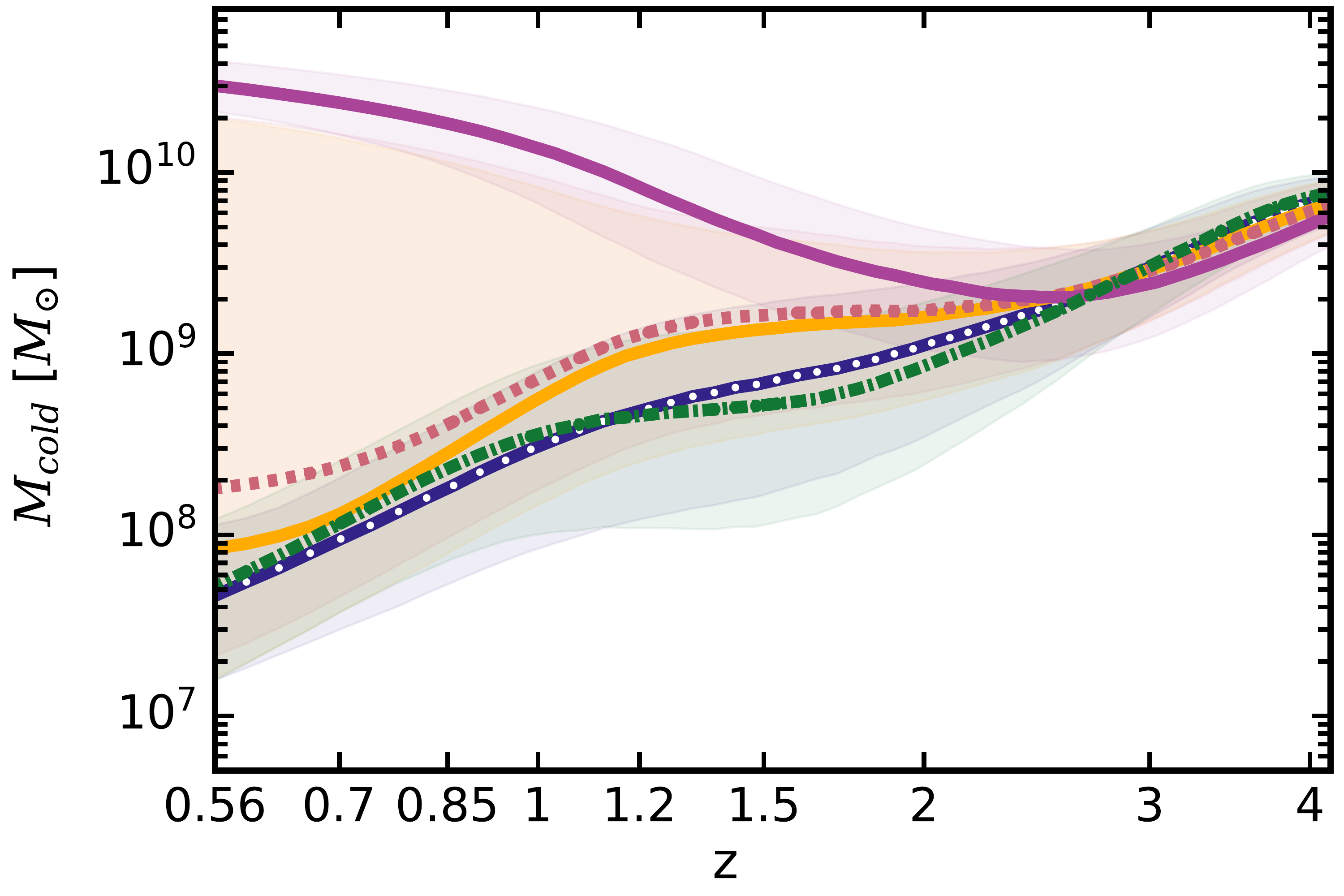}\vspace{-0.3cm}%
	\caption{The redshift evolution of median values of the cold gas mass, \Mcold\ -- the fraction of gas available to be converted into stars -- for the different sub-samples. The figure utilises the same colour scheme, line style keys, and statistical methods as in \hyperref[fig:res_zevol_mvir_M2]{\Fig{fig:res_zevol_mvir_M2}}.}\label{fig:res_zevol_mcold_M2}\vspace{-0.4cm}
\end{figure}

\begin{figure}
	\centering
	\includegraphics[width=\columnwidth,angle=0]{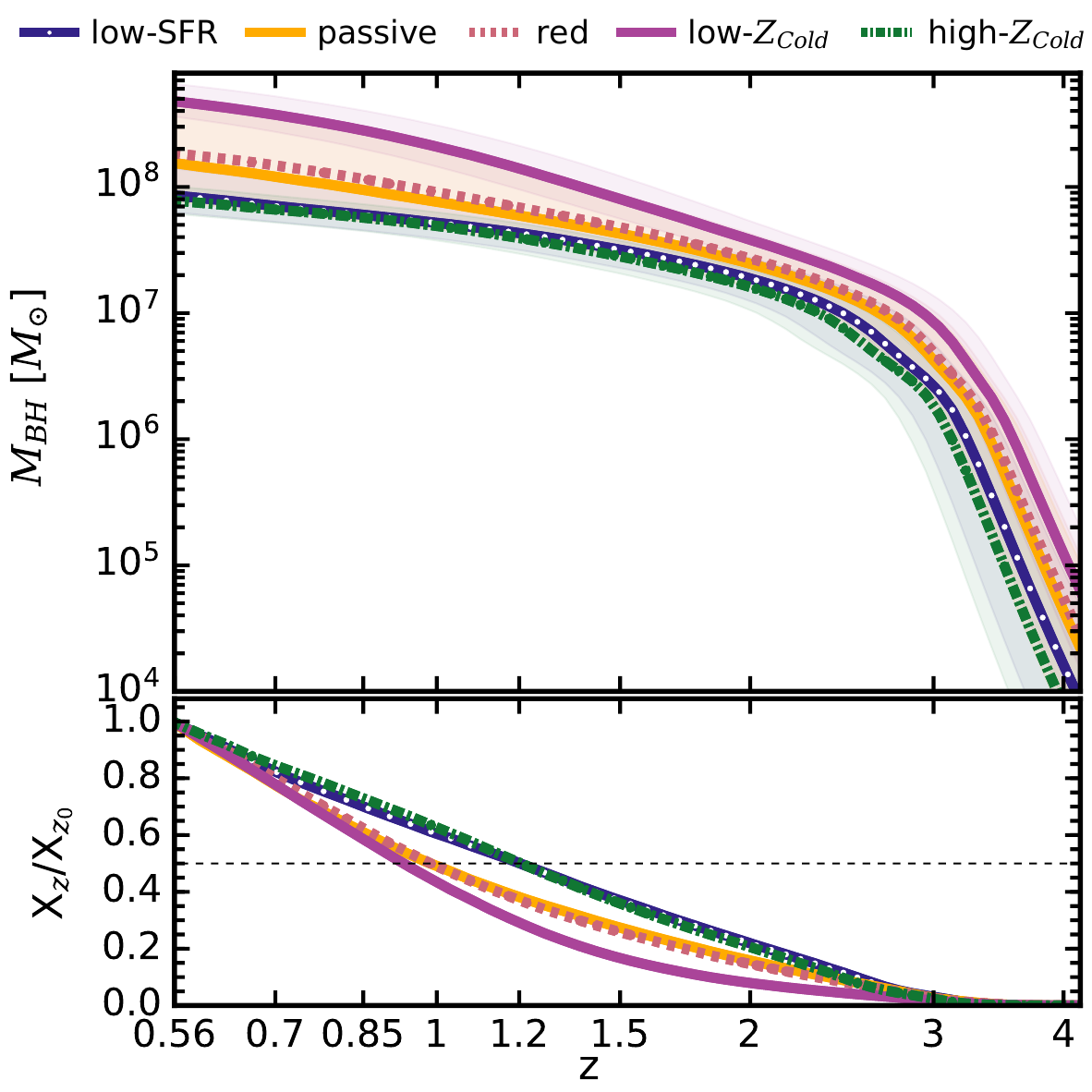}\vspace{-0.3cm}%
	\caption{In the upper panel the redshift evolution of median values of the galaxy's super-massive black hole, \Mbh\, is shown. The corresponding mass growth history relative to the reference redshift $\zstart=0.56$ of our study, is displayed in the lower panel, following the same definitions as in \hyperref[fig:res_zevol_mvir_M2]{\Fig{fig:res_zevol_mvir_M2}}. The figure utilises the same colour scheme, line style keys, and statistical methods as in \hyperref[fig:res_zevol_mvir_M2]{\Fig{fig:res_zevol_mvir_M2}}.}\label{fig:res_zevol_mbh_M2}\vspace{-0.4cm}
\end{figure}

After examining the evolution of stellar and halo masses, the next step is to investigate the assembly histories of the corresponding cold gas, cold-gas fraction (CGF), and central black hole masses. In \hyperref[fig:res_zevol_mcold_M2]{\Fig{fig:res_zevol_mcold_M2}} we show the redshift evolution of the cold gas mass, \Mcold, which represents the gas available for conversion into stars. The steady growth in halo and stellar mass is supported by a consistently declining supply of \Mcold\ and a decreasing cold gas fraction, CGF$~=~\Mcold/\Mstar$, towards later cosmic times for all sub-samples except \lowZ. \lowZ\ galaxies exhibit a constant CGF and maintain an extensive reservoir of cold gas. Notably, during their late-time evolution after $z\sim2$, they were able to accumulate additional cold gas, resulting in a larger reservoir at lower redshifts compared to higher redshifts. This suggests that these galaxies are gaining additional fuel through their merger activity and smooth accretion from the cosmic web. This scenario is consistent with the evolution of their black hole masses, \Mbh, as shown in \hyperref[fig:res_zevol_mbh_M2]{\Fig{fig:res_zevol_mbh_M2}}. In other words, the most massive galaxies also exhibit the highest \Mbh\ and possess the largest reservoir of \Mcold\ to sustain their continued star formation. Conversely, galaxies with lower \Mvir, \Mstar, and \Mbh\ consume their cold gas at higher redshifts. These galaxies were among the first and fastest to assemble half of their stellar and black hole masses (see, e.g., the \highZ\ and \low\ samples in the lower panel of \hyperref[fig:res_zevol_mbh_M2]{\Fig{fig:res_zevol_mbh_M2}}). The following sections will explore potential reasons for the significant difference in evolution observed in the \Zcold\ sub-samples.

\subsection{Redshift evolution of intrinsic galaxy properties}\label{sec:res_intr}

\begin{figure}
	\centering
	\includegraphics[width=\columnwidth,angle=0]{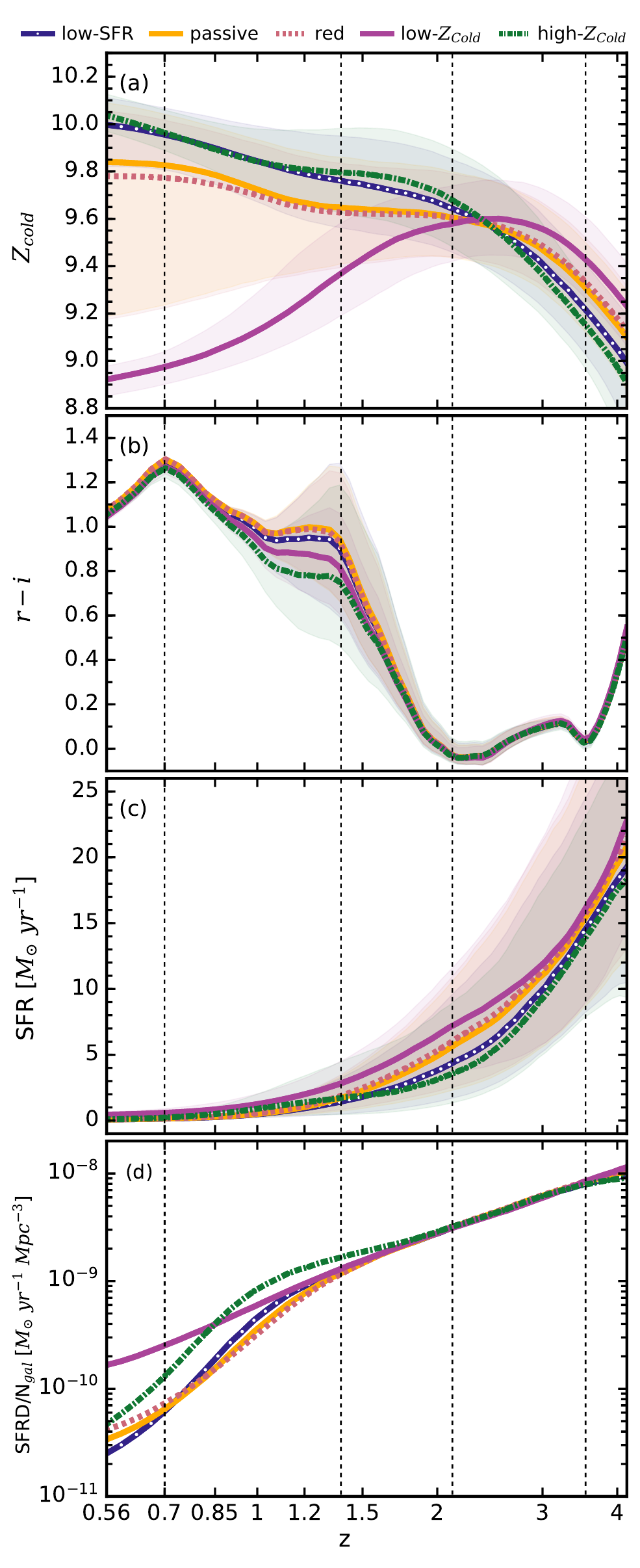}\vspace{-0.35cm}%
	\caption{From top to bottom, we present the redshift evolution of the following galaxy properties: $(a)$ the cold gas-phase metallicity, \Zcold, $(b)$ the observed colour, \ri, $(c)$ the star formation rate, \SFR, and $(d)$ the star formation rate density, \SFRD, normalised by the number of galaxies in each sub-sample at a particular redshift. The vertical thin dashed lines mark redshifts of prominent line features in \ri. The figure utilises the same colour scheme, line style keys, and statistical methods as in \hyperref[fig:res_zevol_mvir_M2]{\Fig{fig:res_zevol_mvir_M2}}.}\label{fig:res_more_props}\vspace{-0.4cm}
\end{figure}

Until now, the discussion has centred on the assembly and growth history of mass-related properties. Here the focus shifts to additional properties such as star formation, metallicity, and colour. Therefore, we show in \hyperref[fig:res_more_props]{\Fig{fig:res_more_props}}, from top to bottom, the redshift evolution is presented for: $(a)$ the cold gas-phase metallicity, \Zcold, $(b)$ the observed \sdss\ colour \ri, $(c)$ the star formation rate, \SFR, and $(d)$ the star formation rate density, \SFRD, normalised by the number of galaxies in each sub-sample at a particular redshift. Thin vertical dashed lines mark the redshifts $z=[0.7,1.4,2.1,3.5]$, highlighting prominent features in the \ri\ colour evolution in panel $(b)$.
	
At low redshift, \highZ\ galaxies are the most metal-enriched, as expected from their selection criteria. Galaxies of the \low\ sample exhibit comparable metallicities to \highZ\ galaxies, while those in the \red\ and \passive\ samples show the second highest gas-phase metallicities. As anticipated, the \lowZ\ galaxies are the least metal-enriched at \zstart. Interestingly, at $z\sim2.5$, this trend reverses, with \lowZ\ galaxies becoming the most metal-enriched, the highest \Zcold. Moreover, the \highZ\ and \low\ samples exhibit rapid metal production at higher redshift, as indicated by their steeper slopes in the \Zcold\ evolution between $2<z<3$ in panel $(a)$ of \hyperref[fig:res_more_props]{\Fig{fig:res_more_props}}, compared to the other sub-samples. After this period, the production rate slows down between $z\sim2$ and \zstart. In contrast, \lowZ\ galaxies show a peak in \Zcold\ between $2\lesssim z\lesssim3$, followed by a continuous decline at later times.

In comparison to \Zcold, we find little variation in the evolution of \ri-colour with cosmic time among our considered sub-samples, as shown in panel $(b)$ of \hyperref[fig:res_more_props]{\Fig{fig:res_more_props}}. The only exception is a short time interval of $1\lesssim z\lesssim1.5$ where the \low, \passive, and \red\ samples exhibit colours with \ri\ 0.25 redder than the \lowZ\ and \highZ\ samples. This is puzzling, as their colour evolution is otherwise similar before and after this period. A similar behaviour in colour evolution has been suggested in another study using the same galaxy formation model (private communication with Tancara, in prep.). The cause of this gap between sub-samples is currently unclear, but Tancara et al. discuss this aspect in more detail in their upcoming work. Moreover, during this time interval, the colour evolution remains constant across all sub-samples. It is also noteworthy that the \ri\ evolution demonstrates four prominent features, highlighted by vertical dashed lines: a maximum at $z\sim0.7$, the edge of a constant colour evolution interval from $1\lesssim z\lesssim1.5$ as mentioned above, and two minima at $z\sim2$ and $z\sim3.5$, respectively.

The first minimum in the colour evolution occurs around $z\sim3.5$ as a short, rapid drop, followed by a prominent minimum at $z\sim2$ (panel $(b)$), coinciding with the ``cosmic noon'' -- the peak of star formation in cosmic history \citep{Madau+Dickinson14}. During this period, all galaxies discussed in this study reached their bluest colours. Simultaneously, the \lowZ\ sample shows a peak in \Zcold\ (panel $(a)$). Interestingly, until $z\sim1.5$, all galaxy samples exhibit the same median colour evolution and minima in both \ri\ and \gi\ colours. However, their median star formation rates (SFRs) differ, as shown in panel $(c)$. The \lowZ, \passive, and \red\ samples display higher SFRs of approximately $sim$3-5 \Msunyr, compared to the lower rates seen in \low\ and \highZ\ galaxies. After $z\sim2$, the galaxies undergo constant reddening until $z\sim1.5$, followed by a period of no significant evolution until $z\sim1$. This epoch aligns with the time when half of the stellar and halo masses were assembled in all samples, except for \lowZ. Furthermore, their corresponding mass growth functions reverse their curvatures -- from growing more rapidly to more slowly -- at this redshift (see the lower panel in \hyperref[fig:res_zevol_mvir_M2]{\Fig{fig:res_zevol_mvir_M2}}).

At the last feature, a prominent peak in \ri\ at $z\sim0.7$, all galaxies reached their reddest colours in \ri\ independently of their sub-sample assignment. A similar evolution from bluer to redder colours until $z\sim0.7$ was also reported by \citet[][]{Maraston09} for modelled luminous red galaxies. In addition, the same peak can be observed in the \gi\ colour at the same redshift for the \lowZ\ and \highZ\ samples, while for the rest of the sub-samples, this peak occurs slightly earlier in cosmic time, at $z\sim0.85$.

The star formation rate density (\SFRD), normalised by the galaxy number count (N$_{\mathrm{gal}}$) in each sub-sample, is shown in the panel $(d)$ of the same figure and does not indicate strong variations between the samples at cosmic noon. However, slightly higher \SFRD/N$_{\mathrm{gal}}$ values are measured for the \highZ\ galaxies around $z\sim1.5$, coinciding with the edge of the no-evolution period in colour seen in panel $(b)$. This trend is also reflected in the specific star formation rate (\sSFR), where \highZ\ galaxies consistently show higher, or at least similar, median \sSFRs\ as other sub-samples at $z>0.85$. Interestingly, the \SFRD/N$_{\mathrm{gal}}$ of our sub-samples does not show the expected peak around cosmic noon. It is important to note that we normalised the \SFRD\ by the comoving volume and the number of galaxies in each sub-sample to ensure an unbiased comparison across sub-samples. This results in lower values than the cosmic star formation rate density presented by \citet{Madau+Dickinson14}. As shown in the figure, the curves for all sub-samples indicate a moderate increase in evolution at higher redshift. Unfortunately, our ability to explore this further is limited by the available merger tree data, which only tracks galaxies consistently up to $z=4.15$. Nonetheless, an earlier peak in the cosmic star formation rate density\footnote{Note that the cosmic star formation rate density is the cumulative sum of star formation rates normalised by the physical volume, but not additionally normalised by the number density of galaxies in each sub-sample.} of all galaxies in the \galacticus\ model, as shown in Fig.\ 4 of the \MDG\ release paper \citep{Knebe17_MD}, between $z\sim3-4$ supports this hypothesis and may explain why we do not observe a peak at cosmic noon.

To summarise this section, our results indicate that \lowZ\ galaxies consistently exhibit significantly higher \Mstar, \Mvir, as well as the largest black hole mass \Mbh\ in comparison to other sub-samples, including \highZ\ galaxies, as shown in \hyperref[fig:res_zevol_mvir_M2]{\Figsstart{fig:res_zevol_mvir_M2}}-\hyperref[fig:res_zevol_mbh_M2]{\Figsend{fig:res_zevol_mbh_M2}}. These galaxies also accumulate large reservoirs of cold gas mass, \Mcold, (unlike all other sub-samples) and terminate their evolution with more \Mcold\ than they possessed initially at high redshift. Furthermore, they assemble their mass later and more rapidly, and produce stars more efficiently due to their abundant supply of \Mcold\ compared to \highZ\ galaxies.

The \highZ\ galaxies, on the other hand, sit on the lower mass end of the spectrum for \Mvir, \Mstar, as well as \Mbh\ and therefore completed their mass assembly earlier and more continuously than their \lowZ\ counterparts. That is reflected in their lower cold gas fraction in comparison to \lowZ\ galaxies. In our companion paper S19 it was noted that there is a prominent bimodality in \Zcold\ at the initial redshift of our study $\zstart=0.56$. As discussed above, the authors could map this bimodality in \lowZ\ and \highZ\ galaxies on specific galaxy and halo properties. We confirmed their hypothesis that \highZ\ and \lowZ\ galaxies form via distinct pathways and correspond to two separate and distinguishable samples of galaxies that present the overall population of luminous and massive galaxies at \zstart. We aim at understanding what drives their distinct evolution via studying their clustering and location in the large-scale environment of the cosmic web.

\subsection{Galaxy clustering}\label{sec:res_clustering}

In this section, we study the galaxy clustering of our different sub-samples of galaxies through the real-space two-point correlation function (\twoPCF), $\xi(r)$. We use the \textsc{Corrfunc} software package\footnote{\url{http://corrfunc.readthedocs.io/en/master/index.html}\label{ftn:corrfunc}} from \citet{Sinha+Lehman17_Corrfunc} and the standard \citet{Landy+Szalay93} estimator. We calculate \twoPCFs\ with 25 logarithmic-spaced bins in the range of $0.5<r~[\Mpc]~<150$ assuming periodic boundary conditions. As stated previously, the galaxy samples have the same number density but different mean values for the stellar and halo masses which will be reflected in the clustering.

\begin{figure}
	\includegraphics[width=\columnwidth,angle=0]{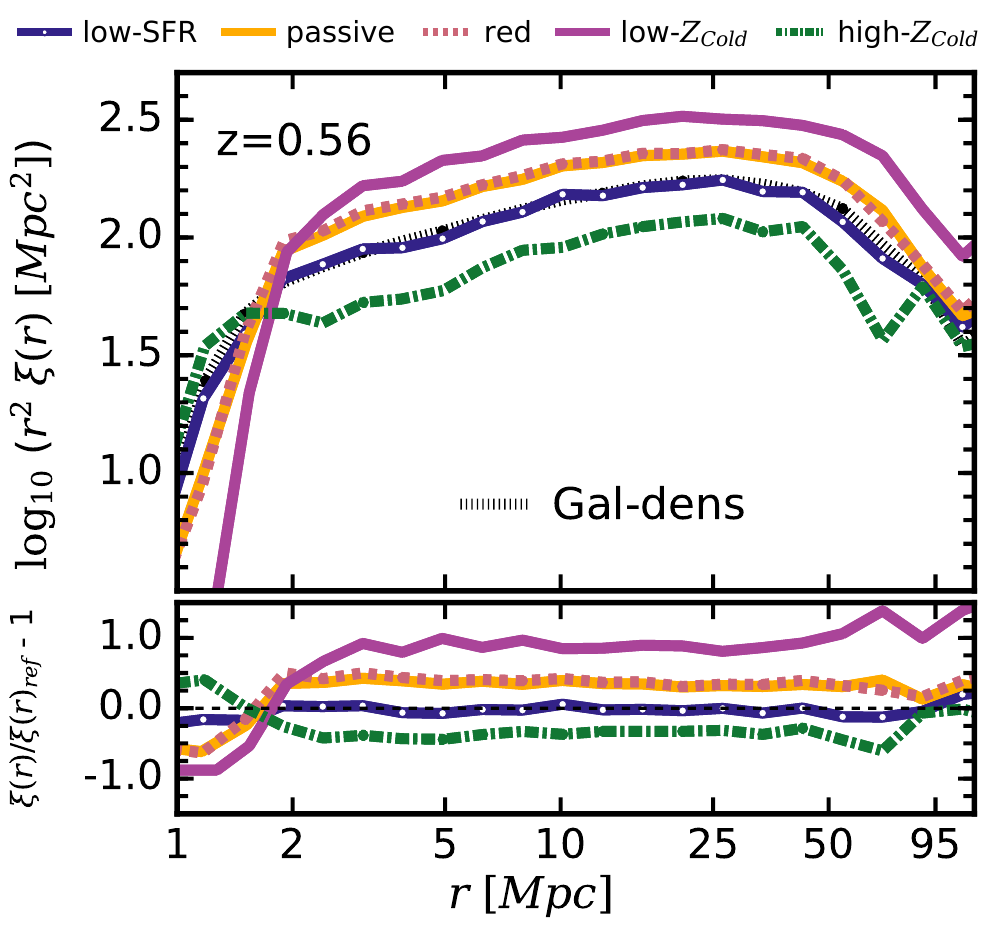}\vspace{-0.2cm}%
	\caption{In the upper panel, we show the real-space two-point correlation function, $\xi(r)$, at redshift $\zstart=0.56$ for all sub-samples (using their respective colour scheme and line style keys as in \hyperref[fig:res_zevol_mvir_M2]{\Fig{fig:res_zevol_mvir_M2}}) and the SAM-CMASS mock-galaxies parent sample, \den\ (short-dashed black line). In the lower panel we display the fractional difference between the clustering function of each sub-sample ($\xi(r)$) and that of \den\ ($\xi(r)_{\mathrm{ref}}$).}\vspace{-0.4cm} \label{fig:res_xi_M2}
\end{figure}

In the upper panel of \hyperref[fig:res_xi_M2]{\Fig{fig:res_xi_M2}} we show the \twoPCF\ of central galaxies at redshift $\zstart=0.56$ for all sub-samples (using the same colour and line style choices as in previous sections). We also include the result from the entire sample of SAM-CMASS mock-galaxies, \den,  as a reference (short-dashed black line). In the lower panel of the same figure, we plot the fractional difference of $\xi(r)$ for each sub-sample with respect to the function of \den, $\xi(r)_{\mathrm{ref}}$. 

As our results indicate, the \low\ sample and the entire sample of SAM-CMASS mock-galaxies, \den, have near-identical correlation \twoPCFs. This means that the 20\% of low-starforming galaxies can mimic the clustering of the entire population of SAM-CMASS mock-galaxies. As we previously described, \den\ is the parent sample from which all sub-samples were extracted following a certain selection criterion listed in \hyperref[tab:subsel]{\Tab{tab:subsel}}. This may largely be coincidental, as predictions for galaxy and halo properties from both the parent sample and the \low\ sub-sample at \zstart\ as well as their subsequent evolution show no comparable trends.

Additionally, the clustering functions of our \passive\ and \red\ samples exhibit very similar \twoPCFs, while \lowZ\ galaxies show a significantly higher clustering amplitude. In contrast, the least clustered sample is \highZ, which displays the lowest amplitudes except for very small separations. A slight turnover in the clustering strength is observed at scales smaller than $r<2$ \Mpc, where \highZ\ (\lowZ) galaxies cluster more (less) strongly. Interestingly, the \lowZ\ and the \highZ\ samples represent the upper and lower limits in the total clustering strengths. These differences in the clustering are driven by the mean halo masses, with \passive, \red, and \lowZ\ galaxies typically residing in the most massive and consequently the most clustered halos.

We also investigate the redshift evolution of the real-space clustering function, though we do not dedicate a separate plot to it, as the clustering strength shows only mild variation with redshift. Galaxies in the \lowZ\ (\highZ) sample are always more (less) strongly clustered while the \low\ sample shows an intermediate strength between \lowZ\ and \highZ. As expected, these findings with the cosmic evolution of the halo masses of each galaxy sub-sample. At smaller separations, the clustering signal for \lowZ\ galaxies decreases rapidly with increasing redshift, whereas \low\ and \highZ\ galaxy pairs remain detectable at $r<2~\Mpc$ or $r<5~\Mpc$ at $z=0.7$ or $z=3.51$, respectively. Due to the given limitation in particle resolution and simulation box side-length, the \twoPCFs\ exhibit growing uncertainties at larger separations.

\subsection{Galaxy properties and the large-scale environment}\label{sec:res_envr}

\begin{figure*}
	\centering
	\includegraphics[width=\textwidth,angle=0]{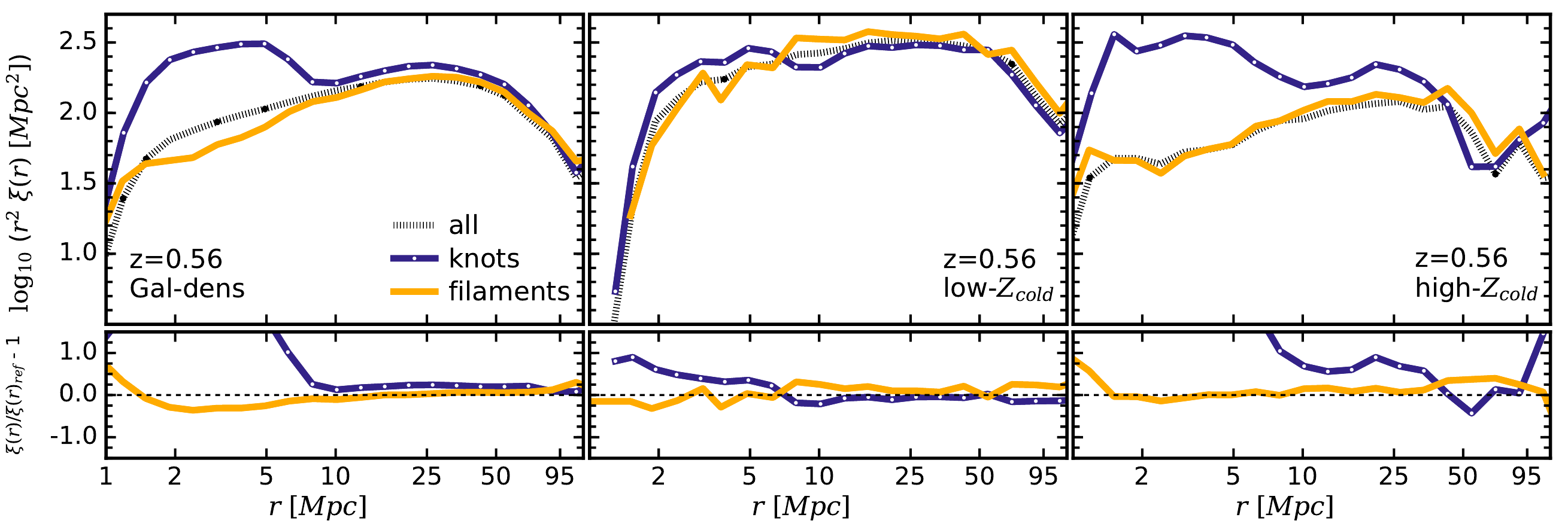}\vspace{-0.2cm}%
	\caption{In the upper panels, we show the real-space two-point correlation function, $\xi(r)$, at $z=0.56$ for galaxies in the parent sample, \den\ (left panel), and for sub-samples \lowZ\ (middle panel) and \highZ\ (right panel). In each panel we show the clustering for all galaxies in the sample (dashed black line), knot galaxies (solid yellow line), and filament galaxies (solid blue line with white dots). In the lower panel, we present the fractional difference in the clustering function of the filament and knot populations ($\xi(r)$), respectively, with respect to the clustering of the entire sample (``all'', $\xi(r)_{ref}$).}\label{fig:res_zcolds_xi}
\end{figure*}

We have shown that the \lowZ\ and \highZ\ galaxy samples represent the upper and lower limits to the parameter space of various galaxy properties (see e.g.\ \hyperref[fig:ssfr2mstar]{\Fig{fig:ssfr2mstar}}, \hyperref[fig:g-i2mhalo]{\Fig{fig:g-i2mhalo}}, \hyperref[fig:res_zevol_mvir_M2]{\Fig{fig:res_zevol_mvir_M2}}, or \hyperref[fig:res_zevol_mbh_M2]{\Fig{fig:res_zevol_mbh_M2}}). In this section, we revisit our analysis by classifying the galaxies in our defined sub-samples according to their large-scale environmental affiliation of the cosmic web, categorised as ``more-dense'' or ``less-dense'' regions. To this extent, we apply the {\sc Vweb} code \citep{Hoffman2012, Libeskind2012, Libeskind2013, Carlesi2014, Cui18a, Cui19} applying it to the dark matter catalogue that underpins our galaxy formation model. The code determines the environmental affiliation of the dark matter halos in which the galaxies reside according to ``knots'', ``filaments'', ``sheets'', and ``voids'' as already discussed in our companion paper S19 (see Appendix A for details). In this analysis, we adopt the categories knots as ``more-dense'' and filaments as ``less-dense'' regions.

The upper panels of \hyperref[fig:res_zcolds_xi]{\Fig{fig:res_zcolds_xi}} show the \twoPCF\ from left to right for \den, \lowZ, and \highZ\ galaxies in knots (dark blue lines with white dots) and filaments (light yellow lines) as well as for all galaxies regardless of the environment (short-dashed black lines). The lower panels show the fractional difference between the knot and filament populations relative to the corresponding full sample. As expected, we observe a variation in clustering strength depending on separation length and large-scale environment, as demonstrated for the entire SAM-CMASS mock-galaxy sample (left panel). Specifically, at smaller separations, $r<10~\Mpc$, knot galaxies (filament galaxies) cluster more (less) strongly, while at larger separations, the clustering strength becomes similar across both environments. The same behaviour is observed for \highZ\ galaxies up to $r<25~\Mpc$. However, unexpectedly, \lowZ\ galaxies do not follow the same trend; instead, the clustering strength of knot, filament, and the overall population of \lowZ\ galaxies is relatively similar. Notably, at the separation length of $r\sim10~\Mpc$, we find a dip in the clustering for knot galaxies, but this feature is absent for filaments galaxies. Moreover, this dip is more pronounced in \highZ\ galaxies compared to \lowZ\ ones.

To complement our galaxy clustering analysis, we compute the median values of the galaxy and halo properties for \lowZ\ and \highZ\ galaxies in filaments and knots shown in \hyperref[tab:zcolds_props]{\Tab{tab:zcolds_props}}. As previously reported by S19, we find a clear tendency for the halo mass, \Mvir, to correlate with the environment. In more detail, the knot (filament) population of \highZ\ galaxies exhibits median halo masses within the $32^{\mathrm{th}}$ and the $68^{\mathrm{th}}$ percentile of \logT(\Mvir\ [\Msun])~=~13.09$_{-0.17}^{+0.17}$ (\logT(\Mvir\ [\Msun])~=~12.97$_{-0.15}^{+0.15}$) compared to the \lowZ\ galaxies of \logT(\Mvir\ [\Msun])~=~3.93$_{-0.14}^{+0.15}$ (\logT(\Mvir\ [\Msun])~=~13.70$_{-0.12}^{+0.14}$), respectively. A visually demonstration and further description can be found in \hyperref[fig:zcolds_mhalo]{\Fig{fig:zcolds_mhalo}} in \hyperref[app:zcolds_mhalo]{\App{app:zcolds_mhalo}}. 

As expected, \lowZ\ galaxies have significantly higher halo masses and are predominantly located in knots (60\%) and, to a lesser extent, in filaments (38\%)\footnote{We do not consider the remaining 2\% of sheet and void galaxies since their pair counts are negligible.}. In contrast to \highZ\ galaxies which are primarily found in filaments (62\%) and in knots (24\%)\footnote{In comparison to \lowZ\ galaxies, the population of ``sheet'' galaxies is larger for the \highZ\ sub-sample, reaching around 14\%. Although that amount is considerably larger than for \lowZ, further discussion on this topic is beyond the scope of this paper and will be explored in future work.}. 

For \lowZ\ galaxies, we observe a clear environmental correlation where knot galaxies (compared to filament galaxies) possess larger (smaller) values for \Mvir, \Mstar, \Mbh, and \Tcons, but lower (higher) values for \SFR, \sSFR, and \con. Here, $\Tcons = \Mcold / \SFR / 10^9$ in \Gyr\ is the cold gas consumption (or depletion) time, representing the efficiency with which the galaxy converts its cold gas into stars based on its current star formation rate, while \con\ denotes the concentration of the Navarro–Frenk–White dark matter halo profile as defined by \citet{Navarro97}. Notably, \lowZ\ galaxies generally contain an order of magnitude more gas and metals in their hot halo (\Mhot\ and \Mzhot, respectively), as well as more metals in stars (\Mzstars), but hold three orders of magnitude more cold gas (\Mcold).

Furthermore, \highZ\ galaxies exhibit higher stellar-to-halo mass ratios (SHMR) and lower cold gas fractions, \cgf, (s)SFRs. The cold gas consumption time for \highZ is approximately 0.5 \Gyr\ for \highZ, while the measurement exceeds the age of the Universe multiple times. Interestingly, despite the fundamental differences in \Zcold\ between the \highZ\ and \lowZ\ sub-samples, both hold identical stellar metallicity values, $Z_*$\footnote{We note that the stellar metallicity is calculated similarly to the cold gas-phase metallicity, except that the mass of metals in the cold gas is replaced by the mass of metals in stars, \Mzstars, normalised by the total stellar mass, \Mstar\ (see \hyperref[sec:method]{\Sec{sec:method}}).}.

The black-hole-to-halo mass ratio, \bheff, also known as the black hole efficiency, a diagnostic tool that indicates how responsive a galaxy might be to AGN activity and how efficiently it can transport gas to its central region, which fuels the central black hole \citep[see e.g.][]{Ferrarese02_bheff,Croton09_bheff}. The relation between \Mbh\ and \Mvir\ is assumed to be tight and redshift dependent (as we will discuss in the next section) \citep[see e.g.][]{Wyithe+Padmanabhan06_bheff,Booth+Schaye11_bheff}, with their evolution being closely linked \citep[see][and the citations therein]{Powell22_mbh-mhalo}. We find that our considered sub-samples, as well as the parent sample, show similar values for \bheff, but with generally lower efficiencies in knots compared to filaments. Along with the colour indices, \gi\ and \ri, the stellar metallicity, $Z_*$; these properties do not show a dependency with environment as demonstrated in \hyperref[tab:zcolds_props]{\Tab{tab:zcolds_props}}.

A significant difference between \lowZ\ and \highZ\ galaxies is observed in the mass of cold gas, \Mcold, the cold gas fraction, \cgf, the gas consumption time, \Tcons, and the specific angular momentum of the baryons, \jbar, being the sum of the angular momenta of the stellar spheroid and stellar disc components normalised by the sum of the total stellar and cold gas masses: $\jbar=(J_{\rm{*,disk}}+J_{\rm{*,bulge}})/(\Mstar+\Mcold)$. Generally, we observe values for \jbar\ ranging approximately between $\logT(\jbar\ [kpc \kms])\sim3-4$, consistent with results from other models and observations \citet[see e.g.][]{Mancera-Pina21_jbar,Rodriguez-Gomez22_TNG_angMom} for all sub-samples except \lowZ. In addition, those galaxies also exhibit a slight environmental dependence: \lowZ\ galaxies in knots exhibit higher \jbar\ than those in filaments. Although the \lowZ\ sample includes the most massive ones, their angular momenta, around $\logT(\jbar\ [kpc \kms])\sim5$, exceed the expected range based on the well-known power-law relation \citet[e.g.][]{Fall83,Romanowsky+Fall12_angMom} by one order of magnitude. Furthermore, throughout their history, \lowZ\ galaxies consistently exhibit higher angular momenta than other sub-samples; however, at $z\sim1.3$, their momenta experience a significant boost. While the exact cause of this deviation remains uncertain, we speculate that merger events may be responsible for this outcome. It is noteworthy mentioning, that all of these properties are linked to the cold gas mass, which ultimately influences the cold gas-phase metallicity, \Zcold, so these differences are expected.

\begin{table*}
	\begin{center}
        \centering\caption{Median values of galaxy and halo properties in large-scale environments.}\vspace{-0.3cm}
		\begin{tabular}{
			M{0.03\linewidth}|M{0.22\linewidth}||M{0.1\linewidth}|M{0.1\linewidth}||M{0.1\linewidth}|M{0.1\linewidth}||M{0.1\linewidth}|M{0.1\linewidth}}
			\hline
			\parbox[t]{10mm}{\multirow{2}{*}{\rotatebox[origin=c]{90}{\parbox{1.5cm}{\centering \tiny{property type}}}}} & sub-sample name& \multicolumn{2}{c||}{\multirow{1}{*}{\parbox{0.15\linewidth}{\centering \den}}} & \multicolumn{2}{c||}{\multirow{1}{*}{\parbox{0.15\linewidth}{\centering \highZ}}} & \multicolumn{2}{c}{\multirow{1}{*}{\parbox{0.15\linewidth}{\centering \lowZ}}}\\
			\cline{2-8} 
			& environment & filaments\par (52\%) & knots\par (41\%) & filaments\par (62\%) & knots\par (24\%) & filaments (32\%) & knots\par (67\%)\\
			\hline
			\hline
\parbox[t]{5mm}{\multirow{2}{*}{\rotatebox[origin=c]{90}{halo}}} & $\log_{10}(M_{\mathrm{vir}}$ [$M_{\odot}$])&	13.27$_{-0.18}^{+0.17}$	&	13.61$_{-0.20}^{+0.19}$	&	12.97$_{-0.15}^{+0.15}$	&	13.10$_{-0.17}^{+0.17}$	&	13.70$_{-0.12}^{+0.14}$	&	13.93$_{-0.14}^{+0.15}$	\\\cline{2-8}   
& c$_{\mathrm{NFW}}$&	5.11$_{-0.15}^{+0.17}$	&	4.82$_{-0.14}^{+0.16}$	&	5.41$_{-0.15}^{+0.16}$	&	5.28$_{-0.16}^{+0.18}$	&	4.76$_{-0.10}^{+0.09}$	&	4.59$_{-0.09}^{+0.09}$	\\\hline
\parbox[t]{5mm}{\multirow{19}{*}{\rotatebox[origin=c]{90}{galaxy}}} &$\log_{10}(M_{*}$ [$M_{\odot}$])&	11.14$_{-0.10}^{+0.09}$	&	11.28$_{-0.10}^{+0.10}$	&	10.97$_{-0.08}^{+0.07}$	&	11.01$_{-0.08}^{+0.07}$	&	11.37$_{-0.06}^{+0.08}$	&	11.46$_{-0.08}^{+0.09}$	\\\cline{2-8}
&$\log_{10}(M_{\mathrm{BH}}$ [$M_{\odot}$])&	8.20$_{-0.17}^{+0.18}$	&	8.43$_{-0.18}^{+0.18}$	&	7.88$_{-0.10}^{+0.11}$	&	7.92$_{-0.11}^{+0.12}$	&	8.61$_{-0.11}^{+0.12}$	&	8.72$_{-0.12}^{+0.13}$	\\\cline{2-8}
&$\log_{10}(M_{\mathrm{cold}}$ [$M_{\odot}$])&	9.31$_{-0.60}^{+0.50}$	&	9.95$_{-0.47}^{+0.31}$	&	7.64$_{-0.51}^{+0.37}$	&	7.75$_{-0.49}^{+0.36}$	&	10.39$_{-0.15}^{+0.13}$	&	10.52$_{-0.14}^{+0.13}$	\\\cline{2-8}
&$\log_{10}(M_{\mathrm{hot}}$ [$M_{\odot}$])&	11.92$_{-0.29}^{+0.26}$	&	12.37$_{-0.28}^{+0.24}$	&	11.43$_{-0.27}^{+0.27}$	&	11.64$_{-0.33}^{+0.30}$	&	12.46$_{-0.19}^{+0.20}$	&	12.76$_{-0.19}^{+0.19}$	\\\cline{2-8}
&$\log_{10}(M_{Z_{\mathrm{cold}}}$ [$M_{\odot}$])&	8.24$_{-0.37}^{+0.29}$	&	8.59$_{-0.27}^{+0.20}$	&	7.13$_{-0.45}^{+0.32}$	&	7.23$_{-0.43}^{+0.31}$	&	8.74$_{-0.15}^{+0.14}$	&	8.86$_{-0.14}^{+0.14}$	\\\cline{2-8}
&$\log_{10}(M_{Z_{*}}$ [$M_{\odot}$])&	9.91$_{-0.10}^{+0.09}$	&	10.05$_{-0.10}^{+0.10}$	&	9.76$_{-0.08}^{+0.07}$	&	9.80$_{-0.08}^{+0.08}$	&	10.14$_{-0.07}^{+0.08}$	&	10.22$_{-0.08}^{+0.09}$	\\\cline{2-8}
&$\log_{10}(M_{Z_{\mathrm{hot,halo}}}$ [$M_{\odot}$])&	8.95$_{-0.28}^{+0.24}$	&	9.37$_{-0.25}^{+0.22}$	&	8.46$_{-0.28}^{+0.26}$	&	8.66$_{-0.30}^{+0.28}$	&	9.48$_{-0.17}^{+0.18}$	&	9.73$_{-0.17}^{+0.17}$	\\\cline{2-8}
&$\log_{10}$(CGF)&	-1.83$_{-0.51}^{+0.41}$	&	-1.35$_{-0.38}^{+0.24}$	&	-3.33$_{-0.49}^{+0.34}$	&	-3.27$_{-0.46}^{+0.33}$	&	-1.00$_{-0.13}^{+0.11}$	&	-0.95$_{-0.11}^{+0.10}$	\\\cline{2-8}
&$Z_{\mathrm{cold}}$&	9.55$_{-0.26}^{+0.23}$	&	9.25$_{-0.17}^{+0.24}$	&	10.05$_{-0.06}^{+0.09}$	&	10.04$_{-0.06}^{+0.08}$	&	8.94$_{-0.07}^{+0.05}$	&	8.93$_{-0.07}^{+0.06}$	\\\cline{2-8}
&$Z_{*}$&	9.34$_{-0.01}^{+0.01}$	&	9.33$_{-0.01}^{+0.01}$	&	9.35$_{-0.01}^{+0.01}$	&	9.35$_{-0.01}^{+0.00}$	&	9.33$_{-0.01}^{+0.01}$	&	9.33$_{-0.01}^{+0.01}$	\\\cline{2-8}
&$Z_{\mathrm{cold}}$-$Z_{*}$&	0.21$_{-0.25}^{+0.22}$	&	-0.08$_{-0.17}^{+0.24}$	&	0.71$_{-0.06}^{+0.08}$	&	0.69$_{-0.06}^{+0.08}$	&	-0.39$_{-0.07}^{+0.05}$	&	-0.40$_{-0.07}^{+0.06}$	\\\cline{2-8}
&\gi&	2.46$_{-0.14}^{+0.12}$	&	2.54$_{-0.13}^{+0.09}$	&	2.67$_{-0.06}^{+0.04}$	&	2.68$_{-0.06}^{+0.04}$	&	2.56$_{-0.07}^{+0.07}$	&	2.63$_{-0.07}^{+0.05}$	\\\cline{2-8}
&\ri&	1.00$_{-0.04}^{+0.03}$	&	1.02$_{-0.04}^{+0.03}$	&	1.05$_{-0.02}^{+0.02}$	&	1.06$_{-0.02}^{+0.01}$	&	1.03$_{-0.02}^{+0.02}$	&	1.05$_{-0.02}^{+0.02}$	\\\cline{2-8}
&$\log_{10}(\SFR$ [$M_{\odot}$ yr$^{-1}$])&	-0.18$_{-0.28}^{+0.21}$	&	-0.25$_{-0.24}^{+0.22}$	&	-1.10$_{-0.43}^{+0.31}$	&	-1.05$_{-0.40}^{+0.28}$	&	-0.24$_{-0.16}^{+0.13}$	&	-0.41$_{-0.17}^{+0.17}$	\\\cline{2-8}
&$\log_{10}(\sSFR$ [yr$^{-1}$])&	-11.36$_{-0.29}^{+0.24}$	&	-11.53$_{-0.31}^{+0.27}$	&	-12.10$_{-0.41}^{+0.30}$	&	-12.10$_{-0.37}^{+0.28}$	&	-11.59$_{-0.22}^{+0.17}$	&	-11.86$_{-0.26}^{+0.24}$	\\\cline{2-8}
&$\log_{10}(T_{\mathrm{cons}}$ [Gyr])&	0.31$_{-0.34}^{+0.42}$	&	0.95$_{-0.48}^{+0.49}$	&	-0.29$_{-0.11}^{+0.12}$	&	-0.25$_{-0.12}^{+0.14}$	&	1.60$_{-0.22}^{+0.25}$	&	1.91$_{-0.27}^{+0.29}$	\\\cline{2-8}
&$\log_{10}(j_{\mathrm{bar}}$ [kpc kms$^{-1}$])&	3.83$_{-0.34}^{+0.42}$	&	4.47$_{-0.48}^{+0.49}$	&	3.22$_{-0.11}^{+0.12}$	&	3.26$_{-0.12}^{+0.14}$	&	5.11$_{-0.22}^{+0.25}$	&	5.43$_{-0.27}^{+0.29}$	\\\hline
\parbox[t]{10mm}{\multirow{2}{*}{\rotatebox[origin=c]{90}{\parbox{1.2cm}{\centering \tiny{combi- nation}}}}}&$\log_{10}$(SHMR)&	-2.12$_{-0.11}^{+0.11}$	&	-2.30$_{-0.12}^{+0.12}$	&	-2.00$_{-0.11}^{+0.10}$	&	-2.07$_{-0.13}^{+0.12}$	&	-2.32$_{-0.10}^{+0.09}$	&	-2.46$_{-0.10}^{+0.10}$	\\\cline{2-8}
&$\log_{10}(M_{\mathrm{BH}}/M_{\mathrm{vir}}$)&	-5.06$_{-0.13}^{+0.13}$	&	-5.17$_{-0.13}^{+0.13}$	&	-5.07$_{-0.15}^{+0.15}$	&	-5.14$_{-0.16}^{+0.16}$	&	-5.11$_{-0.12}^{+0.11}$	&	-5.21$_{-0.12}^{+0.11}$	\\\hline
			\hline
			(i)				& (ii) & (iii) & (iv) & (v)	& (vi) & (vii) & (viii)\\
			\hline
		\end{tabular}
		\tablefoot{The median values spanning between the $32^{\mathrm{nd}}$ and the $68^{\mathrm{th}}$ percentile around the median. Column (I) indicates the corresponding property type: ``halo'', ``galaxy'', or a ``combination'' of both. Column (ii) lists the galaxy property name, while, columns (iii) and (iv) present the values for the knot and filament populations of the parent sample, \den. Columns (v) and (vi) display the values for the \highZ; and columns (vii) and (viii) for the \lowZ\ sub-sample, respectively. Additionally, the second row specifies the percentage of galaxies in each environment and sub-sample relative to the total number of galaxies in this sub-sample at $\zstart=0.56$.}\label{tab:zcolds_props}\vspace{-0.4cm}
	\end{center}
\end{table*}

\subsection{Redshift evolution of knot and filament galaxies}\label{sec:res_envr_evol}

\begin{figure*}
	\centering
	\includegraphics[width=0.75\textwidth,angle=0]{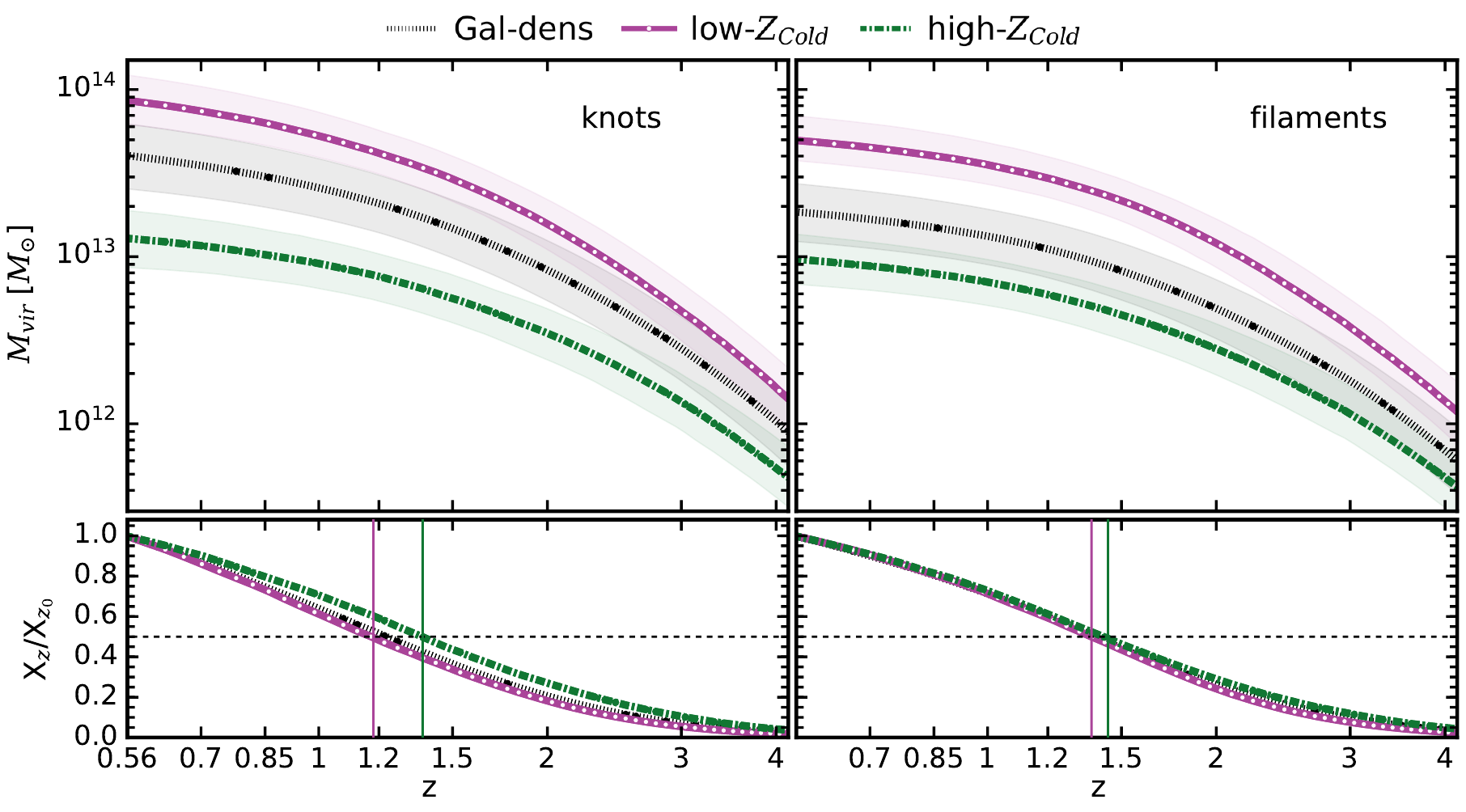}\\
    \includegraphics[width=0.75\textwidth,angle=0]{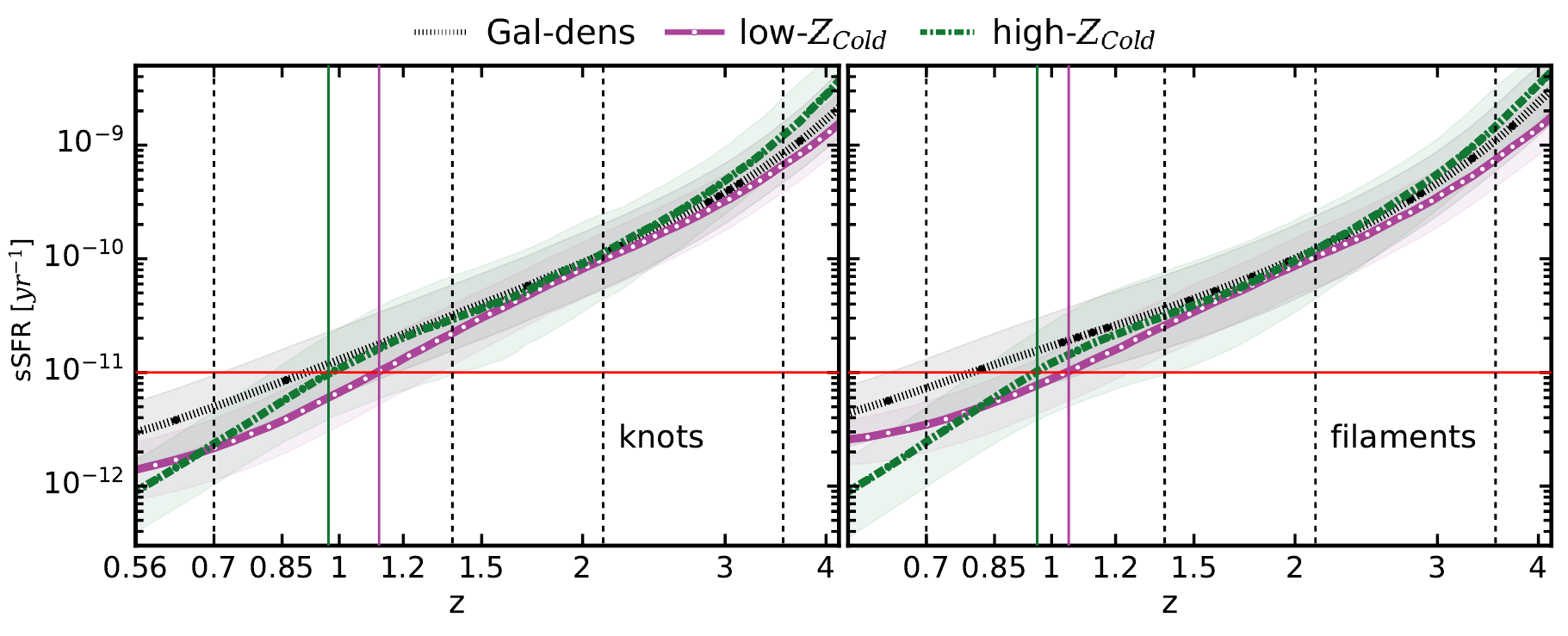}\vspace{-0.35cm}
	\caption{In the upper figure, we present the redshift evolution or the median values of halo mass, \Mvir, (upper panels) and the corresponding mass growth history, relative to the reference redshift of our study, $\zstart = 0.56$, (lower panels). Meanwhile, the lower figure illustrates the evolution of the specific star formation rate, \sSFR. In both figures, the left panels display the evolution for knot environments, while the right panels show the corresponding evolution for filament environments. We include the parent sample, \den\ (short-dashed black line), alongside the \lowZ\ (solid magenta line with white dots) and \highZ\ (dashed-dotted green line) sub-samples. The shaded regions represent the range between the $32^{\mathrm{nd}}$ and $68^{\mathrm{th}}$ percentiles around the median. In the upper figure, vertical solid lines mark the redshift at which 50\% of the halo mass was assembled, known as the half-mass assembly time. In the lower figure, vertical solid lines indicate the redshifts where the \lowZ\ (magenta) and \highZ\ (green) galaxies drop below the classic star formation/quenching threshold, $\sSFR \sim 10^{-11}$ \yrmo, as marked by the horizontal red line \citep{Franx08}. Thin vertical black dashed lines highlight the redshifts of prominent line features in \ri\ shown in panel $(b)$ of \hyperref[fig:res_more_props]{\Fig{fig:res_more_props}}.}\label{fig:res_zevol_envr_mhalo_ssfr}\vspace{-0.4cm}
\end{figure*}

We also tracked the redshift and assembly history of galaxies across different environments, presenting selected results for the parent sample, \den, as well as for the \lowZ\ and \highZ\ sub-samples in this section. In the upper panels of the upper figure of  \hyperref[fig:res_zevol_envr_mhalo_ssfr]{\Fig{fig:res_zevol_envr_mhalo_ssfr}}, we present the assembly histories, and in the lower panels, the growth histories. In the left panel we plot the knot populations and in the right panel for the filament populations, respectively. We use the following colour and line style keys: short-dashed black lines for \den, solid magenta line with dots for \lowZ, and dashed-dotted green line for \highZ.

We find that, in general, all filament populations exhibit flatter growth functions at lower redshift compared to the knot populations. Furthermore, 50\% of the halo mass is assembled at $z\lesssim1.2$ ($z\gtrsim1.2$) in \lowZ\ galaxies (\highZ\ galaxies) located in knots. No significant difference is detected in the half-mass assembly times for the same sub-samples in filaments (refer to the vertical solid lines in the lower panels of the same figure, following the corresponding colour keys). Interestingly, stellar and black hole masses do not show a clear correlation with environment regarding their half-mass assembly time, as galaxies in both knots and filaments tend to assemble at similar redshifts.

\lowZ\ galaxies generally exhibit greater sensitivity to their large-scale environment compared to \highZ\ galaxies, as demonstrated by the specific star formation rate (\sSFR) in the lower figure of \hyperref[fig:res_zevol_envr_mhalo_ssfr]{\Fig{fig:res_zevol_envr_mhalo_ssfr}}. The left (right) panel shows the \sSFR\ evolution in knots (filaments) for the parent sample, \den, and the \lowZ\ and \highZ\ sub-samples. We further highlight the classic star formation/quenching threshold being $\sSFR\sim10^{-11}~\yrmo$ \citep{Franx08}\footnote{Although the star formation threshold depends on redshift, we adopt the corresponding value at $z\sim0$ to avoid complicating our analysis.} as a horizontal solid red line. The vertical thin solid colour-coded lines mark the redshift where the \sSFR\ drops below this threshold for \lowZ\ galaxies (magenta) and \highZ\ galaxies (green). In general, \lowZ\ galaxies enter passive evolution earlier than \highZ\ galaxies, and transitioning slightly earlier when located in the knots than in the filaments but typically at $z\sim1.1$. In contrast, \highZ\ galaxies become passive later than their \lowZ\ counterparts but approximately at the same redshift in knots and filaments, around $z\sim0.9$. The drop below the threshold coincides with a decline in the \SFRD\ as shown before in panel $(d)$ of \hyperref[fig:res_more_props]{\Fig{fig:res_more_props}}.

We analysed the evolution of all galaxy properties listed in \hyperref[tab:zcolds_props]{\Tab{tab:zcolds_props}}. Properties such as \Zcold, \ri colour-index, \SFR, and \SFRD exhibit only slight variations with respect to the environment. Therefore we do not dedicate separate plots to it. The redshift evolution of \Tcons\ reveals a turnover around $z\sim2.7$, where \highZ\ galaxies initially consume their gas less efficiently than the \lowZ\ galaxies. Around $z\sim3$, their consumption time suddenly drops below $0.2~\Gyr$ and rises moderately to $0.5~\Gyr$ with no notable dependence on the environment. Afterwards, the consumption time in \lowZ\ galaxies rapidly surpasses $1~\Gyr$ and eventually exceeds the age of the Universe after $z\sim1$. The parent sample, \den, occupies an intermediate stage between \lowZ\ and \highZ\ sub-samples in terms of gas consumption evolution. All of these trends align with the evolution of the cold gas fraction, where the predictions differ between the sub-samples but show no environmental dependence.

The black hole efficiency, \bheff, reveals another turnover around $z\sim3$, where \lowZ\ galaxies initially show higher efficiency followed by \highZ\ galaxies. The former experiences a decline in efficiency between $1<z<2$. We do not detect any significant difference between environments, although there is a general trend where knot galaxies exhibit higher efficiencies at first, followed by filament galaxies. Additionally, it is notable that \highZ\ galaxies became bulge-dominated much later ($M_{\mathrm{bulge}}/\Mstar>0.8$ around $z\sim3$) than \lowZ\ galaxies.

\subsection{Redshift evolution of galaxy clustering}\label{sec:res_envr_xi}

\begin{figure*}
	\includegraphics[width=\textwidth,angle=0]{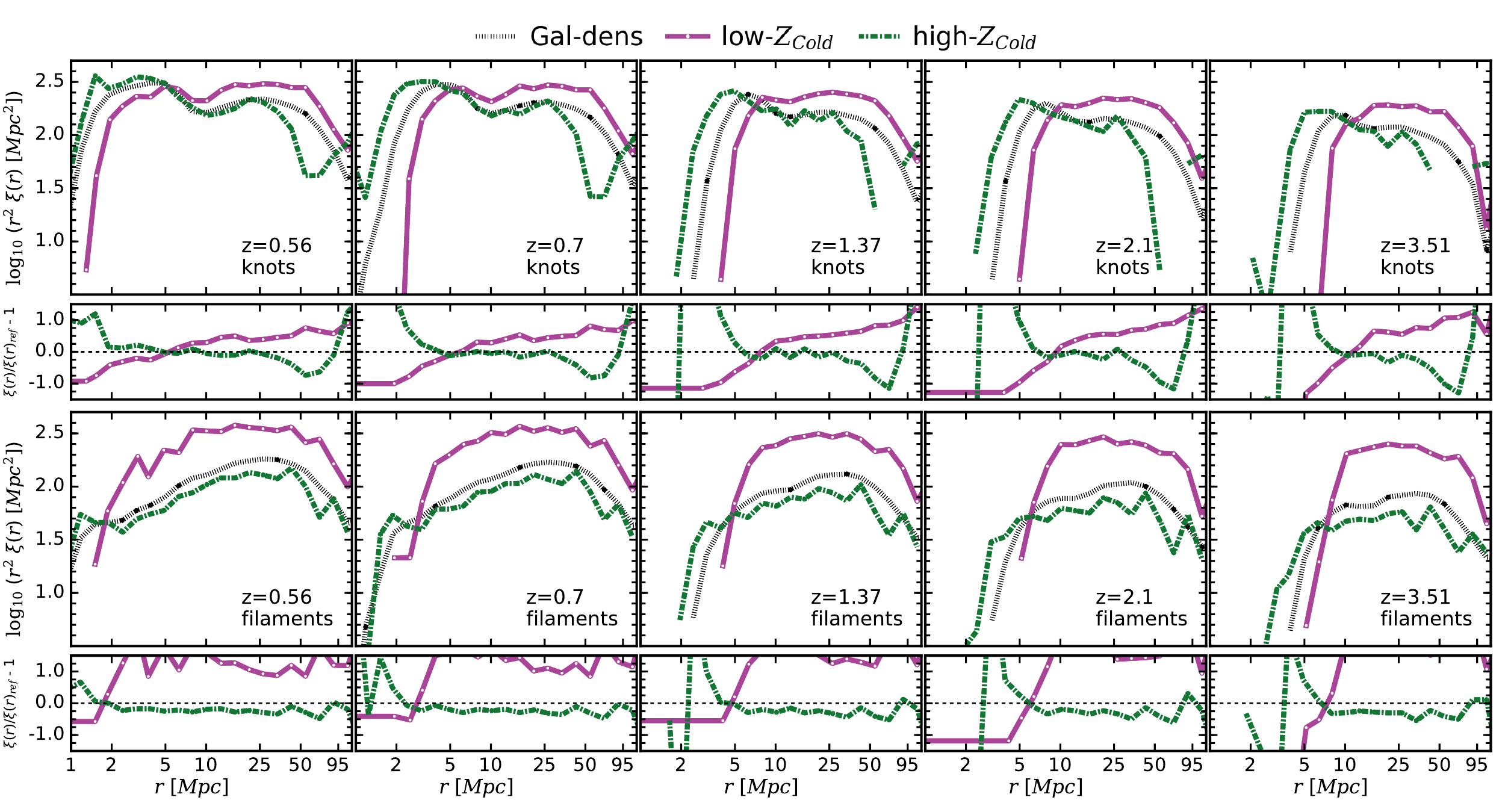}\vspace{-0.2cm}%
	\caption{The redshift evolution of the real-space \twoPCF, $\xi(r)$, for the selected sub-samples: \den, \lowZ\ and \highZ, using the same colour and line style keys as described in \hyperref[fig:res_zevol_envr_mhalo_ssfr]{\Fig{fig:res_zevol_envr_mhalo_ssfr}}. The upper-most panels display clustering for knot galaxies, while the lower-most panels show clustering for filament galaxies. Additionally, the fractional difference of $\xi(r)$ with respect to the clustering of the parent samples, $\xi(r)_{\mathrm{ref}}$ is also presented. From left to right we include the initial redshift of sample selection, $\zstart=0.56$, followed by the same four redshift snapshots marked by vertical lines in \hyperref[fig:res_more_props]{\Fig{fig:res_more_props}}.}\vspace{-0.4cm} \label{fig:res_z-evol_xi_M2}
\end{figure*}

To complete our result section, we show in \hyperref[fig:res_z-evol_xi_M2]{\Fig{fig:res_z-evol_xi_M2}} the clustering evolution of knot galaxies (upper panels) and filament galaxies (lower panels) of our selected sub-samples at various snapshots: $z=[0.56,0.7,1.37,2.1,3.51]$, using the same line style and colour keys as in \hyperref[fig:res_zevol_envr_mhalo_ssfr]{\Fig{fig:res_zevol_envr_mhalo_ssfr}}. For consistency, we chose \zstart\ and the same four redshift snapshots from left to right as indicated by the vertical lines in \hyperref[fig:res_more_props]{\Fig{fig:res_more_props}}. In the two narrow panels, we show again the fractional difference of $\xi(r)$ for each sub-sample relative to the function of \den, $\xi(r)_{\mathrm{ref}}$. As discussed previously, knot and filament populations exhibit different clustering behaviours. However, the evolution with redshift shows a mild trend towards lower clustering strength with increasing redshift. The shift of the separation length, $r$, where the \twoPCFs\ drops to zero for the sub-samples \lowZ\ and \highZ, is a consequence of the different median halo mass range sub-samples hold. \highZ\ galaxies are hosted by lower massive halos which can be found in general at smaller separations than their \lowZ\ counterparts.

It is also notable that the knot galaxies in the \highZ\ sub-sample generally exhibit higher clustering strength at smaller separations whereas their \lowZ\ counterparts do so at higher separations. This leads to a turnover in the clustering strength in comparison to all galaxies between $5 \lesssim~r~[\Mpc]~\lesssim 10$ (see narrow panels in the figure). Moreover, the redshift evolution of filament galaxies is markedly different: \lowZ\ galaxies always cluster significantly stronger than the parent and \highZ\ samples. The \highZ\ galaxies, on the other hand, follow closely the clustering of \den.

\section{Discussion and conclusion}\label{sec:discussion}

The objective of this study was to investigate the redshift evolution and mass assembly history of sub-populations (sub-samples of galaxies) selected from the same overall population of luminous and massive objects at $z\sim0.5$. Thereby we employed a broad spectrum of properties and analysis strategies, including clustering functions and the galaxy-halo connection. In addition to examining the redshift evolution of galaxies within each sub-sample, we also distinguished them based on their location in the large-scale structure of the cosmic web, such as knots (denser regions) and filaments (less dense regions). This approach enabled us to explore their properties as well as the redshift evolution and assembly histories as a function of the environment. In this section, we discuss our findings and critically assess our conclusions.

\textit{Do the sub-samples studied in this work form via distinct formation channels?} We gathered significant evidence suggesting that the sub-samples \highZ\ and \lowZ\ form via distinct formation channels. We cannot confirm a similar conclusion for the rest of our sub-samples (\low, \red, and \passive) since the galaxies in those sub-samples do not clearly map onto a prominent bimodality in the colour, \sSFR, and \Mvir\ spaces as shown in \hyperref[fig:ssfr2mstar]{\Fig{fig:ssfr2mstar}} and \hyperref[fig:g-i2mhalo]{\Fig{fig:g-i2mhalo}}. The galaxies in these sub-samples are mixed populations, comprising galaxies from both peaks. Interestingly, the \low\ sample shares evolutionary tracks with \highZ\ galaxies for nearly all galaxy properties (see e.g.\ \hyperref[fig:res_zevol_mcold_M2]{\Fig{fig:res_zevol_mcold_M2}} and \hyperref[fig:res_zevol_mbh_M2]{\Fig{fig:res_zevol_mbh_M2}}), except for \ri-colour and \cSFRD\ (see \hyperref[fig:res_more_props]{\Fig{fig:res_more_props}}). As shown in \hyperref[fig:res_xi_M2]{\Fig{fig:res_xi_M2}}, these sub-samples cluster very differently, likely due to the fact that they reside in dark matter halos of different masses. Furthermore, \low\ galaxies can serve as a proxy for the overall 2-point correlation function for the entire sample of SAM-CMASS mock-galaxies, \den\ (our parent sample). In addition, we find that the \highZ\ and \lowZ\ sub-samples assembled half of their masses at different epochs (see \hyperref[fig:res_zevol_mvir_M2]{\Fig{fig:res_zevol_mvir_M2}}, \hyperref[fig:res_zevol_mbh_M2]{\Fig{fig:res_zevol_mbh_M2}}, and \hyperref[fig:res_zevol_envr_mhalo_ssfr]{\Fig{fig:res_zevol_envr_mhalo_ssfr}}), exhibit distinct properties, primarily inhabit different environments (see \hyperref[tab:zcolds_props]{\Tab{tab:zcolds_props}} and \hyperref[fig:res_zevol_envr_mhalo_ssfr]{\Fig{fig:res_zevol_envr_mhalo_ssfr}}), and ultimately cluster differently (see \hyperref[fig:res_xi_M2]{\Fig{fig:res_xi_M2}} and \hyperref[fig:res_zcolds_xi]{\Fig{fig:res_zcolds_xi}}). Nonetheless, at intermediate redshifts, they share the same evolution of the colour parameter \ri\ as shown in panel $(b)$ in \hyperref[fig:res_more_props]{\Fig{fig:res_more_props}}.

\textit{To what degree are our results affected by the modelling itself?} Each semi-analytical model comes with its strengths and weaknesses which have been discussed extensively by various authors over the years \citep[see e.g.][]{Knebe15,Knebe17_Carnage}. In this study, we leveraged the strengths of our adopted SAM, \galacticus, particularly its capabilities in modelling the galaxy-halo connection, luminosities, and the evolution of the cosmic star formation rate density \citep[see][]{Knebe17_MD,Cui19,Stoppacher19}. It could be argued that one of the main properties we focused on in this study, \Zcold, is approximated in our model (see the definition in \hyperref[sec:data]{\Sec{sec:data}}), as \galacticus\ does not provide the oxygen abundance directly but predicts only the cold gas mass and masses of metals. We find that the estimated values of \Zcold\ align well with observational data, as demonstrated in the top panel of Fig.\ 7 in \citet{Knebe17_MD}. However, this figure also shows that \galacticus\ continues to exhibit a prominent bimodality in the cold gas-phase metallicity until $z\sim0.1$ and an excess in metallicity beyond $\Zcold\sim9.5$ at $\Mstar\sim10^{10.5}~\Msun$. We assume that this is, in fact, related to the implementation of the AGN-feedback, which directly affects cold-gas-related properties as reported by the same authors. Several studies have reported that the implemented feedback strongly influences the resulting mass-metallicity relation across all redshifts \citep{Torrey14,DeRossi17_metals_EAGLE} while others report the opposite \citep{Taylor+Kobayashi15_AGN,Thorne22_SFH_DEVILS}. Furthermore, the excess detected for our adopted model is relatively moderate within the considered stellar mass range $\Mstar>10^{11}~\Msun$, where most of the SAM-CMASS mock-galaxies can be found. Despite this, we cannot rule out the possibility that the metallicity bimodality in \galacticus\ is a peculiarity of this model. Nonetheless, recent works support the existence of such a bimodality and highlight the complexity of these properties in terms of galaxy evolution, morphology, and environment \citep[e.g.][]{Wotta19_metals,Donnan22_MZR,Pistis22_metals,Omori+Takeuchi22_metalBi}.

\textit{What are the implications of our clustering results?} \hyperref[fig:res_zcolds_xi]{\Fig{fig:res_zcolds_xi}} reveals intriguing insights into the clustering of \highZ\ and \lowZ\ galaxies based on their location within the cosmic web. \highZ\ galaxies in knots and filaments exhibit significantly different clustering patterns, despite residing in halos of similar mass\footnote{The difference in median halo masses is $\sim0.1$ dex in \Msun\ between the knot and filament populations as shown in \hyperref[fig:zcolds_mhalo]{\Fig{fig:zcolds_mhalo}}}. Conversely, \lowZ\ galaxies clustering behaviour in these environments, even though they inhabit halos of different masses. To ensure that these clustering behaviours are not merely a result of differences in stellar or halo masses, we selected four control samples: low (high) stellar mass as $\lowSM<10^{11.3}$~\Msun ($\highSM\geq10^{11.3}$~\Msun) and low (high) halo masses as $\lowHM<10^{13.6}$ \Msun\ ($\highHM\geq10^{13.6}$ \Msun). We found that the assembly histories, redshift evolution, and clustering function of our control samples are different in comparison to those for \lowZ\ and \highZ\ sub-samples reported in this work. In principle, the clustering results reported in \hyperref[fig:res_zcolds_xi]{\Fig{fig:res_zcolds_xi}} could potentially be related to the so-called ``halo'' and ``galaxy assembly bias'', which are secondary dependencies of the halo bias at fixed halo mass manifested in the galaxy population \citep[see e.g.][]{Sheth+Tormen04_ref,Gao05_ref,Wechsler06_ref,Yang06_ref, Croton07_ref}. In \hyperref[app:zcolds_mhalo]{\App{app:zcolds_mhalo}} we conduct statistical tests on the similarity in halo mass distributions of filament and knot galaxies in \highZ. Further investigation will be devoted to addressing this specific connection.

\textit{Are our results in agreement with the fundamental metallicity relation (FMR)?} The FMR relates the gas-phase metallicity and stellar ages based on empirical calibrations of the oxygen-abundance, which in our case corresponds to the property of the cold gas-phase metallicity, \Zcold\ \citep{Mannucci10}. In this framework, galaxies with higher \Zcold\ have undergone stronger metal enrichment processes and are typically more massive. Interestingly, our results suggest the opposite: galaxies that are the most (least) massive exhibit lower (higher) values for \Zcold. We emphasise that the prediction from \galacticus\ on \Zcold\ is in agreement with the FMR. When comparing our results to those in the literature, all the galaxies considered in our sample \lowZ\ and \highZ\ sub-samples are classified as ``high metallicities'' systems, with values of $12 + \logT(\mathrm{O/H}) > 8.5$ (see the values for \Zcold\ in \hyperref[tab:zcolds_props]{\Tab{tab:zcolds_props}}). This corresponds, for example, to the high-metallicity predictions of \citet{Finlator+Dave08}, \citet{Maiolino+Mannucci18_rev}, and \citet{Sanchez-Menguiano19}. Furthermore, the trend that galaxies in denser environments tend to exhibit slightly higher metallicities than those in less dense environments, as reported by studies such as \citet{Mouhcine07} and \citet{Cooper08_OH_env}, is consistent with our sample. In summary, two scenarios could help to interpret the metal enrichment of cold gas in our results: ``quenching by strangulation'' for \highZ\ galaxies \citep[e.g.][]{Peng15}, and ``metal-poor gas accretion'' for \lowZ\ galaxies \citep[e.g.][]{Ceverino16}. 

The former scenario is illustrated in Fig.\ 1 of \citet[][]{Peng15}, and it explains well what we observe when studying the \highZ\ sub-sample: galaxies are cut off from continued gas accretion, but can continue to form stars by recycling of the available enriched interstellar medium. As a consequence, their stellar mass and gas metallicity increase steeply due to the absence of dilution by inflowing metal-poor gas, a process commonly described as a ``closed-box'' model. The same authors predicted an upper stellar mass limit of $\Mstar\sim10^{11}$ \Msun, beyond which this mechanism primarily drives quenching in local quiescent galaxies. Our \highZ\ galaxies fall within the uncertainty range, spanning from the $32^{\mathrm{th}}$ to the $68^{\mathrm{th}}$ percentile around this threshold, as shown in \hyperref[tab:zcolds_props]{\Tab{tab:zcolds_props}}. In addition, \citet{Trussler20} found that the star-forming progenitors of local passive galaxies within the same mass range are principally quenched by starvation on a time-scale of $2~\Gyr$. However, they do not find the environmental trend that we report in this study. 

The latter scenario describes the accretion of metal-poor gas from the intergalactic medium onto the galaxy. We expect that \lowZ\ galaxies gain a significant amount of cold, metal-poor gas through recent gas-rich mergers (e.g.\ due to the incorporation of dwarf satellites) or directly from the intergalactic medium, which suppresses the metal-enrichment process while boosting star formation \citet{Dekel09b,vanDeVoort+Schaye12,Ceverino16}. This scenario was also been described by \citet{Dayal13_dilution} using a simple analytical model. Furthermore, one or more ``hot accretion''-events, likely occurring around massive structures, are also plausible \citep{Sancisi08_rev}. A similar prediction was made by \citet{Yates12} also using a semi-analytical model. This scheme is further supported by the fact that the \lowZ\ population showed higher net metallicity around $z\sim2$ compared to its final value at \zstart\ (see the top panel in \hyperref[fig:res_more_props]{\Fig{fig:res_more_props}}). When considering the growth function of the cold gas reservoir and its metal masses (see \hyperref[fig:res_zevol_mcold_M2]{\Fig{fig:res_zevol_mcold_M2}}), \lowZ\ galaxies demonstrate a steady growth function from $z=1.5$ to \zstart. These findings support our hypothesis that \lowZ\ galaxies continue to be fuelled by cold gas at lower redshift, unlike \highZ\ galaxies, which primarily rely on their initial reservoir of \Mcold\ for star formation. We also expect that \lowZ\ galaxies harbour a younger stellar population compared to the \highZ\ sub-sample. Other authors have also suggested that cold streams could provide sufficient gas to sustain star formation in galaxies hosted by massive halos \citep{Rodriguez-Puebla17} and predicted that at redshifts $z>1$, hot halos can be penetrated by these streams \citep[see e.g.][]{Keres05,Dekel+Birnboim06_ref,Dekel09a_ref}. Investigating the merger histories, particularly the distinction between minor and major mergers, would be a fruitful approach to further illuminate this matter.

\textit{Could the galaxies present in the \lowZ\ sample be ``ultra-luminous massive galaxies'' (ULMGs)?}
The progenitors of \cmass\ galaxies are known to be a passively evolving population of galaxies since $z\sim0.7$. Although \cmass\ galaxies are frequently found in the centre of the most massive clusters and super-clusters \citep{Lietzen12}, it remains unclear which galaxy population they might resemble at lower or higher redshift. ULMGs, believed to be the low-redshift progenitors of the most massive, passively evolving galaxies in the Universe \citep[e.g.][]{Cheema20_MA,Forrest20}, are either very massive and quiescent at $z\sim1.6$ or a population of post-starburst galaxies that were quenched rapidly before $z\sim3-4$. When compared to these studies, even galaxies in our most massive sample, \lowZ, do not exhibit properties consistent with ULMGs, as both stellar and halo masses are still too low at the redshifts we consider. Our findings suggest that SAM-CMASS mock-galaxies constitute a different population and follow an evolutionary path distinct from that of ultra-massive, early-quenched galaxies.

\textit{Does the mass-metallicity relation of our galaxy samples reveal different sub-types of ``compact early-type galaxies'' (cETGs)?} In their study of low-redshift cETGs, \citet{Kim20} identified two environmentally dependent variants of the same population. One sub-population deviates from the classical mass-metallicity relation for early-type galaxies \citep{Chilingarian+Zolotukhin15,Janz16}, suggesting the existence of different formation channels. Although we cannot directly compare our results to their findings due to differences in metallicity proxies, we observe similar trends in our data. Specifically, they found a higher-metallicity, lower-mass population primarily located in filaments, which aligns with the possibility that our \highZ\ population could evolve into cETGs at lower redshift. Furthermore, \citet{Borzyszkowski17_ZOMGI} introduced two populations of halos that have halo masses similar to those hosting our \lowZ\ and \highZ\ galaxies. In a follow-up study, \citet{Romano-Diaz17_ZOMGII} discussed the detection of halo assembly bias, which resonates with our clustering results for the \lowZ\ and \highZ\ populations across different environments. These findings further support a potential link between halo assembly bias and the formation of distinct galaxy sub-populations.

\textit{Why do we not compare our results to observations?} Comparing the redshift evolution of observation with data from models is generally challenging and involves significant modelling on the observational side. For instance, generating star formation histories for observed galaxies typically relies on ``spectral energy distribution fitting'' techniques available as computational packages \citep[see e.g.][]{Alarcon23_Diffstar}. These techniques depend on stellar synthesis population models such as developed by e.g.\ \citet[][]{Conroy09} or \citet[][]{Vazdekis10_MILES}. Moreover, this work focuses on the intrinsic evolution of galaxies, which cannot be easily compared to observations since we cannot trace observed galaxies across time. Therefore, we chose not to include observational data in this work, as such a comparison would lack meaning. In a follow-up study, we will address the challenge of comparing to observations and how to generate the redshift evolution for observed galaxy populations of galaxies in more detail. We are working towards developing a complementary approach that involves tracking progenitors, inspired by observational data selection, and will incorporate observational data for reference. This work is currently in preparation.

\section{Summary and outlook}\label{sec:summary}

In this study, we analyse the assembly history and redshift evolution of semi-analytically modelled massive and luminous red galaxies within the redshift range $0.5\lesssim z\lesssim 4$. In particular, we use the total population of the SAM-CMASS mock-galaxy sample, \den, introduced and described in \citet[][referred to as S19]{Stoppacher19}, which replicates the selection of the well-studied observational \boss-\cmass\ sample. \den\ serves as our parent sample and we extract sub-samples, from which we extract sub-samples based on typical selection criteria for galaxy properties at from it using typical selection criteria based on galaxy properties at the initial redshift of our study $\zstart=0.56$ (see \hyperref[tab:subsel]{\Tab{tab:subsel}}). The galaxy properties used for selection are \gi-colour (\red\ sub-sample), star formation activity (\low\ and \passive\ sub-samples), and the cold gas-phase metallicity, \Zcold, (\lowZ\ and \highZ\ sub-samples).

We subsequently tracked the progenitors of those galaxies that entered our defined sub-samples at \zstart\ to high redshift in order to reproduce the redshift evolution and growth histories starting at $z=4.15$ to $z=0.56$ (see illustration in \hyperref[fig:cartoon_M2]{\Fig{fig:cartoon_M2}}). In the first part of this work, we present key properties, including halo masses (\Mvir, \hyperref[fig:res_zevol_mvir_M2]{\Fig{fig:res_zevol_mvir_M2}}), stellar masses (\Mstar), cold gas masses (\Mcold, \hyperref[fig:res_zevol_mcold_M2]{\Fig{fig:res_zevol_mcold_M2}}), black hole masses (\Mbh, \hyperref[fig:res_zevol_mbh_M2]{\Fig{fig:res_zevol_mbh_M2}}), as well as the cold gas-phase metallicities (\Zcold), the observed colours \ri, the star formation rate (densities) (\SFR\ and \SFRD) in \hyperref[fig:res_more_props]{\Fig{fig:res_more_props}} as well as their clustering functions (\hyperref[fig:res_xi_M2]{\Fig{fig:res_xi_M2}}). In the second part, we focus specifically on \lowZ\ and \highZ\ sub-samples, examining their clustering functions (see \hyperref[fig:res_zcolds_xi]{\Fig{fig:res_zcolds_xi}}) and evolutionary histories in different large-scale environments, such as knots (denser regions) and filaments (less dense regions), as detailed in \hyperref[fig:res_zevol_envr_mhalo_ssfr]{\Fig{fig:res_zevol_envr_mhalo_ssfr}},  \hyperref[fig:res_z-evol_xi_M2]{\Fig{fig:res_z-evol_xi_M2}} and \hyperref[tab:zcolds_props]{\Tab{tab:zcolds_props}}.

We summarise our most important results as the following:

\begin{enumerate}
    \item The galaxies in each sub-sample exhibit distinct median evolution trends for their galaxy and halo properties, alongside varying as well as distinct growth and clustering functions. Notably, the most massive galaxies are the least metal-enriched at $z=0.56$. Members of the \lowZ\ sub-sample consistently exhibit higher median values for \Mstar, \Mvir, and \Mbh. In contrast, galaxies with intermediate to lower masses for \Mvir, \Mstar, or \Mbh\ (corresponding to the \passive, \red, and \highZ\ sub-samples) tend to consume their cold gas at higher redshift. The latter two samples are the earliest and quickest to assemble half of their \Mstar\ and \Mbh. At higher/intermediate redshift they exhibit lower \Zcold, \SFR, and \cSFRD, compared to other sub-samples (see \hyperref[fig:res_zevol_mvir_M2]{\Figsstart{fig:res_zevol_mvir_M2}}-\hyperref[fig:res_more_props]{\Figsend{fig:res_more_props}}).

    \item We observe that the sub-samples \lowZ\ and \highZ\ follow distinct evolutionary pathways and occupy different environments (see \hyperref[tab:zcolds_props]{\Tab{tab:zcolds_props}}). Specifically, \lowZ\ galaxies are predominantly found in denser regions like knots (60\%). Despite their later formation, they transition to passive evolution than their \highZ\ counterparts. \highZ\ galaxies mostly inhabit less dense regions such as filaments (62\%). \highZ\ filament and knot galaxies exhibit different clustering functions in both shape and amplitude, whereas the \lowZ\ filament and knot populations show similar clustering behaviour (see \hyperref[fig:res_zcolds_xi]{\Fig{fig:res_zcolds_xi}}).

    \item Furthermore, the redshift evolution and assembly histories of \lowZ\ and \highZ\ galaxies are influenced by their environment. \highZ\ galaxies formed half of their halo mass significantly earlier in knots compared to filaments (see upper figure in \hyperref[fig:res_zevol_envr_mhalo_ssfr]{\Fig{fig:res_zevol_envr_mhalo_ssfr}}). In contrast, \lowZ\ galaxies formed at essentially the same redshift in both environments. Regarding the quenching of star formation, \highZ\ galaxies generally transitioned to passive evolution later than \lowZ\ galaxies in both environments. However, \lowZ\ galaxies quenched slightly earlier in knots than in filaments (see lower figure in \hyperref[fig:res_zevol_envr_mhalo_ssfr]{\Fig{fig:res_zevol_envr_mhalo_ssfr}}).
    
\end{enumerate}

In this work, we discussed, among other results, the evolution of galaxy properties using the progenitors of galaxies assigned to well-defined sub-samples at a particular redshift. That means in particular that until now we examined how the properties of galaxies, such as those that were the reddest or most metal-enriched at \zstart, evolve over time. As shown in panel $(a)$ of \hyperref[fig:res_more_props]{\Fig{fig:res_more_props}}, galaxies meeting a specific selection criterion at \zstart\ (in this case low or high cold gas-phase metallicity, \Zcold) do not necessarily maintain that criterion throughout their redshift evolution. This is expected and reflected in almost all galaxy properties as they undergo substantial evolution and change during their assembly history. In addition, in this particular study, we focus on the intrinsic evolution of galaxies throughout cosmic history which provides valuable insights into the true evolution of a given galaxy population. As noted in the discussion section, understanding this intrinsic evolution is crucial. Moving forward, we aim to develop a complementary approach that integrates observational data to trace redshift evolution. This will enhance our ability to compare predictions from models with observations, which is increasingly relevant given the extensive sky surveys from recent and upcoming observational missions.

\begin{acknowledgements}
DS is funded by the Spanish Ministry of Universities and the European Next Generation Fond under the Margarita Salas Fellowship CA1/RSUE/2021-00720. DS also wants to thank Brant Robertson for granting her time to work on this project and Jos\'e O\~norbe for his guidance and support on bureaucratic issues as well as as the team from Koblischek-byKatrin for serving excellent coffee in an amazing atmosphere in Weiz and further Dr.\ Rosa Maria La\ss nig for her kind and professional support during the last years. The \textsc{CosmoSim}-database used in this paper is a service by the Leibniz-Institute for Astrophysics Potsdam (AIP). The MultiDark database was developed in cooperation with the Spanish MultiDark Consolider Project CSD2009-00064. In addition, this work has benefited from the publicly available software tools and packages: \textsc{Matplotlib} \citep{Matplotlib}; Python Software Foundation (\url{https://www.python.org}) 1990-2023, version 2.7 \& 3; \textsc{Anaconda} (\url{https://www.anaconda.com}); \textsc{pyenv} (\url{https://github.com/pyenv/pyenv}); \textsc{Seaborn} (\url{https://seaborn.pydata.org/}) and \textsc{Statsmodels} \citep{seaborn2010statsmodels}; \textsc{AstroPy} (\url{https://www.astropy.org/}) \citep{astropy13,astropy18}; \textsc{Jupyter Notebook} (\url{https://jupyter-notebook.readthedocs.io/en/latest/}); The CentOS Project (\url{https://www.centos.org}), The Fedora Project (\url{https://fedoraproject.org/}), \textsc{Topcat} \citep{Taylor13_Topcat}. We used OpenAI's \texttt{GPT-4} language model for assistance with language refinement and drafting parts of the manuscript. ADMD thanks Fondecyt for financial support through the Fondecyt Regular 2021 grant 1210612. AK is supported by the Ministerio de Ciencia e Innovaci\'{o}n (MICINN), Spain under research grant PID2021-122603NB-C21 and further thanks The Me In You for girl in amour. NDP acknowledges support from a RAICES grant from the Ministerio de Ciencia, Tecnolog\'ia e Innovaci\'on, Argentina. 
\end{acknowledgements}

\bibliographystyle{aa}
\bibliography{archive}

\begin{thebibliography}{185}
\expandafter\ifx\csname natexlab\endcsname\relax\def\natexlab#1{#1}\fi

\bibitem[{{Alam} {et~al.}(2015){Alam}, {Albareti}, {Allende Prieto}, {Anders}, {Anderson}, \& {Anderton}}]{Alam15_SDSS}
{Alam}, S., {Albareti}, F.~D., {Allende Prieto}, C., {et~al.} 2015, \apjs, 219, 12

\bibitem[{{Alarcon} {et~al.}(2023){Alarcon}, {Hearin}, {Becker}, \& {Chaves-Montero}}]{Alarcon23_Diffstar}
{Alarcon}, A., {Hearin}, A.~P., {Becker}, M.~R., \& {Chaves-Montero}, J. 2023, \mnras, 518, 562

\bibitem[{{Allende Prieto} {et~al.}(2001){Allende Prieto}, {Lambert}, \& {Asplund}}]{Allende01}
{Allende Prieto}, C., {Lambert}, D.~L., \& {Asplund}, M. 2001, \apjl, 556, L63

\bibitem[{{Argudo-Fern{\'a}ndez} {et~al.}(2018){Argudo-Fern{\'a}ndez}, {Lacerna}, \& {Duarte Puertas}}]{Argudo18}
{Argudo-Fern{\'a}ndez}, M., {Lacerna}, I., \& {Duarte Puertas}, S. 2018, \aap, 620, A113

\bibitem[{{Asplund} {et~al.}(2009){Asplund}, {Grevesse}, {Sauval}, \& {Scott}}]{Asplund09}
{Asplund}, M., {Grevesse}, N., {Sauval}, A.~J., \& {Scott}, P. 2009, \araa, 47, 481

\bibitem[{{Astropy Collaboration} {et~al.}(2018){Astropy Collaboration}, {Price-Whelan}, {Sip{\H{o}}cz}, {G{\"u}nther}, {Lim}, {Crawford}, {Conseil}, {Shupe}, {Craig}, {Dencheva}, {Ginsburg}, {VanderPlas}, {Bradley}, {P{\'e}rez-Su{\'a}rez}, {de Val-Borro}, {Aldcroft}, {Cruz}, {Robitaille}, {Tollerud}, {Ardelean}, {Babej}, {Bach}, {Bachetti}, {Bakanov}, {Bamford}, {Barentsen}, {Barmby}, {Baumbach}, {Berry}, {Biscani}, {Boquien}, {Bostroem}, {Bouma}, {Brammer}, {Bray}, {Breytenbach}, {Buddelmeijer}, {Burke}, {Calderone}, {Cano Rodr{\'\i}guez}, {Cara}, {Cardoso}, {Cheedella}, {Copin}, {Corrales}, {Crichton}, {D'Avella}, {Deil}, {Depagne}, {Dietrich}, {Donath}, {Droettboom}, {Earl}, {Erben}, {Fabbro}, {Ferreira}, {Finethy}, {Fox}, {Garrison}, {Gibbons}, {Goldstein}, {Gommers}, {Greco}, {Greenfield}, {Groener}, {Grollier}, {Hagen}, {Hirst}, {Homeier}, {Horton}, {Hosseinzadeh}, {Hu}, {Hunkeler}, {Ivezi{\'c}}, {Jain}, {Jenness}, {Kanarek}, {Kendrew}, {Kern}, {Kerzendorf}, {Khvalko}, {King}, {Kirkby}, {Kulkarni},
  {Kumar}, {Lee}, {Lenz}, {Littlefair}, {Ma}, {Macleod}, {Mastropietro}, {McCully}, {Montagnac}, {Morris}, {Mueller}, {Mumford}, {Muna}, {Murphy}, {Nelson}, {Nguyen}, {Ninan}, {N{\"o}the}, {Ogaz}, {Oh}, {Parejko}, {Parley}, {Pascual}, {Patil}, {Patil}, {Plunkett}, {Prochaska}, {Rastogi}, {Reddy Janga}, {Sabater}, {Sakurikar}, {Seifert}, {Sherbert}, {Sherwood-Taylor}, {Shih}, {Sick}, {Silbiger}, {Singanamalla}, {Singer}, {Sladen}, {Sooley}, {Sornarajah}, {Streicher}, {Teuben}, {Thomas}, {Tremblay}, {Turner}, {Terr{\'o}n}, {van Kerkwijk}, {de la Vega}, {Watkins}, {Weaver}, {Whitmore}, {Woillez}, {Zabalza}, \& {Astropy Contributors}}]{astropy18}
{Astropy Collaboration}, {Price-Whelan}, A.~M., {Sip{\H{o}}cz}, B.~M., {et~al.} 2018, \aj, 156, 123

\bibitem[{{Astropy Collaboration} {et~al.}(2013){Astropy Collaboration}, {Robitaille}, {Tollerud}, {Greenfield}, {Droettboom}, {Bray}, {Aldcroft}, {Davis}, {Ginsburg}, {Price-Whelan}, {Kerzendorf}, {Conley}, {Crighton}, {Barbary}, {Muna}, {Ferguson}, {Grollier}, {Parikh}, {Nair}, {Unther}, {Deil}, {Woillez}, {Conseil}, {Kramer}, {Turner}, {Singer}, {Fox}, {Weaver}, {Zabalza}, {Edwards}, {Azalee Bostroem}, {Burke}, {Casey}, {Crawford}, {Dencheva}, {Ely}, {Jenness}, {Labrie}, {Lim}, {Pierfederici}, {Pontzen}, {Ptak}, {Refsdal}, {Servillat}, \& {Streicher}}]{astropy13}
{Astropy Collaboration}, {Robitaille}, T.~P., {Tollerud}, E.~J., {et~al.} 2013, \aap, 558, A33

\bibitem[{{Baugh}(2006)}]{Baugh06}
{Baugh}, C.~M. 2006, Rep. Prog. Phys., 69, 3101

\bibitem[{{Baugh}(2013)}]{Baugh12}
{Baugh}, C.~M. 2013, in The Intriguing Life of Massive Galaxies, Vol. 295, 191--199

\bibitem[{{Begelman}(2014)}]{Begelman14_BHs}
{Begelman}, M.~C. 2014, arXiv e-prints, arXiv:1410.8132

\bibitem[{{Behroozi} {et~al.}({2013}{\natexlab{a}}){Behroozi}, {Wechsler}, \& {Wu}}]{Behroozi13a}
{Behroozi}, P.~S., {Wechsler}, R.~H., \& {Wu}, H.-Y. {2013}{\natexlab{a}}, \apj, 762, 109

\bibitem[{{Behroozi} {et~al.}({2013}{\natexlab{b}}){Behroozi}, {Wechsler}, {Wu}, {Busha}, {Klypin}, \& {Primack}}]{Behroozi13b}
{Behroozi}, P.~S., {Wechsler}, R.~H., {Wu}, H.-Y., {et~al.} {2013}{\natexlab{b}}, \apj, 763, 18

\bibitem[{{Benson}(2010)}]{Benson10}
{Benson}, A.~J. 2010, \physrep, 495, 33

\bibitem[{{Benson}(2012)}]{Benson12}
{Benson}, A.~J. 2012, \na, 17, 175

\bibitem[{{Benson} \& {Bower}(2010)}]{Benson+Bower10}
{Benson}, A.~J. \& {Bower}, R. 2010, \mnras, 405, 1573

\bibitem[{{Benson} {et~al.}(2002){Benson}, {Frenk}, {Lacey}, {Baugh}, \& {Cole}}]{Benson02}
{Benson}, A.~J., {Frenk}, C.~S., {Lacey}, C.~G., {Baugh}, C.~M., \& {Cole}, S. 2002, \mnras, 333, 177

\bibitem[{Blanton \& Berlind(2007)}]{Blanton07}
Blanton, M.~R. \& Berlind, A.~A. 2007, \apj, 664, 791–803

\bibitem[{{Booth} \& {Schaye}(2011)}]{Booth+Schaye11_bheff}
{Booth}, C.~M. \& {Schaye}, J. 2011, \mnras, 413, 1158

\bibitem[{{Borzyszkowski} {et~al.}(2017){Borzyszkowski}, {Porciani}, {Romano-D{\'\i}az}, \& {Garaldi}}]{Borzyszkowski17_ZOMGI}
{Borzyszkowski}, M., {Porciani}, C., {Romano-D{\'\i}az}, E., \& {Garaldi}, E. 2017, \mnras, 469, 594

\bibitem[{{Bose} \& {Loeb}(2021)}]{Bose+Loeb21_HA_model}
{Bose}, S. \& {Loeb}, A. 2021, \apj, 912, 114

\bibitem[{{Bryan} \& {Norman}(1998)}]{Bryan+Norman98}
{Bryan}, G.~L. \& {Norman}, M.~L. 1998, \apj, 495, 80

\bibitem[{{Carlesi} {et~al.}(2014){Carlesi}, {Knebe}, {Lewis}, {Wales}, \& {Yepes}}]{Carlesi2014}
{Carlesi}, E., {Knebe}, A., {Lewis}, G.~F., {Wales}, S., \& {Yepes}, G. 2014, \mnras, 439, 2943

\bibitem[{{Ceverino} {et~al.}(2016){Ceverino}, {S{\'a}nchez Almeida}, {Mu{\~n}oz Tu{\~n}{\'o}n}, {Dekel}, {Elmegreen}, {Elmegreen}, \& {Primack}}]{Ceverino16}
{Ceverino}, D., {S{\'a}nchez Almeida}, J., {Mu{\~n}oz Tu{\~n}{\'o}n}, C., {et~al.} 2016, \mnras, 457, 2605

\bibitem[{{Chabrier}(2003)}]{Chabrier03}
{Chabrier}, G. 2003, \pasp, 115, 763

\bibitem[{{Cheema} {et~al.}(2020){Cheema}, {Sawicki}, {Arcila-Osejo}, {Golob}, {Moutard}, {Arnouts}, \& {Coupon}}]{Cheema20_MA}
{Cheema}, G.~K., {Sawicki}, M., {Arcila-Osejo}, L., {et~al.} 2020, \mnras, 494, 804

\bibitem[{{Chilingarian} \& {Zolotukhin}(2015)}]{Chilingarian+Zolotukhin15}
{Chilingarian}, I. \& {Zolotukhin}, I. 2015, Science, 348, 418

\bibitem[{{Chuang} {et~al.}(2016){Chuang}, {Prada}, {Pellejero-Ibanez}, \& et~al.}]{Chuang16}
{Chuang}, C.-H., {Prada}, F., {Pellejero-Ibanez}, M., \& et~al. 2016, \mnras, 461, 3781

\bibitem[{{Cole} {et~al.}(2001){Cole}, {Norberg}, {Baugh}, {Frenk}, {Bland-Hawthorn}, {Bridges}, {Cannon}, {Colless}, {Collins}, {Couch}, {Cross}, {Dalton}, {De Propris}, {Driver}, {Efstathiou}, {Ellis}, {Glazebrook}, {Jackson}, {Lahav}, {Lewis}, {Lumsden}, {Maddox}, {Madgwick}, {Peacock}, {Peterson}, {Sutherland}, \& {Taylor}}]{Cole01_2dF}
{Cole}, S., {Norberg}, P., {Baugh}, C.~M., {et~al.} 2001, MNRAS, 326, 255

\bibitem[{Conroy(2013)}]{Conroy13_rev}
Conroy, C. 2013, \araa, 51, 393–455

\bibitem[{{Conroy} {et~al.}(2009){Conroy}, {Gunn}, \& {White}}]{Conroy09}
{Conroy}, C., {Gunn}, J.~E., \& {White}, M. 2009, \apj, 699, 486

\bibitem[{{Conselice}(2014)}]{Conselice14_evo_rev}
{Conselice}, C.~J. 2014, \araa, 52, 291

\bibitem[{Contini {et~al.}(2020)Contini, Gu, Ge, Rhee, Yi, \& Kang}]{Contini20}
Contini, E., Gu, Q., Ge, X., {et~al.} 2020, \apj, 889, 156

\bibitem[{{Cooper} {et~al.}(2008){Cooper}, {Tremonti}, {Newman}, \& {Zabludoff}}]{Cooper08_OH_env}
{Cooper}, M.~C., {Tremonti}, C.~A., {Newman}, J.~A., \& {Zabludoff}, A.~I. 2008, \mnras, 390, 245

\bibitem[{{Costantin} {et~al.}(2021){Costantin}, {P{\'e}rez-Gonz{\'a}lez}, {M{\'e}ndez-Abreu}, {Huertas-Company}, {Dimauro}, {Alcalde-Pampliega}, {Buitrago}, {Ceverino}, {Daddi}, {Dom{\'\i}nguez-S{\'a}nchez}, {Espino-Briones}, {Hern{\'a}n-Caballero}, {Koekemoer}, \& {Rodighiero}}]{Costantin21}
{Costantin}, L., {P{\'e}rez-Gonz{\'a}lez}, P.~G., {M{\'e}ndez-Abreu}, J., {et~al.} 2021, \apj, 913, 125

\bibitem[{{Croton}(2009)}]{Croton09_bheff}
{Croton}, D.~J. 2009, \mnras, 394, 1109

\bibitem[{{Croton}(2013)}]{Croton13_h}
{Croton}, D.~J. 2013, \pasa, 30, e052

\bibitem[{{Croton} {et~al.}(2007){Croton}, {Gao}, \& {White}}]{Croton07_ref}
{Croton}, D.~J., {Gao}, L., \& {White}, S. D.~M. 2007, \mnras, 374, 1303

\bibitem[{{Croton} {et~al.}(2006){Croton}, {Springel}, {White}, {De Lucia}, {Frenk}, {Gao}, {Jenkins}, {Kauffmann}, {Navarro}, \& {Yoshida}}]{Croton06}
{Croton}, D.~J., {Springel}, V., {White}, S.~D.~M., {et~al.} 2006, \mnras, 365, 11

\bibitem[{{Cuesta} {et~al.}(2016){Cuesta}, {Vargas-Maga{\~n}a}, {Beutler}, \& et~al.}]{Cuesta16}
{Cuesta}, A.~J., {Vargas-Maga{\~n}a}, M., {Beutler}, F., \& et~al. 2016, \mnras, 457, 1770

\bibitem[{{Cui} {et~al.}(2019){Cui}, {Knebe}, {Libeskind}, {Planelles}, {Yang}, {Cui}, {Dav{\'e}}, {Kang}, {Mostoghiu}, {Staveley-Smith}, {Wang}, {Wang}, \& {Yepes}}]{Cui19}
{Cui}, W., {Knebe}, A., {Libeskind}, N.~I., {et~al.} 2019, \mnras, 557

\bibitem[{{Cui} {et~al.}(2018){Cui}, {Knebe}, {Yepes}, {Yang}, {Borgani}, {Kang}, {Power}, \& {Staveley-Smith}}]{Cui18a}
{Cui}, W., {Knebe}, A., {Yepes}, G., {et~al.} 2018, \mnras, 473, 68

\bibitem[{{Dav{\'e}} {et~al.}(2020){Dav{\'e}}, {Crain}, {Stevens}, {Narayanan}, {Saintonge}, {Catinella}, \& {Cortese}}]{Dave20_OH_Eagle}
{Dav{\'e}}, R., {Crain}, R.~A., {Stevens}, A. R.~H., {et~al.} 2020, \mnras, 497, 146

\bibitem[{{Davies} {et~al.}(2021){Davies}, {Crain}, \& {Pontzen}}]{Davies21_Eagle}
{Davies}, J.~J., {Crain}, R.~A., \& {Pontzen}, A. 2021, \mnras, 501, 236

\bibitem[{{Dayal} {et~al.}(2013){Dayal}, {Ferrara}, \& {Dunlop}}]{Dayal13_dilution}
{Dayal}, P., {Ferrara}, A., \& {Dunlop}, J.~S. 2013, \mnras, 430, 2891

\bibitem[{{de Jong} \& {Lacey}(2000)}]{deJong+Lacey00_SB-SDM}
{de Jong}, R.~S. \& {Lacey}, C. 2000, \apj, 545, 781

\bibitem[{{De Lucia} {et~al.}(2006){De Lucia}, {Springel}, {White}, {Croton}, \& {Kauffmann}}]{DeLucia06}
{De Lucia}, G., {Springel}, V., {White}, S.~D.~M., {Croton}, D., \& {Kauffmann}, G. 2006, \mnras, 366, 499

\bibitem[{{De Rossi} {et~al.}(2017){De Rossi}, {Bower}, {Font}, {Schaye}, \& {Theuns}}]{DeRossi17_metals_EAGLE}
{De Rossi}, M.~E., {Bower}, R.~G., {Font}, A.~S., {Schaye}, J., \& {Theuns}, T. 2017, \mnras, 472, 3354

\bibitem[{{Dekel} \& {Birnboim}(2006)}]{Dekel+Birnboim06_ref}
{Dekel}, A. \& {Birnboim}, Y. 2006, \mnras, 368, 2

\bibitem[{{Dekel} {et~al.}(2009{\natexlab{a}}){Dekel}, {Birnboim}, {Engel}, {Freundlich}, {Goerdt}, {Mumcuoglu}, {Neistein}, {Pichon}, {Teyssier}, \& {Zinger}}]{Dekel09a_ref}
{Dekel}, A., {Birnboim}, Y., {Engel}, G., {et~al.} 2009{\natexlab{a}}, \nat, 457, 451

\bibitem[{{Dekel} {et~al.}(2009{\natexlab{b}}){Dekel}, {Sari}, \& {Ceverino}}]{Dekel09b}
{Dekel}, A., {Sari}, R., \& {Ceverino}, D. 2009{\natexlab{b}}, \apj, 703, 785

\bibitem[{{Dolfi} {et~al.}(2023){Dolfi}, {G{\'o}mez}, {Monachesi}, {Varela-Lavin}, {Tissera}, {Sif{\'o}n}, \& {Galaz}}]{Dolfi23_AH_lop}
{Dolfi}, A., {G{\'o}mez}, F.~A., {Monachesi}, A., {et~al.} 2023, \mnras, 526, 567

\bibitem[{{Donnan} {et~al.}(2022){Donnan}, {Tojeiro}, \& {Kraljic}}]{Donnan22_MZR}
{Donnan}, C.~T., {Tojeiro}, R., \& {Kraljic}, K. 2022, at, 6, 599

\bibitem[{{Dressler}(1980)}]{Dressler80}
{Dressler}, A. 1980, \apj, 236, 351

\bibitem[{{Dutta} {et~al.}(2020){Dutta}, {Fumagalli}, {Fossati}, {Lofthouse}, {Prochaska}, {Arrigoni Battaia}, {Bielby}, {Cantalupo}, {Cooke}, {Murphy}, \& {O'Meara}}]{Dutta20}
{Dutta}, R., {Fumagalli}, M., {Fossati}, M., {et~al.} 2020, \mnras, 499, 5022

\bibitem[{{Efstathiou} {et~al.}(1982){Efstathiou}, {Lake}, \& {Negroponte}}]{Efstathiou82}
{Efstathiou}, G., {Lake}, G., \& {Negroponte}, J. 1982, \mnras, 199, 1069

\bibitem[{{Efstathiou} \& {Rees}(1988)}]{Efstathiou+Rees88_clustering_ref}
{Efstathiou}, G. \& {Rees}, M.~J. 1988, \mnras, 230, 5p

\bibitem[{{Eisenstein} {et~al.}(2011){Eisenstein}, {Weinberg}, {Agol}, {Aihara}, {Allende Prieto}, {Anderson}, {Arns}, {Aubourg}, {Bailey}, {Balbinot}, \& et~al.}]{Eisenstein11_SDSS3}
{Eisenstein}, D.~J., {Weinberg}, D.~H., {Agol}, E., {et~al.} 2011, \aj, 142, 72

\bibitem[{{Fall}(1983)}]{Fall83}
{Fall}, S.~M. 1983, in IAU Symposium, Vol. 100, Internal Kinematics and Dynamics of Galaxies, ed. E.~{Athanassoula}, 391--398

\bibitem[{{Favole} {et~al.}(2016){Favole}, {McBride}, {Eisenstein}, {Prada}, {Swanson}, {Chuang}, \& {Schneider}}]{Favole16b}
{Favole}, G., {McBride}, C.~K., {Eisenstein}, D.~J., {et~al.} 2016, \mnras, 462, 2218

\bibitem[{{Ferland} {et~al.}(2013){Ferland}, {Porter}, {van Hoof}, {Williams}, {Abel}, {Lykins}, {Shaw}, {Henney}, \& {Stancil}}]{Ferland13}
{Ferland}, G.~J., {Porter}, R.~L., {van Hoof}, P.~A.~M., {et~al.} 2013, \rmxaa, 49, 137

\bibitem[{{Ferrara} {et~al.}(1999){Ferrara}, {Bianchi}, {Cimatti}, \& {Giovanardi}}]{Ferrara99}
{Ferrara}, A., {Bianchi}, S., {Cimatti}, A., \& {Giovanardi}, C. 1999, \apjs, 123, 437

\bibitem[{{Ferrarese}(2002)}]{Ferrarese02_bheff}
{Ferrarese}, L. 2002, \apj, 578, 90

\bibitem[{{Filho} {et~al.}(2015){Filho}, {S{\'a}nchez Almeida}, {Mu{\~n}oz-Tu{\~n}{\'o}n}, {Nuza}, {Kitaura}, \& {He{\ss}}}]{Filho15_lowZ_envr}
{Filho}, M.~E., {S{\'a}nchez Almeida}, J., {Mu{\~n}oz-Tu{\~n}{\'o}n}, C., {et~al.} 2015, \apj, 802, 82

\bibitem[{{Finlator} \& {Dav{\'e}}(2008)}]{Finlator+Dave08}
{Finlator}, K. \& {Dav{\'e}}, R. 2008, \mnras, 385, 2181

\bibitem[{{Forrest} {et~al.}(2020){Forrest}, {Marsan}, {Annunziatella}, {Wilson}, {Muzzin}, {Marchesini}, {Cooper}, {Chan}, {McConachie}, {Gomez}, {Kado-Fong}, {La Barbera}, {Lange-Vagle}, {Nantais}, {Nonino}, {Saracco}, {Stefanon}, \& {van der Burg}}]{Forrest20}
{Forrest}, B., {Marsan}, Z.~C., {Annunziatella}, M., {et~al.} 2020, \apj, 903, 47

\bibitem[{{Franx} {et~al.}(2008){Franx}, {van Dokkum}, {F{\"o}rster Schreiber}, {Wuyts}, {Labb{\'e}}, \& {Toft}}]{Franx08}
{Franx}, M., {van Dokkum}, P.~G., {F{\"o}rster Schreiber}, N.~M., {et~al.} 2008, \apj, 688, 770

\bibitem[{{Fukugita} {et~al.}(1996){Fukugita}, {Ichikawa}, {Gunn}, {Doi}, {Shimasaku}, \& {Schneider}}]{Fukugita96_ugriz}
{Fukugita}, M., {Ichikawa}, T., {Gunn}, J.~E., {et~al.} 1996, \aj, 111, 1748

\bibitem[{{Gao} {et~al.}(2005){Gao}, {Springel}, \& {White}}]{Gao05_ref}
{Gao}, L., {Springel}, V., \& {White}, S. D.~M. 2005, \mnras, 363, L66

\bibitem[{{Gnedin} {et~al.}(2004){Gnedin}, {Kravtsov}, {Klypin}, \& {Nagai}}]{Gnedin04}
{Gnedin}, O.~Y., {Kravtsov}, A.~V., {Klypin}, A.~A., \& {Nagai}, D. 2004, \apj, 616, 16

\bibitem[{{Guo} {et~al.}(2018){Guo}, {Yang}, \& {Lu}}]{Guo18}
{Guo}, H., {Yang}, X., \& {Lu}, Y. 2018, \apj, 858, 30

\bibitem[{{Gupta} {et~al.}(2020){Gupta}, {Tran}, {Cohn}, {Alcorn}, {Yuan}, {Rodriguez-Gomez}, {Harshan}, {Forrest}, {Kewley}, {Glazebrook}, {Straatman}, {Kacprzak}, {Nanayakkara}, {Labb{\'e}}, {Papovich}, \& {Cowley}}]{Gupta20_SFH_obs}
{Gupta}, A., {Tran}, K.-V., {Cohn}, J., {et~al.} 2020, \apj, 893, 23

\bibitem[{{Harada} {et~al.}(2023){Harada}, {Yajima}, \& {Abe}}]{Harada23_metal_outflow}
{Harada}, N., {Yajima}, H., \& {Abe}, M. 2023, \mnras, 525, 5868

\bibitem[{{H{\"a}ring} \& {Rix}(2004)}]{Haring+Rix2004}
{H{\"a}ring}, N. \& {Rix}, H.-W. 2004, ApJL, 604, L89

\bibitem[{{Hashemizadeh} {et~al.}(2021){Hashemizadeh}, {Driver}, {Davies}, {Robotham}, {Bellstedt}, {Windhorst}, {Bremer}, {Phillipps}, {Jarvis}, {Holwerda}, {Lagos}, {Koushan}, {Siudek}, {Maddox}, {Thorne}, \& {Elahi}}]{Hashemizadeh21}
{Hashemizadeh}, A., {Driver}, S.~P., {Davies}, L. J.~M., {et~al.} 2021, \mnras, 505, 136

\bibitem[{{Hawarden} {et~al.}(1979){Hawarden}, {van Woerden}, {Mebold}, {Goss}, \& {Peterson}}]{Hawarden79_reju_ref}
{Hawarden}, T.~G., {van Woerden}, H., {Mebold}, U., {Goss}, W.~M., \& {Peterson}, B.~A. 1979, \aap, 76, 230

\bibitem[{{Hernquist}(1990)}]{Hernquist90}
{Hernquist}, L. 1990, \apj, 356, 359

\bibitem[{{Hoffman} {et~al.}(2012){Hoffman}, {Metuki}, {Yepes}, {Gottl{\"o}ber}, {Forero-Romero}, {Libeskind}, \& {Knebe}}]{Hoffman2012}
{Hoffman}, Y., {Metuki}, O., {Yepes}, G., {et~al.} 2012, \mnras, 425, 2049

\bibitem[{{Hopkins} {et~al.}(2014){Hopkins}, {Kere{\v s}}, {O{\~n}orbe}, {Faucher-Gigu{\`e}re}, {Quataert}, {Murray}, \& {Bullock}}]{Hopkins14}
{Hopkins}, P.~F., {Kere{\v s}}, D., {O{\~n}orbe}, J., {et~al.} 2014, \mnras, 445, 581

\bibitem[{{Hubble}(1936)}]{Hubble36}
{Hubble}, E.~P. 1936, {Realm of the Nebulae}

\bibitem[{{Hughes} {et~al.}(2013){Hughes}, {Cortese}, {Boselli}, {Gavazzi}, \& {Davies}}]{Hughes13_OH}
{Hughes}, T.~M., {Cortese}, L., {Boselli}, A., {Gavazzi}, G., \& {Davies}, J.~I. 2013, \aap, 550, A115

\bibitem[{{Hunter}(2007)}]{Matplotlib}
{Hunter}, J.~D. 2007, Computing in Science and Engineering, 9, 90

\bibitem[{{Inagaki} {et~al.}(2015){Inagaki}, {Lin}, {Huang}, {Hsieh}, \& {Sugiyama}}]{Inagaki15_MAH-cluster}
{Inagaki}, T., {Lin}, Y.-T., {Huang}, H.-J., {Hsieh}, B.-C., \& {Sugiyama}, N. 2015, \mnras, 446, 1107

\bibitem[{{Janz} {et~al.}(2016){Janz}, {Norris}, {Forbes}, {Huxor}, {Romanowsky}, {Frank}, {Escudero}, {Faifer}, {Forte}, {Kannappan}, {Maraston}, {Brodie}, {Strader}, \& {Thompson}}]{Janz16}
{Janz}, J., {Norris}, M.~A., {Forbes}, D.~A., {et~al.} 2016, \mnras, 456, 617

\bibitem[{{Johnston} {et~al.}(2022){Johnston}, {H{\"a}u{\ss}ler}, {Jegatheesan}, {Fraser-McKelvie}, {Coccato}, {Cortesi}, {Jaff{\'e}}, {Galaz}, {Mora}, \& {Ordenes-Brice{\~n}o}}]{Johnston22_SFH_obs}
{Johnston}, E.~J., {H{\"a}u{\ss}ler}, B., {Jegatheesan}, K., {et~al.} 2022, \mnras, 514, 6141

\bibitem[{{Kaiser}(1984)}]{Kaiser84_clustering_ref}
{Kaiser}, N. 1984, \apjl, 284, L9

\bibitem[{{Kalinova} {et~al.}(2021){Kalinova}, {Colombo}, {S{\'a}nchez}, {Kodaira}, {Garc{\'\i}a-Benito}, {Gonz{\'a}lez Delgado}, {Rosolowsky}, \& {Lacerda}}]{Kalinova21_quench}
{Kalinova}, V., {Colombo}, D., {S{\'a}nchez}, S.~F., {et~al.} 2021, \aap, 648, A64

\bibitem[{{Kere{\v{s}}} {et~al.}(2005){Kere{\v{s}}}, {Katz}, {Weinberg}, \& {Dav{\'e}}}]{Keres05}
{Kere{\v{s}}}, D., {Katz}, N., {Weinberg}, D.~H., \& {Dav{\'e}}, R. 2005, \mnras, 363, 2

\bibitem[{{Kim} {et~al.}(2020){Kim}, {Jeong}, {Rey}, {Lee}, {Lee}, {Joo}, \& {Kim}}]{Kim20}
{Kim}, S., {Jeong}, H., {Rey}, S.-C., {et~al.} 2020, \apj, 903, 65

\bibitem[{{Klypin} {et~al.}(2016){Klypin}, {Yepes}, {Gottl{\"o}ber}, {Prada}, \& {He{\ss}}}]{Klypin16_MD}
{Klypin}, A., {Yepes}, G., {Gottl{\"o}ber}, S., {Prada}, F., \& {He{\ss}}, S. 2016, \mnras, 457, 4340

\bibitem[{{Knebe} {et~al.}(2018{\natexlab{a}}){Knebe}, {Pearce}, {Gonzalez-Perez}, {Thomas}, {Benson}, {Asquith}, {Blaizot}, {Bower}, {Carretero}, {Castander}, {Cattaneo}, {Cora}, {Croton}, {Cui}, {Cunnama}, {Devriendt}, {Elahi}, {Font}, {Fontanot}, {Gargiulo}, {Helly}, {Henriques}, {Lee}, {Mamon}, {Onions}, {Padilla}, {Power}, {Pujol}, {Ruiz}, {Srisawat}, {Stevens}, {Tollet}, {Vega-Mart{\'\i}nez}, \& {Yi}}]{Knebe17_Carnage}
{Knebe}, A., {Pearce}, F.~R., {Gonzalez-Perez}, V., {et~al.} 2018{\natexlab{a}}, \mnras, 475, 2936

\bibitem[{{Knebe} {et~al.}(2015){Knebe}, {Pearce}, {Thomas}, {Benson}, {Blaizot}, {Bower}, {Carretero}, {Castander}, \& {et al.}}]{Knebe15}
{Knebe}, A., {Pearce}, F.~R., {Thomas}, P.~A., {et~al.} 2015, \mnras, 451, 4029

\bibitem[{{Knebe} {et~al.}(2018{\natexlab{b}}){Knebe}, {Stoppacher}, {Prada}, {Behrens}, {Benson}, {Cora}, {Croton}, {Padilla}, {Ruiz}, {Sinha}, {Stevens}, {Vega-Mart{\'\i}nez}, {Behroozi}, {Gonzalez-Perez}, {Gottl{\"o}ber}, {Klypin}, {Yepes}, {Enke}, {Libeskind}, {Riebe}, \& {Steinmetz}}]{Knebe17_MD}
{Knebe}, A., {Stoppacher}, D., {Prada}, F., {et~al.} 2018{\natexlab{b}}, \mnras, 474, 5206

\bibitem[{{Kormendy}(1979)}]{Kormendy79_morph_ref}
{Kormendy}, J. 1979, \apj, 227, 714

\bibitem[{{Koyama} {et~al.}(2013){Koyama}, {Smail}, {Kurk}, {Geach}, {Sobral}, {Kodama}, {Nakata}, {Swinbank}, {Best}, {Hayashi}, \& {Tadaki}}]{Koyama13}
{Koyama}, Y., {Smail}, I., {Kurk}, J., {et~al.} 2013, \mnras, 434, 423

\bibitem[{{Krajnovi{\'c}} {et~al.}(2020){Krajnovi{\'c}}, {Ural}, {Kuntschner}, {Goudfrooij}, {Wolfe}, {Cappellari}, {Davies}, {de Zeeuw}, {Duc}, {Emsellem}, {Karick}, {McDermid}, {Mei}, \& {Naab}}]{Krajnovic20}
{Krajnovi{\'c}}, D., {Ural}, U., {Kuntschner}, H., {et~al.} 2020, \aap, 635, A129

\bibitem[{{Krumholz} {et~al.}(2009){Krumholz}, {McKee}, \& {Tumlinson}}]{Krumholz09}
{Krumholz}, M.~R., {McKee}, C.~F., \& {Tumlinson}, J. 2009, \apj, 699, 850

\bibitem[{{Lackner} {et~al.}(2012){Lackner}, {Cen}, {Ostriker}, \& {Joung}}]{Lackner12_MA_envr_hydro}
{Lackner}, C.~N., {Cen}, R., {Ostriker}, J.~P., \& {Joung}, M.~R. 2012, \mnras, 425, 641

\bibitem[{{Landy} \& {Szalay}(1993)}]{Landy+Szalay93}
{Landy}, S.~D. \& {Szalay}, A.~S. 1993, \apj, 412, 64

\bibitem[{{Li} \& {White}(2009)}]{Li+White09_SMF}
{Li}, C. \& {White}, S. D.~M. 2009, \mnras, 398, 2177

\bibitem[{{Libeskind} {et~al.}(2013){Libeskind}, {Hoffman}, {Forero-Romero}, {Gottl{\"o}ber}, {Knebe}, {Steinmetz}, \& {Klypin}}]{Libeskind2013}
{Libeskind}, N.~I., {Hoffman}, Y., {Forero-Romero}, J., {et~al.} 2013, \mnras, 428, 2489

\bibitem[{{Libeskind} {et~al.}(2012){Libeskind}, {Hoffman}, {Knebe}, {Steinmetz}, {Gottl{\"o}ber}, {Metuki}, \& {Yepes}}]{Libeskind2012}
{Libeskind}, N.~I., {Hoffman}, Y., {Knebe}, A., {et~al.} 2012, \mnras, 421, L137

\bibitem[{{Lietzen} {et~al.}(2012){Lietzen}, {Tempel}, {Hein{\"a}m{\"a}ki}, {Nurmi}, {Einasto}, \& {Saar}}]{Lietzen12}
{Lietzen}, H., {Tempel}, E., {Hein{\"a}m{\"a}ki}, P., {et~al.} 2012, \aap, 545, A104

\bibitem[{{Lin} {et~al.}(2016){Lin}, {Mandelbaum}, {Huang}, {Huang}, {Dalal}, {Diemer}, {Jian}, \& {Kravtsov}}]{Lin16_assembly}
{Lin}, Y.-T., {Mandelbaum}, R., {Huang}, Y.-H., {et~al.} 2016, \apj, 819, 119

\bibitem[{{Liu} {et~al.}(2016){Liu}, {Lu}, {Xie}, {Chen}, \& {Zhao}}]{Lui16_LRGs}
{Liu}, G.~C., {Lu}, Y.~J., {Xie}, L.~Z., {Chen}, X.~L., \& {Zhao}, Y.~H. 2016, \aap, 585, A52

\bibitem[{{Luparello} {et~al.}(2015){Luparello}, {Lares}, {Paz}, {Yaryura}, {Lambas}, \& {Padilla}}]{Luparello15}
{Luparello}, H.~E., {Lares}, M., {Paz}, D., {et~al.} 2015, \mnras, 448, 1483

\bibitem[{{Lutz} {et~al.}(2020){Lutz}, {Kilborn}, {Catinella}, {Cortese}, {Brown}, \& {Koribalski}}]{Lutz20_OH}
{Lutz}, K.~A., {Kilborn}, V., {Catinella}, B., {et~al.} 2020, \aap, 635, A69

\bibitem[{{Madau} \& {Dickinson}(2014)}]{Madau+Dickinson14}
{Madau}, P. \& {Dickinson}, M. 2014, \araa, 52, 415

\bibitem[{{Maiolino} \& {Mannucci}(2019)}]{Maiolino+Mannucci18_rev}
{Maiolino}, R. \& {Mannucci}, F. 2019, \aapr, 27, 3

\bibitem[{{Mancera Pi{\~n}a} {et~al.}(2021){Mancera Pi{\~n}a}, {Posti}, {Pezzulli}, {Fraternali}, {Fall}, {Oosterloo}, \& {Adams}}]{Mancera-Pina21_jbar}
{Mancera Pi{\~n}a}, P.~E., {Posti}, L., {Pezzulli}, G., {et~al.} 2021, \aap, 651, L15

\bibitem[{{Mannucci} {et~al.}(2010){Mannucci}, {Cresci}, {Maiolino}, {Marconi}, \& {Gnerucci}}]{Mannucci10}
{Mannucci}, F., {Cresci}, G., {Maiolino}, R., {Marconi}, A., \& {Gnerucci}, A. 2010, \mnras, 408, 2115

\bibitem[{{Maraston}(2005)}]{Maraston05}
{Maraston}, C. 2005, \mnras, 362, 799

\bibitem[{{Maraston} {et~al.}(2013){Maraston}, {Pforr}, {Henriques}, {Thomas}, {Wake}, {Brownstein}, {Capozzi}, {Tinker}, \& {et al.}}]{Maraston13}
{Maraston}, C., {Pforr}, J., {Henriques}, B.~M., {et~al.} 2013, \mnras, 435, 2764

\bibitem[{{Maraston} {et~al.}(2009){Maraston}, {Str{\"o}mb{\"a}ck}, {Thomas}, {Wake}, \& {Nichol}}]{Maraston09}
{Maraston}, C., {Str{\"o}mb{\"a}ck}, G., {Thomas}, D., {Wake}, D.~A., \& {Nichol}, R.~C. 2009, \mnras, 394, L107

\bibitem[{{Masters} {et~al.}(2011){Masters}, {Maraston}, {Nichol}, \& et~al.}]{Masters11}
{Masters}, K.~L., {Maraston}, C., {Nichol}, R.~C., \& et~al. 2011, \mnras, 418, 1055

\bibitem[{{Montero-Dorta} {et~al.}(2016){Montero-Dorta}, {Bolton}, {Brownstein}, {Swanson}, {Dawson}, {Prada}, {Eisenstein}, {Maraston}, \& {et al.}}]{Montero-Dorta16}
{Montero-Dorta}, A.~D., {Bolton}, A.~S., {Brownstein}, J.~R., {et~al.} 2016, \mnras, 461, 1131

\bibitem[{{Montero-Dorta} {et~al.}(2017{\natexlab{a}}){Montero-Dorta}, {Bolton}, \& {Shu}}]{Montero-Dorta17}
{Montero-Dorta}, A.~D., {Bolton}, A.~S., \& {Shu}, Y. 2017{\natexlab{a}}, \mnras, 468, 47

\bibitem[{{Montero-Dorta} {et~al.}(2017{\natexlab{b}}){Montero-Dorta}, {P{\'e}rez}, {Prada}, {Rodr{\'\i}guez-Torres}, {Favole}, {Klypin}, {Cid Fernandes}, {Gonz{\'a}lez Delgado}, {Dom{\'\i}nguez}, {Bolton}, {Garc{\'\i}a-Benito}, {Jullo}, \& {Niemiec}}]{Montero-Dorta17_assembly}
{Montero-Dorta}, A.~D., {P{\'e}rez}, E., {Prada}, F., {et~al.} 2017{\natexlab{b}}, \apj, 848, L2

\bibitem[{{Mouhcine} {et~al.}(2007){Mouhcine}, {Baldry}, \& {Bamford}}]{Mouhcine07}
{Mouhcine}, M., {Baldry}, I.~K., \& {Bamford}, S.~P. 2007, \mnras, 382, 801

\bibitem[{{Mueller} {et~al.}(2018){Mueller}, {Percival}, {Linder}, {Alam}, {Zhao}, {S{\'a}nchez}, {Beutler}, \& {Brinkmann}}]{Mueller18}
{Mueller}, E.-M., {Percival}, W., {Linder}, E., {et~al.} 2018, \mnras, 475, 2122

\bibitem[{{Musso} {et~al.}(2018){Musso}, {Cadiou}, {Pichon}, {Codis}, {Kraljic}, \& {Dubois}}]{Musso18}
{Musso}, M., {Cadiou}, C., {Pichon}, C., {et~al.} 2018, \mnras, 476, 4877

\bibitem[{{Navarro} {et~al.}(1997){Navarro}, {Frenk}, \& {White}}]{Navarro97}
{Navarro}, J.~F., {Frenk}, C.~S., \& {White}, S.~D.~M. 1997, \apj, 490, 493

\bibitem[{{Niemiec} {et~al.}(2018){Niemiec}, {Jullo}, {Montero-Dorta}, {Prada}, {Rodriguez-Torres}, {Perez}, {Klypin}, {Erben}, {Makler}, {Moraes}, {Pereira}, \& {Shan}}]{Niemiec18}
{Niemiec}, A., {Jullo}, E., {Montero-Dorta}, A.~D., {et~al.} 2018, \mnras, 477, L1

\bibitem[{{Norberg} {et~al.}(2002){Norberg}, {Cole}, {Baugh}, {Frenk}, {Baldry}, {Bland-Hawthorn}, {Bridges}, {Cannon}, \& {et al.}}]{Norberg02}
{Norberg}, P., {Cole}, S., {Baugh}, C.~M., {et~al.} 2002, \mnras, 336, 907

\bibitem[{{Omori} \& {Takeuchi}(2022)}]{Omori+Takeuchi22_metalBi}
{Omori}, K.~C. \& {Takeuchi}, T.~T. 2022, \aap, 660, A145

\bibitem[{{Oser} {et~al.}(2010){Oser}, {Ostriker}, {Naab}, {Johansson}, \& {Burkert}}]{Oser10_ref}
{Oser}, L., {Ostriker}, J.~P., {Naab}, T., {Johansson}, P.~H., \& {Burkert}, A. 2010, \apj, 725, 2312

\bibitem[{{Ostriker} {et~al.}(2010){Ostriker}, {McKee}, \& {Leroy}}]{Ostriker10_eq-model}
{Ostriker}, E.~C., {McKee}, C.~F., \& {Leroy}, A.~K. 2010, \apj, 721, 975

\bibitem[{{Pandey} \& {Sarkar}(2020)}]{Pandey+Sarkar20}
{Pandey}, B. \& {Sarkar}, S. 2020, \mnras, 498, 6069

\bibitem[{{Pandya} {et~al.}(2017){Pandya}, {Brennan}, {Somerville}, {Choi}, {Barro}, {Wuyts}, {Taylor}, {Behroozi}, {Kirkpatrick}, {Faber}, {Primack}, {Koo}, {McIntosh}, {Kocevski}, {Bell}, {Dekel}, {Fang}, {Ferguson}, {Grogin}, {Koekemoer}, {Lu}, {Mantha}, {Mobasher}, {Newman}, {Pacifici}, {Papovich}, {van der Wel}, \& {Yesuf}}]{Pandya17_reju}
{Pandya}, V., {Brennan}, R., {Somerville}, R.~S., {et~al.} 2017, \mnras, 472, 2054

\bibitem[{{Peng} {et~al.}(2015){Peng}, {Maiolino}, \& {Cochrane}}]{Peng15}
{Peng}, Y., {Maiolino}, R., \& {Cochrane}, R. 2015, \nat, 521, 192

\bibitem[{{Peng} {et~al.}(2010){Peng}, {Lilly}, {Kova{\v{c}}}, \& et~al.}]{Peng10}
{Peng}, Y.-j., {Lilly}, S.~J., {Kova{\v{c}}}, K., \& et~al. 2010, \apj, 721, 193

\bibitem[{{Pistis} {et~al.}(2022){Pistis}, {Pollo}, {Scodeggio}, {Figueira}, {Durkalec}, {Ma{\l}ek}, {Iovino}, {Vergani}, \& {Salim}}]{Pistis22_metals}
{Pistis}, F., {Pollo}, A., {Scodeggio}, M., {et~al.} 2022, \aap, 663, A162

\bibitem[{{Pizagno} {et~al.}(2007){Pizagno}, {Prada}, {Weinberg}, {Rix}, {Pogge}, {Grebel}, {Harbeck}, {Blanton}, {Brinkmann}, \& {Gunn}}]{Pizagno07_Tully-Fisher}
{Pizagno}, J., {Prada}, F., {Weinberg}, D.~H., {et~al.} 2007, \aj, 134, 945

\bibitem[{{Planck Collaboration} {et~al.}(2016){Planck Collaboration}, {Ade}, {Aghanim}, {Arnaud}, {Ashdown}, {Aumont}, {Baccigalupi}, {Banday}, {Barreiro}, {Bartlett}, \& et~al.}]{Planck15}
{Planck Collaboration}, {Ade}, P.~A.~R., {Aghanim}, N., {et~al.} 2016, \aap, 594, A13

\bibitem[{{Powell} {et~al.}(2022){Powell}, {Allen}, {Caglar}, {Cappelluti}, {Harrison}, {Irving}, {Koss}, {Mantz}, {Oh}, {Ricci}, {Shaper}, {Stern}, {Trakhtenbrot}, {Urry}, \& {Wong}}]{Powell22_mbh-mhalo}
{Powell}, M.~C., {Allen}, S.~W., {Caglar}, T., {et~al.} 2022, \apj, 938, 77

\bibitem[{{Reid} {et~al.}(2016){Reid}, {Ho}, {Padmanabhan}, \& et~al.}]{Reid16_BOSS_DR12_LSS}
{Reid}, B., {Ho}, S., {Padmanabhan}, N., \& et~al. 2016, \mnras, 455, 1553

\bibitem[{{Reid} {et~al.}(2010){Reid}, {Percival}, {Eisenstein}, {Verde}, {Spergel}, {Skibba}, {Bahcall}, {Budavari}, {Frieman}, {Fukugita}, {Gott}, {Gunn}, {Ivezi{\'c}}, {Knapp}, {Kron}, {Lupton}, {McKay}, {Meiksin}, {Nichol}, {Pope}, {Schlegel}, {Schneider}, {Stoughton}, {Strauss}, {Szalay}, {Tegmark}, {Vogeley}, {Weinberg}, {York}, \& {Zehavi}}]{Reid10_LRG_clustering_cosmology}
{Reid}, B.~A., {Percival}, W.~J., {Eisenstein}, D.~J., {et~al.} 2010, \mnras, 404, 60

\bibitem[{{Remus} \& {Kimmig}(2023)}]{Remus23_reju}
{Remus}, R.-S. \& {Kimmig}, L.~C. 2023, Submitted to ApJ, arXiv:2310.16089

\bibitem[{{Rodriguez-Gomez} {et~al.}(2022){Rodriguez-Gomez}, {Genel}, {Fall}, {Pillepich}, {Huertas-Company}, {Nelson}, {P{\'e}rez-Monta{\~n}o}, {Marinacci}, {Pakmor}, {Springel}, {Vogelsberger}, \& {Hernquist}}]{Rodriguez-Gomez22_TNG_angMom}
{Rodriguez-Gomez}, V., {Genel}, S., {Fall}, S.~M., {et~al.} 2022, \mnras, 512, 5978

\bibitem[{{Rodr{\'\i}guez-Puebla} {et~al.}(2017){Rodr{\'\i}guez-Puebla}, {Primack}, {Avila-Reese}, \& {Faber}}]{Rodriguez-Puebla17}
{Rodr{\'\i}guez-Puebla}, A., {Primack}, J.~R., {Avila-Reese}, V., \& {Faber}, S.~M. 2017, \mnras, 470, 651

\bibitem[{{Rodr{\'{\i}}guez-Torres} {et~al.}(2016){Rodr{\'{\i}}guez-Torres}, {Chuang}, {Prada}, {Guo}, {Klypin}, {Behroozi}, {Hahn}, {Comparat}, \& {et al.}}]{Rodriguez-Torres16}
{Rodr{\'{\i}}guez-Torres}, S.~A., {Chuang}, C.-H., {Prada}, F., {et~al.} 2016, \mnras, 460, 1173

\bibitem[{{Romano-D{\'\i}az} {et~al.}(2017){Romano-D{\'\i}az}, {Garaldi}, {Borzyszkowski}, \& {Porciani}}]{Romano-Diaz17_ZOMGII}
{Romano-D{\'\i}az}, E., {Garaldi}, E., {Borzyszkowski}, M., \& {Porciani}, C. 2017, \mnras, 469, 1809

\bibitem[{{Romanowsky} \& {Fall}(2012)}]{Romanowsky+Fall12_angMom}
{Romanowsky}, A.~J. \& {Fall}, S.~M. 2012, \apjs, 203, 17

\bibitem[{{Rosas-Guevara} {et~al.}(2022){Rosas-Guevara}, {Tissera}, {Lagos}, {Paillas}, \& {Padilla}}]{Rosas-Guevara22_EAGLE_voids}
{Rosas-Guevara}, Y., {Tissera}, P., {Lagos}, C. d.~P., {Paillas}, E., \& {Padilla}, N. 2022, \mnras, 517, 712

\bibitem[{{Ross} {et~al.}(2017){Ross}, {Banik}, {Avila}, {Percival}, {Dodelson}, {Garcia-Bellido}, {Crocce}, {Elvin-Poole}, {Giannantonio}, {Manera}, \& {Sevilla-Noarbe}}]{Ross17}
{Ross}, A.~J., {Banik}, N., {Avila}, S., {et~al.} 2017, \mnras, 472, 4456

\bibitem[{{Saito} {et~al.}(2016){Saito}, {Leauthaud}, {Hearin}, {Bundy}, {Zentner}, {Behroozi}, {Reid}, {Sinha}, {Coupon}, {Tinker}, {White}, \& {Schneider}}]{Saito16}
{Saito}, S., {Leauthaud}, A., {Hearin}, A.~P., {et~al.} 2016, \mnras, 460, 1457

\bibitem[{{S{\'a}nchez-Menguiano} {et~al.}(2019){S{\'a}nchez-Menguiano}, {S{\'a}nchez Almeida}, {Mu{\~n}oz-Tu{\~n}{\'o}n}, {S{\'a}nchez}, {Filho}, {Hwang}, \& {Drory}}]{Sanchez-Menguiano19}
{S{\'a}nchez-Menguiano}, L., {S{\'a}nchez Almeida}, J., {Mu{\~n}oz-Tu{\~n}{\'o}n}, C., {et~al.} 2019, \apj, 882, 9

\bibitem[{{Sancisi} {et~al.}(2008){Sancisi}, {Fraternali}, {Oosterloo}, \& {van der Hulst}}]{Sancisi08_rev}
{Sancisi}, R., {Fraternali}, F., {Oosterloo}, T., \& {van der Hulst}, T. 2008, \aapr, 15, 189

\bibitem[{{Santucho} {et~al.}(2020){Santucho}, {Luparello}, {Lares}, {Lambas}, {Ruiz}, \& {Sgr{\'o}}}]{Santucho20}
{Santucho}, V., {Luparello}, H.~E., {Lares}, M., {et~al.} 2020, \mnras, 494, 3227

\bibitem[{{Sarkar} \& {Pandey}(2020)}]{Sarkar+Pandey20}
{Sarkar}, S. \& {Pandey}, B. 2020, \mnras, 497, 4077

\bibitem[{{Sawicki} {et~al.}(2020){Sawicki}, {Arcila-Osejo}, {Golob}, {Moutard}, {Arnouts}, \& {Cheema}}]{Sawicki20_MA_obs}
{Sawicki}, M., {Arcila-Osejo}, L., {Golob}, A., {et~al.} 2020, \mnras, 494, 1366

\bibitem[{{Schaye} {et~al.}(2015){Schaye}, {Crain}, {Bower}, {Furlong}, {Schaller}, {Theuns}, {Dalla Vecchia}, {Frenk}, \& {et al.}}]{Schaye15_EAGLE}
{Schaye}, J., {Crain}, R.~A., {Bower}, R.~G., {et~al.} 2015, \mnras, 446, 521

\bibitem[{Seabold \& Perktold(2010)}]{seaborn2010statsmodels}
Seabold, S. \& Perktold, J. 2010, in 9th Python in Science Conference

\bibitem[{{Shakura} \& {Sunyaev}(1973)}]{Shakura+Sunyaev73_disc_acc}
{Shakura}, N.~I. \& {Sunyaev}, R.~A. 1973, \aap, 24, 337

\bibitem[{{Shandarin} {et~al.}(2010){Shandarin}, {Habib}, \& {Heitmann}}]{Shandarin09_cosmic_web_rev}
{Shandarin}, S., {Habib}, S., \& {Heitmann}, K. 2010, \prd, 81, 103006

\bibitem[{{Sheth} {et~al.}(2001){Sheth}, {Mo}, \& {Tormen}}]{Sheth+Tormen01_bias}
{Sheth}, R.~K., {Mo}, H.~J., \& {Tormen}, G. 2001, \mnras, 323, 1

\bibitem[{{Sheth} \& {Tormen}(2004)}]{Sheth+Tormen04_ref}
{Sheth}, R.~K. \& {Tormen}, G. 2004, \mnras, 350, 1385

\bibitem[{{Sinha} \& {Garrison}(2017)}]{Sinha+Lehman17_Corrfunc}
{Sinha}, M. \& {Garrison}, L. 2017, {Corrfunc: Blazing fast correlation functions on the CPU}

\bibitem[{{Somerville} \& {Dav{\'e}}(2015)}]{Somerville+Dave15_rev}
{Somerville}, R.~S. \& {Dav{\'e}}, R. 2015, \araa, 53, 51

\bibitem[{{Song} {et~al.}(2021){Song}, {Laigle}, {Hwang}, {Devriendt}, {Dubois}, {Kraljic}, {Pichon}, {Slyz}, \& {Smith}}]{Song21_MA_envr}
{Song}, H., {Laigle}, C., {Hwang}, H.~S., {et~al.} 2021, \mnras, 501, 4635

\bibitem[{{Spavone} {et~al.}(2021){Spavone}, {Krajnovi{\'c}}, {Emsellem}, {Iodice}, \& {den Brok}}]{Spavone21_MA_obs}
{Spavone}, M., {Krajnovi{\'c}}, D., {Emsellem}, E., {Iodice}, E., \& {den Brok}, M. 2021, \aap, 649, A161

\bibitem[{{Stoppacher} {et~al.}(2019){Stoppacher}, {Prada}, {Montero-Dorta}, {Rodr{\'{\i}}guez-Torres}, {Knebe}, {Favole}, {Cui}, {Benson}, {Behrens}, \& {Klypin}}]{Stoppacher19}
{Stoppacher}, D., {Prada}, F., {Montero-Dorta}, A.~D., {et~al.} 2019, \mnras, 486, 1316

\bibitem[{{Sullivan} {et~al.}(2017){Sullivan}, {Wiegand}, \& {Eisenstein}}]{Sullivan17}
{Sullivan}, J.~M., {Wiegand}, A., \& {Eisenstein}, D.~J. 2017, submitted, arXiv:1711.09899

\bibitem[{{Sureshkumar} {et~al.}(2021){Sureshkumar}, {Durkalec}, {Pollo}, {Bilicki}, {Loveday}, {Farrow}, {Holwerda}, {Hopkins}, {Liske}, {Pimbblet}, {Taylor}, \& {Wright}}]{Sureshkumar21_2pCF}
{Sureshkumar}, U., {Durkalec}, A., {Pollo}, A., {et~al.} 2021, \aap, 653, A35

\bibitem[{{Taylor}(2013)}]{Taylor13_Topcat}
{Taylor}, M. 2013, Starlink User Note, 253

\bibitem[{{Taylor} \& {Kobayashi}(2015)}]{Taylor+Kobayashi15_AGN}
{Taylor}, P. \& {Kobayashi}, C. 2015, \mnras, 448, 1835

\bibitem[{{Thomas} {et~al.}(2010){Thomas}, {Maraston}, {Schawinski}, {Sarzi}, \& {Silk}}]{Thomas10_alpha_envr}
{Thomas}, D., {Maraston}, C., {Schawinski}, K., {Sarzi}, M., \& {Silk}, J. 2010, MNRAS, 404, 1775

\bibitem[{{Thorne} {et~al.}(2022){Thorne}, {Robotham}, {Bellstedt}, {Davies}, {Cook}, {Cortese}, {Holwerda}, {Phillipps}, \& {Siudek}}]{Thorne22_SFH_DEVILS}
{Thorne}, J.~E., {Robotham}, A. S.~G., {Bellstedt}, S., {et~al.} 2022, \mnras, 517, 6035

\bibitem[{{Torrey} {et~al.}(2014){Torrey}, {Vogelsberger}, {Genel}, {Sijacki}, {Springel}, \& {Hernquist}}]{Torrey14}
{Torrey}, P., {Vogelsberger}, M., {Genel}, S., {et~al.} 2014, \mnras, 438, 1985

\bibitem[{{Trussler} {et~al.}(2020){Trussler}, {Maiolino}, {Maraston}, {Peng}, {Thomas}, {Goddard}, \& {Lian}}]{Trussler20}
{Trussler}, J., {Maiolino}, R., {Maraston}, C., {et~al.} 2020, \mnras, 491, 5406

\bibitem[{{van de Voort} \& {Schaye}(2012)}]{vanDeVoort+Schaye12}
{van de Voort}, F. \& {Schaye}, J. 2012, \mnras, 423, 2991

\bibitem[{{Vazdekis} {et~al.}(2010){Vazdekis}, {S{\'a}nchez-Bl{\'a}zquez}, {Falc{\'o}n-Barroso}, {Cenarro}, {Beasley}, {Cardiel}, {Gorgas}, \& {Peletier}}]{Vazdekis10_MILES}
{Vazdekis}, A., {S{\'a}nchez-Bl{\'a}zquez}, P., {Falc{\'o}n-Barroso}, J., {et~al.} 2010, \mnras, 404, 1639

\bibitem[{{Wang} \& {Lilly}(2021)}]{Wang+Lilly21_OH}
{Wang}, E. \& {Lilly}, S.~J. 2021, \apj, 910, 137

\bibitem[{Wang {et~al.}(2018)Wang, Norberg, Brough, Brown, da~Cunha, Davies, Driver, Holwerda, Hopkins, Lara-Lopez, \& et~al.}]{WangL18}
Wang, L., Norberg, P., Brough, S., {et~al.} 2018, \aap, 618, A1

\bibitem[{{Wechsler} \& {Tinker}(2018)}]{Wechsler+Tinker18_rev}
{Wechsler}, R.~H. \& {Tinker}, J.~L. 2018, \araa, 56, 435

\bibitem[{{Wechsler} {et~al.}(2006){Wechsler}, {Zentner}, {Bullock}, {Kravtsov}, \& {Allgood}}]{Wechsler06_ref}
{Wechsler}, R.~H., {Zentner}, A.~R., {Bullock}, J.~S., {Kravtsov}, A.~V., \& {Allgood}, B. 2006, \apj, 652, 71

\bibitem[{{Weinmann} {et~al.}(2006){Weinmann}, {van den Bosch}, {Yang}, \& {Mo}}]{Weinmann06_SDSS_groups}
{Weinmann}, S.~M., {van den Bosch}, F.~C., {Yang}, X., \& {Mo}, H.~J. 2006, \mnras, 366, 2

\bibitem[{{White} \& {Frenk}(1991)}]{White+Frenk91}
{White}, S.~D.~M. \& {Frenk}, C.~S. 1991, \apj, 379, 52

\bibitem[{{Wotta} {et~al.}(2019){Wotta}, {Lehner}, {Howk}, {O'Meara}, {Oppenheimer}, \& {Cooksey}}]{Wotta19_metals}
{Wotta}, C.~B., {Lehner}, N., {Howk}, J.~C., {et~al.} 2019, \apj, 872, 81

\bibitem[{{Wyithe} \& {Padmanabhan}(2006)}]{Wyithe+Padmanabhan06_bheff}
{Wyithe}, J. S.~B. \& {Padmanabhan}, T. 2006, \mnras, 366, 1029

\bibitem[{{Yang} {et~al.}(2006){Yang}, {Mo}, \& {van den Bosch}}]{Yang06_ref}
{Yang}, X., {Mo}, H.~J., \& {van den Bosch}, F.~C. 2006, \apjl, 638, L55

\bibitem[{{Yates} {et~al.}(2012){Yates}, {Kauffmann}, \& {Guo}}]{Yates12}
{Yates}, R.~M., {Kauffmann}, G., \& {Guo}, Q. 2012, \mnras, 422, 215

\bibitem[{{Zehavi} {et~al.}(2005){Zehavi}, {Zheng}, {Weinberg}, {Frieman}, {Berlind}, {Blanton}, {Scoccimarro}, \& {SDSS Collaboration}}]{Zehavi05}
{Zehavi}, I., {Zheng}, Z., {Weinberg}, D.~H., {et~al.} 2005, \apj, 630, 1

\bibitem[{{Zhai} {et~al.}(2023){Zhai}, {Tinker}, {Banerjee}, {DeRose}, {Guo}, {Mao}, {McLaughlin}, {Storey-Fisher}, \& {Wechsler}}]{Zhai23_Aemulus}
{Zhai}, Z., {Tinker}, J.~L., {Banerjee}, A., {et~al.} 2023, \apj, 948, 99

\bibitem[{{Zhang} {et~al.}(2023){Zhang}, {Li}, {Leja}, {Whitaker}, {Nersesian}, {Bezanson}, \& {van der Wel}}]{Zhang23_reju}
{Zhang}, J., {Li}, Y., {Leja}, J., {et~al.} 2023, \apj, 952, 6

\bibitem[{{Zwicky} {et~al.}(1961){Zwicky}, {Herzog}, {Wild}, {Karpowicz}, \& {Kowal}}]{Zwicky61}
{Zwicky}, F., {Herzog}, E., {Wild}, P., {Karpowicz}, M., \& {Kowal}, C.~T. 1961, {Catalogue of galaxies and of clusters of galaxies, Vol. I}

\end{thebibliography}

\begin{appendix}

\section{Information on the adopted galaxy catalogue: \MDgal}\label{app:SAM}

As \galacticus\ is primarily described in \citet{Benson12}, with additional features outlined in \citet[Sec.~2.2 and Tab.~1][]{Knebe17_MD} as part of the \MDG\ catalogues, we summarise only its aspects most relevant to this work. The \galacticus\ semi-analytic model incorporates a stellar population synthesis model from \citet{Conroy09}, an initial mass function from \citet{Chabrier03}, and a dust model of \citet{Ferrara99}. The definition of the dark matter halo mass is given by:

\vspace{-0.4cm}\begin{equation}
 M_{\rm ref}(<R_{\rm ref}) = \Delta_{\rm ref} \rho_{\rm c} \frac{4\pi}{3} R_{\rm ref}^3, \label{eq:mass_def}
\end{equation} \vspace{-0.4cm}

\noindent where $\Delta_{\rm ref}=\Delta_{\rm BN98}$ for \MBN\ with $\Delta_{\rm BN98}$ being the virial factor as given by the Eq. (6) of \citet{Bryan+Norman98}; $\rho_{\rm c}$ denotes the critical density of the Universe, and $R_{\rm ref}$ is the corresponding halo radius at which the interior mean density matches the desired value as specified on the right-hand side of \hyperref[eq:mass_def]{\Eq{eq:mass_def}}. We note that \rockstar\ offers various halo mass definitions; however, this version of \galacticus\ was run using this one.

The parameters for galaxy formation physics in \galacticus\ were determined through a manual search of the parameter space, aiming to match a variety of observational data. These include the $z=0$ stellar mass function of galaxies \citep{Li+White09_SMF}, $z=0$ K and b$_{\rm J}$-band luminosity functions \citep{Cole01_2dF,Norberg02}, the local Tully-Fisher relation \citep{Pizagno07_Tully-Fisher}, the colour-magnitude distribution of galaxies in the local Universe \citep{Weinmann06_SDSS_groups}, the distribution of disc sizes at $z=0$ \citep{deJong+Lacey00_SB-SDM}, the black hole mass to bulge mass relation \citep{Haring+Rix2004}, and the star formation history of the Universe \citep{Hopkins14}. 

This version of \galacticus\ employs a simple accretion model where gas accretes from the intergalactic medium onto the dark matter halo, as outlined by \citet[Eq.~(35) in][]{Benson02}. Cooling rates from the hot halo are computed using the traditional cooling radius approach from \citet{White+Frenk91}. Metallicity-dependent cooling curves are calculated with \texttt{CLOUDY} \citep[v13.01,][]{Ferland13}. The disc is modelled as either a radiatively efficient, geometrically thin, \citet{Shakura+Sunyaev73_disc_acc}-type disc (if the accretion rate being between 0.01 and 0.3 $\dot{M_{\rm Edd}}$, where $\dot{M_{\rm Edd}}$ being the Eddington accretion rate) or otherwise as an advection-dominated flow-accretion thick disc, following \citep{Begelman14_BHs}. The model dynamically switches between these two modes.

Star formation is modelled using the prescription of \citet[][i.e. their Eq.~(1) for the star formation rate surface density, and Eq.~(2) for the molecular fraction]{Krumholz09}, assuming that the cold gas of each galaxy follows an exponential radial distribution. The scale length of this distribution is determined from the disc's angular momentum by solving for the equilibrium radius within the gravitational potential of the disc+bulge+dark matter halo system \citep[][]{Gnedin04}. Metal enrichment is tracked using the instantaneous recycling approximation, with a recycled fraction of 0.46 and yield of 0.035. Metals are assumed to be fully mixed in all phases, thereby tracing all mass flows between phases.

The supernova feedback is implemented using a wind mass loading factor, $\beta$, computed as $\beta = (V_{\rm disc}/250 {\rm km/s})^{-3.5}$ where $V_{\rm disc}$ is the circular velocity at the disc's scale radius. Gas expelled from the galaxy by winds is retained in a reservoir for outflowed gas, which gradually leaks mass back into the hot halo on a timescale of $t_{\rm dyn}/5$, where $t_{\rm dyn}$ is the dynamical time of the halo at the virial radius. Material is transferred from the disc to the spheroid on an instability timescale, which is defined by an instability parameter as described in \citet{Efstathiou82} \citep[see also Eq.~(1) in][]{Knebe17_MD}.

\galacticus\ does not include a specific starburst mode. Instead, star formation in the spheroid occurs at a rate $\dot{M}_\star = 0.04 M_{\rm gas}/t_{\rm dyn} (V/200 {\rm km/s})^{-2}$, where $t_{\rm dyn}$ is the dynamical time of the spheroid at its half mass radius, and $V$ its circular velocity at the same radius. The model tracks the mass and spin of black holes in detail, assuming an initial seed mass of 100 \Msun. AGN feedback is incorporated in both the ``radio'' mode \citep[see][]{Benson+Bower10} and the ``quasar'' mode \citep[see][]{Ostriker10_eq-model} with a black hole wind efficiency of 0.0024 being implemented. The model uses the standard spheroid implementation adopting a spheroid density profile (Hernquist-profile \citep{Hernquist90}), which is described by a single-length scale where stars trace the gas density.

If the (baryonic) mass ratio of two merging galaxies exceeds 1:4, a ``major'' merger is assumed. In this scenario, the merging galaxies are transformed into a spheroidal remnant. Otherwise, a ''minor`` merger occurs, where the less massive galaxy is incorporated into the spheroid of the more massive galaxy, leaving the disc of the larger galaxy unaffected.

\section{Tracing modelled galaxies across cosmic history}\label{app:M2}

\begin{figure}
	\centering
	\vspace{-0.3cm}\includegraphics[angle=-90,width=0.9\columnwidth]{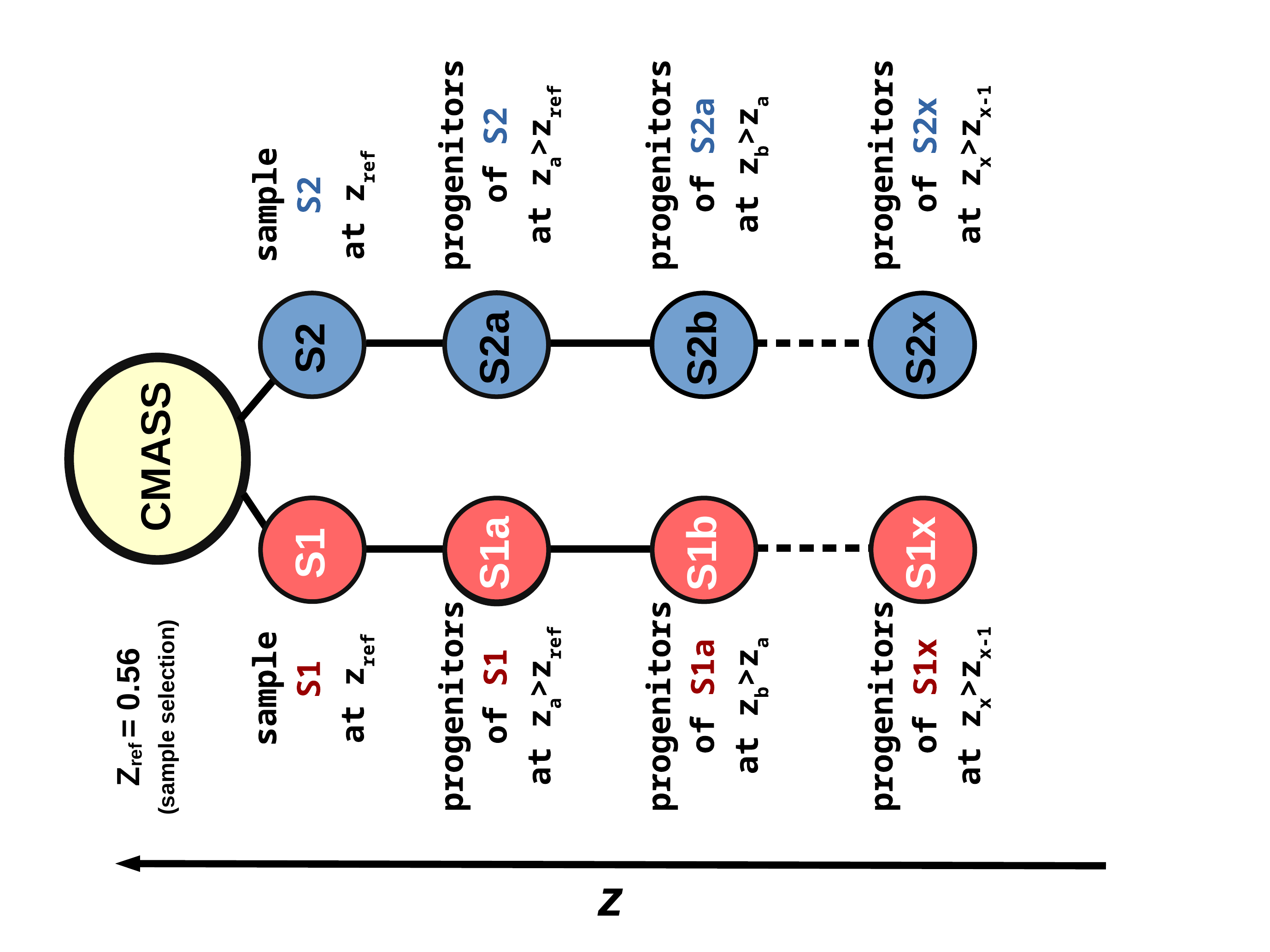}\vspace{-1.3cm}
	\caption{Schematic representation of the selection procedure and tracking of progenitor halos: The selection criteria are applied only once at the $\zstart=0.56$, resulting in the S1 and S2 from the entire population of CMASS mock-galaxies catalogue. The progenitors of these samples are then identified at the subsequent simulation snapshot using their unique identification number, resulting in the progenitor sub-samples S1a corresponding to S1-sample and S2a to S2-sample. This step is repeated and we subsequently move towards higher redshifts (S1x and S2x). This process is repeated as we progress to higher redshifts, generating S1x and S2x. This approach represents the conventional method for extracting information on the redshift evolution of galaxies in simulations.}\label{fig:cartoon_M2}\vspace{-0.4cm}%
\end{figure}

\hyperref[fig:cartoon_M2]{\Fig{fig:cartoon_M2}} provides a schematic representation of the selection procedure and tracking of progenitor halos. This is achieved by using the unique identification numbers -- \texttt{parentIndex} -- of central dark matter halos hosting the galaxy of interest to trace their main progenitor halos on their merger trees back in time. This information is provided by the halo finder and the corresponding tree builder algorithm. By definition, the merger tree main branch of the algorithm applied to the data used in this study (e.g.\ \rockstar\ and \consistenttree), is the most massive progenitor branch and can be traced on the leftmost side of each sub-tree -- the lowest \texttt{mainLeafId} -- in the friend-of-friends (FoF) groups. That means in practice that the main progenitor halo, the one of interest, always has the lowest ID (denoted by the \texttt{satelliteNodeIndex}) for the same \texttt{parentIndex} and is also the most massive progenitor. That is particularly useful given that the \MDgal\ catalogue includes millions of galaxies and satellite galaxies which need to be filtered and traced to find the specific one of interest. It is important to note that while the \galacticus\ model uses the same IDs, it employs a different naming convention that may not be immediately recognisable to those familiar with \rockstar\ terminology. For additional information, readers are referred to the \textsc{Cosmosim}-database.

It is noted that while it may seem excessive to dedicate an additional figure solely to the tracking of progenitor halos, the companion paper in preparation will demonstrate that the sub-sample selection procedure does not necessarily need to be conducted at a specific redshift. This flexibility will lead to a different approach for target selection.

\section{Illustration halo mass dependency on environment}\label{app:zcolds_mhalo}

\begin{figure}
	\centering
	\includegraphics[width=\columnwidth,angle=0]{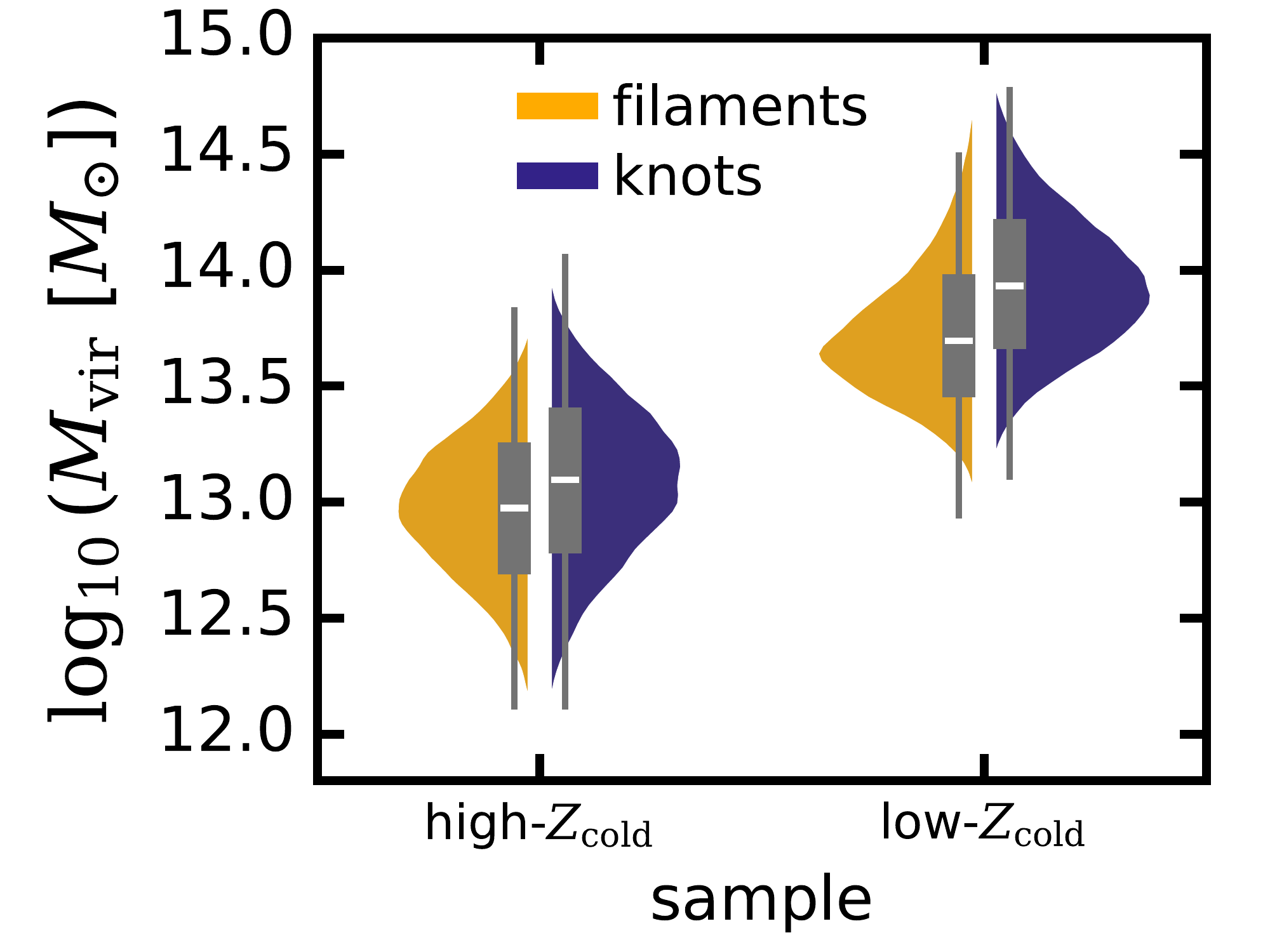}\vspace{-0.2cm}
	\caption{The violin plot visualises the distribution of halo mass (\Mvir) for the sample, with the \highZ\ galaxies shown on the left and the \lowZ\ galaxies on the right. Each violin displays the distribution of filament and knot galaxies: the left side of each violin represents the filament galaxy distribution, while the right side shows the knot galaxy distribution. We also overlay box plots in grey, with the median values marked in white. This visual representation confirms the results of our Kolmogorov-Smirnov (KS) test: filament and knot galaxies in the \highZ\ sub-sample are statistically closer, while filament and knot galaxies in the \lowZ\ exhibit greater differences.}\label{fig:zcolds_mhalo}
\end{figure}

In this appendix, we illustrate and assess the statistical similarity of galaxies in the \lowZ\ and \highZ\ sub-samples based on their large-scale environments (filaments or knots). As shown in \hyperref[fig:res_zcolds_xi]{\Fig{fig:res_zcolds_xi}} \highZ\ filament and knot galaxies exhibit distinct clustering functions, even though the median halo masses for these galaxies differ by only 0.13 dex in \Msun. Conversely, the median halo masses for \lowZ\ filament and knot galaxies are separated by 0.23 dex in \Msun, yet they show similar clustering functions in both shape and amplitude at $z=0.56$. In \hyperref[fig:zcolds_mhalo]{\Fig{fig:zcolds_mhalo}}, we present violin plots showing the distribution of halo mass (\Mvir) for \highZ\ and \lowZ\ galaxies, with overlaid box plots in grey. This plot visually confirms that the knot population (displayed in blue by the left half of the violin) and the filament population (shown in yellow by the right half of the violin) behave consistently with the findings from the clustering analysis.

Their median values are more spread out, closer to the edges of the interquartile range. \highZ\ galaxies, on the other hand, have median values that are more closely aligned. Although the median halo masses of \highZ\ galaxies in both knots and filaments are similar, their clustering behaviour differs (\hyperref[fig:res_zcolds_xi]{\Fig{fig:res_zcolds_xi}}). Conversely, \lowZ\ galaxies cluster similarly in both knots and filaments, despite their statistical distributions differing. To support these observations, we conducted a Kolmogorov-Smirnov (KS) test using the Python package \texttt{SciPy.stats} to compare the filament and knot populations in both sub-samples.

The KS-test is a robust method to assess whether their distributions are statistically similar. The KS statistic outputs a value between 0 and 1 which represents the maximum distance between the cumulative distribution functions (CDFs) of the compared samples. KS statistic values closer to 0 indicate that the distributions are more similar, while values closer to 1 suggest greater divergence between the distributions. For the \highZ\ sample, the KS-test returned a value of 0.15 (with a p-value $< 0.05$\footnote{The p-value indicates the strength of evidence against the null hypothesis, with a lower p-value suggesting that the observed results are less likely to have occurred by chance, thereby lending more credibility to the obtained statistics. Typically, p-values smaller than 0.05 are considered to indicate statistical significance.}) when comparing filament and knot galaxies. This result indicates that the two distributions are relatively similar, with a moderate difference of 0.15 between their CDFs. In the \lowZ\ sub-sample, the KS-test returned a value of 0.3, double that of the \highZ\ sample. This larger value signifies a greater discrepancy between the distributions, meaning they are less similar. We also conducted similar tests in narrow halo mass bins and found consistent results.

\end{appendix}

\end{document}